\documentclass{emulateapj}
\usepackage{apjfonts}
\usepackage{amssymb}
\usepackage{amsmath}

\newcommand{\scaleup}{\epsscale{1.1}}

\newcommand{\plotter}{\plotone}
\newcommand{\plotterr}{\plotone}

\newcommand{\breaker}{}

\newcommand{\etal}{et al.}

\newcommand{\msun}{M_{\sun}}

\newcommand{\fgas}{f_{\rm gas}}

\newcommand{\MPAurl}{{\url{http://www.mpa-garching.mpg.de/millennium/}}}
\newcommand{\Durhamurl}{{\url{http://icc.dur.ac.uk/index.php?content=Millennium/Millennium}}}
\newcommand{\KBurl}{{\url{http://www.astro.utoronto.ca/~bundy/millennium/}}}
\newcommand{\mergercalcurl}{\scriptsize{\path{http://www.cfa.harvard.edu/~phopkins/Site/mergercalc.html}}}

\shorttitle{Uncertainties in Predicted Merger Rates}
\shortauthors{Hopkins \etal}
\slugcomment{Accepted to ApJ, June, 2010}
\begin{document}

\title{Mergers in $\Lambda$CDM: Uncertainties in 
Theoretical Predictions and Interpretations of the Merger Rate}
\author{ 
Philip F.\ Hopkins\altaffilmark{1}, 
Darren Croton\altaffilmark{2},
Kevin Bundy\altaffilmark{1},
Sadegh Khochfar\altaffilmark{3},
Frank van den Bosch\altaffilmark{4},
Rachel S.\ Somerville\altaffilmark{5,6},
Andrew Wetzel\altaffilmark{1},
Dusan Keres\altaffilmark{7},
Lars Hernquist\altaffilmark{3},
Kyle Stewart\altaffilmark{8},
Joshua D.\ Younger\altaffilmark{9},
Shy Genel\altaffilmark{3},
\&\ Chung-Pei Ma\altaffilmark{1}
}
\altaffiltext{1}{Department of Astronomy, University of California 
Berkeley, Berkeley, CA 94720, USA} 
\altaffiltext{2}{Centre for Astrophysics \&\ Supercomputing, Swinburne University 
of Technology, P.O.\ Box 218, Hawthorn, VIC 3122, Australia} 
\altaffiltext{3}{Max Planck Institut 
f{\"u}r Extraterrestrische Physik, Giessenbachstr., D-85748, Garching, Germany}
\altaffiltext{4}{Department of Physics and Astronomy, 
University of Utah, 115 South 1400 East, Salt Lake City, UT 84112-0830, USA}
\altaffiltext{5}{Space Telescope
  Science Institute, 3700 San Martin Dr., Baltimore, MD 21218, USA}
\altaffiltext{6}{Department of Physics and Astronomy, Johns Hopkins
  University, Baltimore, MD 21218, USA}
\altaffiltext{7}{Harvard-Smithsonian Center for Astrophysics, 60
  Garden Street, Cambridge, MA 02138, USA} 
\altaffiltext{8}{NASA Postdoctoral Program Fellow, Jet Propulsion Laboratory, 
Pasadena, CA 91109, USA}
\altaffiltext{9}{Hubble Fellow, 
Institute for Advanced Study, Einstein Drive, Princeton, 
NJ 08540, USA}

\begin{abstract}

Different theoretical methodologies lead to order-of-magnitude variations in 
predicted galaxy-galaxy merger rates. We examine how this arises, and quantify  
the dominant uncertainties. Modeling of dark matter and 
galaxy inspiral/merger times 
contribute factor $\sim2$ uncertainties. 
Different estimates of the halo-halo merger rate, the subhalo ``destruction'' 
rate, and the halo merger rate with some dynamical friction time delay 
for galaxy-galaxy mergers, agree to within this factor $\sim2$, provided proper 
care is taken to define mergers consistently. 
There are some caveats: if halo/subhalo 
masses are not appropriately defined the 
major merger rate can be dramatically suppressed, 
and in models with ``orphan'' galaxies and under-resolved subhalos 
the merger timescale can be severely over-estimated. 
The dominant differences in galaxy-galaxy 
merger rates between models owe to the treatment of the baryonic physics. 
Cosmological hydrodynamic simulations without strong feedback 
and some older semi-analytic models, with known discrepancies in mass functions, 
can be biased by large factors ($\sim5$) in predicted merger rates. 
However, provided that models yield a reasonable match to the total galaxy mass function, 
the differences in properties of {\em central} galaxies are sufficiently small 
to alone contribute small (factor $\sim1.5$) additional systematics to merger rate 
predictions. 
But variations in the baryonic physics of {\em satellite} galaxies
in models can also have a dramatic effect on merger rates. 
The well-known problem of satellite ``over-quenching'' in most current 
semi-analytic models (SAMs) -- whereby 
SAM satellite populations are too efficiently stripped of their gas -- could lead to 
order of magnitude under-estimates of merger rates for low-mass, gas-rich 
galaxies. Models in which the masses of satellites are fixed by observations 
(or SAMs adjusted to resolve this ``over-quenching'') 
tend to predict higher merger rates, but with  
factor $\sim2$ uncertainties stemming from the uncertainty in those observations.
The choice of mass used to defined ``major'' and ``minor'' mergers also matters: 
stellar-stellar major mergers 
can be more or less abundant than halo-halo major mergers by an order of 
magnitude. 
At low masses, most true major mergers (mass ratio defined in terms of their 
baryonic or dynamical mass) will appear to be minor mergers in 
their stellar mass ratio - observations and models using just stellar criteria 
could underestimate major merger rates by factors $\sim3-5$. 
We discuss the uncertainties in relating any merger rate to spheroid formation (in 
observations or theory): 
in order to achieve better than factor $\sim3$ accuracy, 
it is necessary to account for 
the distribution of merger orbital parameters, gas fractions, 
and the full efficiency of merger-induced effects as a function 
of mass ratio. 
\end{abstract}

\keywords{galaxies: formation --- galaxies: evolution --- galaxies: active --- cosmology: theory}

\section{Introduction}
\label{sec:intro}

In the now established $\Lambda$CDM 
cosmology, structure grows hierarchically \citep[e.g.][]{whiterees78}, making 
mergers an inescapable element in galaxy formation. 
The galaxy-galaxy merger rate, as a function of properties 
such as galaxy mass, redshift, and mass ratio, is 
a quantity of fundamental interest. It is critical for
informing models of the growth and assembly of galaxies, 
the distribution of bulge and spheroid mass, 
properties of disks such as their thickness, morphologies, 
and flaring, and the fueling and 
growth of the most luminous starbursts and infrared systems 
as well as massive BHs and quasars. 

The last few years have seen the emergence of 
a ``concordance'' precision $\Lambda$CDM cosmology
\citep[see e.g.][and references therein]{komatsu:wmap5}, with 
remarkable convergence between different probes of 
structure formation. 
Meanwhile, cosmological dark-matter simulations have 
developed the ability to track large populations of dark matter 
halos and substructure within halos over 
cosmological volumes and timescales. Different calculations 
now yield results for e.g.\ the dark matter halo mass function 
that agree at the $\sim5-10\%$ level 
\citep[see e.g.][and references therein]{reed:halo.mfs}. 

Despite these advances, however, there has yet to be a similar 
convergence in theoretical predictions of the {\it galaxy} merger rate. 
Depending on the models used to convert dark matter 
halo growth histories into 
galaxy growth histories, 
different theoretical predictors of the merger rate in the same 
cosmology can differ by more than an order of magnitude 
\citep[see e.g.\ the comparison in][]{jogee:merger.density.08.conf.proc,
bertone:merger.rate.vs.stellar.wind.model,lopezsanjuan:merger.fraction.to.z1}. 
And the differences cannot be trivially attributed to 
e.g.\ the known fact that models predict somewhat different absolute 
galaxy masses and abundances -- in terms of mergers {\em per galaxy}, 
there is similar variation. 

Some of this may owe to the fact that dark matter merger rates 
are -- unlike dark matter mass functions -- derivative 
(instantaneous) quantities and so are more sensitive to 
choices of definition and methodology in simulations. But 
much of the difference owes to the fact that the mapping from 
dark matter merger rates to galaxy merger rates is non-trivial, and 
requires some relation between the history of a given dark matter halo and 
a galaxy inside that halo. 
In models that attempt to predict galaxy formation 
in an {\em a priori} sense, from physical 
prescriptions for gas cooling, star formation, and feedback from 
stars and accreting BHs, the resulting predicted halo to 
galaxy ``map'' and corresponding merger 
rates can be very sensitive to the input prescriptions. 
It has been shown that even in otherwise identical models, small 
changes in the treatment of gas cooling or stellar wind physics 
can lead to order-of-magnitude changes in the predicted 
major merger rates: compare the merger rates in 
\citet{bower:sam} and \citet{font:durham.sam.update} 
(our \S~\ref{sec:compare.hods:satellites:resolution}), 
or those in 
\citet{delucia:sam} and \citet{bertone:stellar.wind.recycling.sam} 
\citep[presented in][]{bertone:merger.rate.vs.stellar.wind.model}.

These quantitative differences in the number of mergers per galaxy 
have led to qualitatively different conclusions regarding galaxy formation. 
Most models, especially those based on empirical halo occupation 
constraints or direct cosmological hydrodynamic simulation, 
have found that there are too many mergers in low mass systems to explain the survival 
and prevalence of disks in the simplest scenario 
\citep{granato:sam,somerville:sam,somerville:new.sam,
koda:disk.survival.prescriptions,stewart:merger.rates,khochfarsilk:new.sam.dry.mergers,
hopkins:disk.survival.cosmo,stewart:disk.survival.vs.mergerrates,
sommerlarsen03:disk.sne.fb,abadi03:disk.structure,governato:disk.formation,
robertson:cosmological.disk.formation,
okamoto:feedback.vs.disk.morphology,scannapieco:fb.disk.sims,
kazantzidis:mw.merger.hist.sim,kazantzidis:thin.disk.thickening,
purcell:minor.merger.thindisk.destruction}. 

More directly, {\em observed} merger rates 
find that $\sim5-10\%$ of low or intermediate-mass 
galaxies ($M_{\ast}\lesssim10^{10}\,\msun$ at redshifts $z\sim0.2-1.2$) are in mergers -- 
strongly morphologically disturbed or in 
``major'' similar-mass pairs about to merge (at small scales and small relative velocities)
\citep{bridge:merger.fractions,bridge:merger.fraction.new,kartaltepe:pair.fractions,
conselice:mgr.pairs.updated,
jogee:merger.density.08,lotz:merger.fraction,lin:mergers.by.type,
robaina:mgrs.make.es.sufficient}. 
This is equal to or higher than the predicted rates from the models above. 
Even without correcting for the short expected 
duty cycles/lifetimes of this merger phase \citep[see e.g.][]{
lotz:merger.selection,conselice:mgr.pairs.updated}, these fractions are {\em already} 
higher than the observed fraction of bulge-dominated galaxies at these 
masses \citep{allen:bulge-disk,
benson:disk.spheroid.mfs,
dominquez:bulge.disk.sample,weinzirl:b.t.dist}. 
Likewise the number density of IR luminous systems apparently in mergers 
at high redshifts is so high as to yield tension with the relic bulge mass 
in the Universe \citep[see e.g.][]{younger:highz.smgs,
genel:smg.numden.vs.mergers,tacconi:smg.mgr.lifetime.to.quiescent,
hopkins:merger.lfs,cimatti:highz.compact.gal.smgs}. 
Typically, some additional physics (often some form of strong stellar feedback)
must be invoked to preserve high gas fractions and suppress 
the efficiency of bulge formation. 
Indeed, this is {\em always} the case 
in hydrodynamic simulations of galaxy formation. 
This has led to the development of various feedback models 
\citep[e.g.][and references therein]{sommerlarsen03:disk.sne.fb,abadi03:disk.structure,
governato:disk.formation,okamoto:feedback.vs.disk.morphology,scannapieco:fb.disk.sims} 
and emphasis on gas-richness as a stabilizing factor 
\citep{robertson:disk.formation,hopkins:disk.survival,
stewart:disk.survival.vs.mergerrates,moster:2010.thin.disk.heating.vs.gas} 
that allows disks to survive such mergers. 
However, some semi-analytic calculations have claimed the opposite: that 
there are not enough mergers at low masses to explain even 
the small bulges observed at these masses 
\citep{parry:sam.merger.vs.morph}.

On the other hand, at $\gtrsim L_{\ast}$, most empirical and semi-analytic 
models predict similar numbers of mergers and agree that 
massive bulges are primarily formed in major mergers 
\citep[see e.g.][and references therein]{hopkins:merger.rates,
parry:sam.merger.vs.morph,cattaneo:new.sam.w.mergerfx}. 
But some simulations have claimed that there 
are more minor and fewer major mergers at these 
high masses \citep{naab:etg.formation}. 

As observations of e.g.\ pair and morphologically disturbed 
fractions improve, it is rapidly becoming possible to empirically 
constrain merger rates and fractions at a level better than the apparent
order-of-magnitude scatter in predictions \citep[see 
e.g.][]{kartaltepe:pair.fractions,lin:mergers.by.type,domingue:2mass.merger.mf,
bundy:merger.fraction.new,conselice:mgr.pairs.updated,
robaina:2009.sf.fraction.from.mergers,
robaina:mgrs.make.es.sufficient,comerford:2009.bh.binaries,
lotz:merger.fraction,bridge:merger.fractions,lopezsanjuan:mgr.rate.pairs}.
However, the conversion of such 
observations to merger rates still requires knowledge of appropriate 
timescales and contamination from 
projection effects and/or minor mergers -- all of which mean that 
interpretation of the observations is somewhat dependent on 
how well or poorly theoretical predictions agree on basic quantities. 
Moreover, an observed pair or morphologically disturbed fraction is 
fundamentally {\em not} invertible at much better than the factor of 
$\sim2$ level, as disturbances and dynamics are non-unique 
and depend on other quantities such as galaxy gas content; as such, 
they must be forward-modeled.

Fundamentally, the galaxy-galaxy merger rate depends on 
two basic quantities: the dark matter halo merger rate 
and the manner in which galaxies populate halos.
However, models arrive at these quantities by various methods. 
In this paper, we attempt to examine in detail different 
theoretical models and identify the sources of apparently large variations
in galaxy merger rates between the models. We are interested both in defining the 
``error budget'' of theoretical predictions, as well as in identifying 
the definitions, physics, and prescriptions that lead to 
substantial differences. 

In the follow sections we show that combining empirical halo occupation constraints 
with dark matter merger histories from high-resolution simulations leads to 
a relatively small (factor $\sim2$)
systematic uncertainty in merger rates (\S~\ref{sec:empirical}). 
We then compare various different substructure based calculations for 
subhalo-subhalo mergers and/or merger time delays (\S~\ref{sec:compare.dark}). 
We highlight several caveats and potential 
problems in adopting these merger rates: without proper care, it is possible to 
obtain apparently order-of-magnitude different dark matter merger rates 
from the same simulation or analytic calculation.

We reveal that most of the differences 
between model predictions owe to the treatment of 
baryons in galaxy formation (\S~\ref{sec:compare.hods}). Specifically, we 
show that cosmological hydrodynamic merger simulations lacking 
feedback do not reproduce the observed 
$M_{\rm gal}(M_{\rm halo})$ relation (not just the normalization of 
this relation, but also its shape) and so do not map robustly between 
halo-halo and galaxy-galaxy mergers 
(\S~\ref{sec:compare.hods:central}). Semi-analytic models, on the other hand, 
perform well for central galaxies, but commonly have 
well-known problems reproducing observed properties of 
satellite galaxies; this accounts for most of the 
discrepancies both between different semi-analytic models 
and between those models and empirical 
approaches/calculations (\S~\ref{sec:compare.hods:satellites}). 
In \S~\ref{sec:definitions}, we discuss the importance of consistent and 
appropriate mass ratio definitions, and the variations between different 
definitions, as well as the range in the effects of a given merger 
assuming changes in
galaxy gas fractions, orbital parameters, and 
structural properties. 
We summarize our conclusions and discuss future 
tests and improvements in \S~\ref{sec:discuss}. 

Except where otherwise specified, we adopt a ``concordance'' cosmology with 
$(\Omega_{\rm M},\,\Omega_{\Lambda},\,h,\,\sigma_{8},\,n_{s})$=
$(0.3,\,0.7,\,0.7,\,0.9,\,1.0)$ 
and a \citet{chabrier:imf} IMF, and appropriately
normalize all observations and models shown.
The choice of IMF
systematically shifts the normalization of stellar masses herein, but
does not otherwise change our comparisons. 

Throughout, we use the notation $M_{\rm gal}$ to denote the baryonic 
(stellar+cold gas) mass of galaxies; the stellar, cold gas, and dark 
matter halo masses 
are denoted $M_{\ast}$, $M_{\rm gas}$, and $M_{\rm halo}$, respectively. 
When we refer to merger mass ratios, we use the same subscripts 
to denote the relevant masses used to define a mass ratio 
(e.g.\ $\mu_{\rm gal} = M_{\rm gal,\, 2}/M_{\rm gal,\,1}$), 
always taken such that $0<\mu<1$ ($M_{\rm gal,\,1}>M_{\rm gal,\,2}$).

\breaker
\section{Overview: Comparing Model Merger Rates}
\label{sec:empirical}

The galaxy merger rate is fundamentally determined by two 
quantities: the halo/subhalo merger rate, and the manner in which galaxies populate 
halos (the halo occupation function). The halo merger rate (the precise 
definition of which will be discussed in \S~\ref{sec:compare.dark:halos}) is
determined in cosmological simulations; there are two basic choices, 
however, for the halo occupation function. Semi-analytic models 
and cosmological hydrodynamic simulations attempt to predict this 
from first principles. Empirical models based on the halo occupation 
formalism simply adopt this quantity based on observational 
constraints.

\subsection{Methodologies}
\label{sec:empirical:methodology}

Semi-analytic models and direct simulations 
are a well known and traditional means of modeling 
galaxy formation. In cosmological hydrodynamic simulations, 
galaxy formation is simulated from 
cosmological initial conditions. Cooling is computed self-consistently, 
but some prescriptions must be adopted for star formation 
(as a function of local gas properties) and 
feedback from star formation and stellar evolution as 
well as black hole (BH) and active galactic nuclei (AGN) fueling, growth and feedback. 
It is well-known however that, in the absence of detailed prescriptions for 
both stellar and AGN feedback, such models do not reproduce 
the observed correlations between galaxy and halo mass -- they 
overpredict the masses of both low and high mass galaxies, 
by a mass-dependent factor \citep[i.e.\ predict a different magnitude and 
shape of $M_{\rm gal}(M_{\rm halo})$; see][]{keres:fb.constraints.from.cosmo.sims,
maller:sph.merger.rates,oppenheimer:recycled.wind.accretion,choi:2009.mf.vs.metalcooling}. As a 
consequence the most successful 
versions of such models have considered high-resolution 
``zoom-in'' models of individual galaxies in formation, 
with extensive feedback prescriptions \citep[e.g.][]{sommerlarsen03:disk.sne.fb,
governato:disk.formation,
okamoto:feedback.vs.disk.morphology,scannapieco:fb.disk.sims}. 
Unfortunately, it remains prohibitively expensive to simulate 
cosmological populations of galaxies 
(needed for the statistics to quantify galaxy-galaxy merger 
rates as a function of mass, mass ratio, redshift, and other properties) 
with the resolution and physical prescriptions necessary 
to follow these feedback and formation models, let alone to widely vary these 
prescriptions. And even many strong-feedback 
simulations overpredict the baryon conversion efficiency 
\citep{guo:2010.hod.constraints,abadi03:disk.structure,
okamoto:feedback.vs.disk.morphology,governato:disk.formation,
scannapieco:fb.disk.sims,piontek:feedback.vs.disk.form.sims}. 

Given this limitation, a more common approach 
to modeling galaxy formation in an {\em a priori} fashion has 
been the construction of semi-analytic models. In these models, 
galaxy formation prescriptions are ``painted onto'' a dark matter 
background. Dark matter halo locations, growth histories, merger rates, 
and other properties are taken from dark-matter only 
cosmological simulations and/or analytic extended Press-Schechter 
theory designed to match those simulations. 
Galaxies are analytically assigned to each halo from the simulation, 
with their evolution followed in a Monte Carlo fashion according to a 
simple set of analytic prescriptions, designed to approximate the 
processes of cooling from halo gas, star formation, and stellar and 
AGN feedback. Such approaches have the advantage that they are 
inexpensive, allowing experimentation with a wide variety of 
formulations for e.g.\ feedback (with various free parameters), constrained  
so as to reproduce observable quantities such as the galaxy 
mass function at $z=0$. However, the prescriptions on which the 
results depend are approximations that usually must be applied to a 
variety of regimes well outside the range where they are 
observationally or physically (from e.g.\ direct simulations) calibrated. 
The results in certain regimes can be unexpected or, in some cases, 
unphysical. In addition, it is well known that there are many degeneracies 
between small changes to different prescriptions -- there is 
no unique solution to reproducing e.g.\ the $z=0$ stellar mass function, 
so the models are not guaranteed to be similar even if they are 
calibrated to the same observations initially 
\citep[see e.g.][and references therein]{neistein:sam.degeneracies}. 

In order to circumvent these physical uncertainties, 
models constructed using an empirical approach 
(the halo occupation formalism) have become popular
as a means to predict the evolutionary paths of galaxies, including their clustering 
evolution, net growth, star formation histories, 
and merger rates \citep{zheng:hod.evolution,yan:clf.evolution,
conroy:mhalo-mgal.evol,conroy:hod.vs.z,perezgonzalez:hod.ell.evol,
tinker:hod,cooray:hod.clf,brown:hod.evol,wetzel:mgr.rate.subhalos,
stewart:merger.rates,hopkins:merger.rates,moster:stellar.vs.halo.mass.to.z1}. By matching 
the observed correlation functions of galaxies versus
stellar mass, redshift, and other properties, one can empirically 
infer which halos the galaxies populate.\footnote{
We should formally distinguish traditional strict HOD models 
from subhalo abundance matching models. In the former,  
one paints galaxies onto host halos with no regard for halo merger 
histories or dark matter substructure, using clustering as a constraint 
\citep[see e.g.][]{yan:clf.evolution,yang:clf,tinker:hod,
kravtsov:subhalo.mfs,wechsler:assembly.bias}. In the latter, subhalo orbits 
are tracked in simulations and galaxies are assigned strictly to 
subhalos, but by matching the observed stellar mass function, so that the 
correlation function is a {\em prediction} of the model \citep[which 
agrees very well; see e.g.][]{conroy:monotonic.hod,valeostriker:monotonic.hod,
lee:2009.uv.lum.vs.mhalo,
wetzel:mgr.rate.subhalos,wetzel:subhalo.disruption,
moster:stellar.vs.halo.mass.to.z1,
behroozi:mgal.mhalo.uncertainties}. For our purposes however, the 
two methods give the same mass functions and clustering/substructure, so 
are treated interchangeably.} 
Independent observational constraints on halo masses (discussed in 
detail in \S~\ref{sec:compare.hods}) such as weak lensing, 
kinematic modeling, X-ray gas measurements, and group velocity 
dispersions yield complementary constraints. 
Populating a cosmological 
simulation of halos+subhalos in this manner, one can evolve it 
forward and determine where and when mergers occur. 

Such approaches are not, of course, models for galaxy formation -- they populate 
galaxies in halos according to what is observed, and then (evolving them 
forward some arbitrarily small amount in time) can be used to predict quantities 
such as the merger rate (or what the merger rate must be if both dark-matter 
clustering simulations and the clustering observations are correct). 
They do not make any explicit statement about how the galaxies formed
in the first place, 
nor the physics that are important. They require, for example, that 
the ratio of halo to stellar mass be high in low and high-mass systems, but 
include no information about what feedback processes (believed to be 
stellar and AGN feedback, respectively) explain this. 
They are also, of course, limited by the observations: there are uncertainties 
in those observations corresponding to a range of $M_{\rm gal}(M_{\rm halo})$ 
allowed, and there are redshifts and masses that the observations
do not probe \citep[see][and references therein]{moster:stellar.vs.halo.mass.to.z1,
behroozi:mgal.mhalo.uncertainties}, as well as degeneracies 
in fitting the model parameters that may hide critical differences between 
different sub-populations (for example, those which will or will not merge 
in a short time \citep[see the discussion in][]{guo:2010.hod.constraints}.
Roughly speaking, the observational constraints themselves are robust 
above $\sim10^{9}\,\msun$ at $z=0$ and $\sim10^{11}\,\msun$ at $z\sim2$, but rapidly 
become more uncertain outside of this range.

\subsection{Semi-Empirical Models: Sources of Uncertainty}
\label{sec:empirical:robustness}

\begin{figure*}
    \centering
    \scaleup
    \plotter{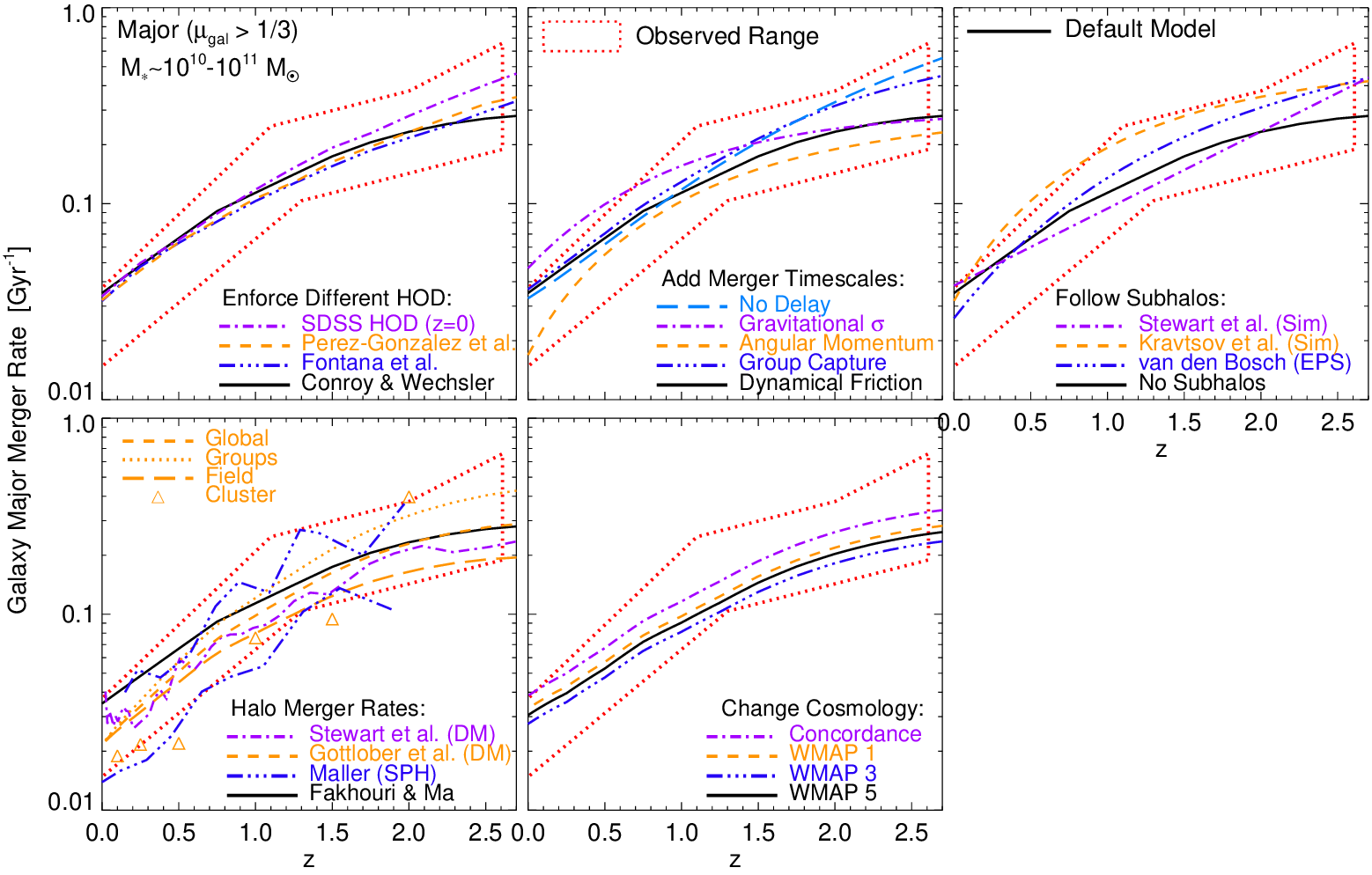}
    \caption{Comparison of the galaxy-galaxy major merger rate 
    (mergers per galaxy per Gyr) for $M_{\rm gal}\sim10^{10}-10^{11}\,\msun$ 
    galaxies as a function of redshift, from different 
    variations in semi-empirical models -- i.e.\ models in which galaxies 
    are placed in halos strictly according to observational constraints. 
    Black line in each case corresponds to the ``default'' model 
    from \citet{hopkins:merger.rates}. Red dotted range is the 
    range allowed by observations compiled in that paper. 
    {\em Top Left:} Changing the halo occupation model 
    (observational constraints used to place galaxies in halos); 
    each of the model choices listed is discussed in detail in the text 
    (\S~\ref{sec:empirical:robustness}). 
    The choices shown bracket the range allowed by a number of 
    independent observational constraints. 
    {\em Top Center:} 
    Changing the merger ``timescale'' -- i.e.\ timescale used to 
    estimate the delay between a halo-halo merger and the 
    resulting galaxy-galaxy merger (during which the satellite no longer 
    grows). 
    {\em Top Right:} 
    Directly following subhalos in simulations, and assigning 
    mergers when those subhalos are destroyed/fully merge into the 
    central halo, instead of using the merger ``timescale'' approach. 
    Results are shown from various methods of following subhalos in simulations 
    and EPS trees. 
    {\em Bottom Left:}
    Adopting estimates of the parent halo-halo merger rate, 
    from different dark matter simulations. 
    For the merger rates from \citet{gottlober:merger.rate.vs.env}, we 
    also show the results using their merger rates in different environments 
    (labeled). 
    We also compare rates using cosmological hydrodynamic simulations 
    to ``tag'' the halo-halo mergers of interest by identifying their 
    resulting galaxy-galaxy mergers. 
    {\em Bottom Center:} 
    Changing the cosmological parameters. 
    The allowed variations in these semi-empirical or HOD models lead to 
    factor $\sim2-3$ differences in the 
    predicted merger rate. 
    \label{fig:mgr.rate.vs.model}}
\end{figure*}

With the advent of dark-matter simulations that can follow 
substructure at high resolution and tight observational constraints on 
the stellar mass functions and clustering of galaxies as a function of mass 
and redshift, semi-empirical approaches have become quite robust. 
\citet{stewart:merger.rates} and \citet{hopkins:merger.rates} 
present predicted merger rates derived from such models, 
based on a variety of different assumptions and observational 
constraints. 
We refer to those papers for model details: the basic methodology is similar.
In \citet{hopkins:merger.rates}, halo-halo merger histories are taken 
from the Millenium dark matter simulation \citep{fakhouri:halo.merger.rates} 
(these are updated with the Millenium-II results in \citealt{fakhouri:millenium.2.merger.rates}, but 
the results for our purposes are identical), 
and in \citet{stewart:merger.rates} from independent dark matter simulations
\footnote{\label{foot:stewartfoot}
The simulation is run with the ART code \citep{kravtsov:1997.ART}, with distinct 
halo identification and subhalo tracking 
algorithms from those used in the Millenium simulations \citep{kravtsov:subhalo.mfs}, and a 
different subhalo `destruction' criteria based on the fractional mass loss 
via stripping (subhalos merged when this falls below a fixed 
fraction of their inflow mass) as opposed to an absolute mass resolution limit. 
}; 
each halo is then populated with galaxies using the abundance-matching 
methodology (although variations and alternative halo occupation 
approaches are attempted), ensuring (by construction) a match to observed 
galaxy mass functions 
and correlation functions at all redshifts. In short time 
intervals, the model halos are evolved forward; when halos merge, either 
the subhalos are followed until they are no longer resolved or a ``dynamical 
friction timescale'' is assigned as calibrated from high-resolution 
simulations \citep{boylankolchin:merger.time.calibration}, and at the end of this time 
the hosted galaxies are assumed to merge. Summing over the entire 
simulation volume, this gives the predicted galaxy-galaxy merger rate. 

The two independent models 
find predicted merger rates that agree reasonably well: typical
variations in the model inputs lead to factor $<2$ changes in the predicted 
merger rates, at least at redshifts $z\lesssim2$ where the halo occupation 
function is reasonably well-constrained by observations. 
From this perspective, 
the advantage of the semi-empirical models is that they do {\em not} 
attempt to model galaxy formation -- as a result, differences between 
various predictions must owe either to the halo occupation distribution 
adopted (i.e.\ constraints on the galaxy mass-halo mass distribution), or 
to the ``background'' dark matter dynamics. 

Figure~\ref{fig:mgr.rate.vs.model} summarizes the comparisons 
from \citet{hopkins:merger.rates}.\footnote{
The merger rates derived from this model, including several of the 
variations discussed here, can be obtained from 
the ``merger rate calculator'' routine publicly available at 
\mergercalcurl.}
The different model choices are discussed 
therein, and we will break each down in detail in what follows. For now, we simply 
illustrate the different effects on the merger rate.\footnote{Henceforth, we take the 
term ``merger rate'' to refer to the galaxy-galaxy merger rate, unless otherwise 
specified.} 
Specifically, we compare the predicted major merger rate (mergers of 
mass ratio $\mu_{\rm gal}>1/3$, where $\mu_{\rm gal}\equiv M_{\rm secondary}/M_{\rm primary} < 1$
in terms of the galaxy baryonic mass $M_{\rm gal}$) 
for intermediate-mass galaxies as a 
function of redshift.
We begin with the model from \citet{hopkins:merger.rates}, and consider the effects 
of the choices below. We take this as our basic HOD or semi-empirical 
model throughout, frequently comparing with the 
model from \citet{stewart:merger.rates} (which uses different 
sets of empirical HOD constraints, dark matter simulations, and methods 
of subhalo tracking to identify mergers), but with many of the variations below considered 
separately (each can be considered its own model). 

{\bf (1)} Cosmology: We re-run the \citet{hopkins:merger.rates} model adopting the cosmological 
parameters from WMAP1 \citep{spergel:wmap1}, WMAP3 
\citep{spergel:wmap3} and WMAP5 \citep{komatsu:wmap5}, as 
well as a ``concordance'' model with 
$(\Omega_{\rm M},\,\Omega_{\Lambda},\,h,\,\sigma_{8},\,n_{s})$=$(0.3,\,0.7,\,0.7,\,0.9,\,1.0)$. 
Because the halo occupation statistics are constrained to reproduce the same stellar 
mass function, most of the difference between cosmologies (for example, 
in the predicted halo mass function) is effectively 
normalized out in this approach. 

{\bf (2)} Halo-Halo Merger Rates: The \citet{hopkins:merger.rates} model 
adopts the halo-halo merger rate fits from \citet{fakhouri:millenium.2.merger.rates}, 
from the Millenium simulation \citep{springel:millenium}.
An alternative dark-matter simulation, of comparable resolution, with 
halo merger rates determined using a different methodology$^{\ref{foot:stewartfoot}}$, 
is described in \citet{stewart:merger.rates}. Another is found 
in \citet{gottlober:merger.rate.vs.env} 
\citep[see also][]{kravtsov:subhalo.mfs,zentner:substructure.sam.hod}; they 
quantify the fit separately to field, group, and cluster environments. 
We also consider the galaxy-galaxy merger rates directly 
determined from cosmological hydrodynamic 
simulations in \citet{maller:sph.merger.rates}. It is well-known that such simulations yield 
galaxies that are overmassive relative to those observed 
(i.e.\ predict an HOD in conflict with that observed) -- therefore we only use these 
mergers as ``markers'' of where the galaxy-galaxy mergers should occur, 
and (in order to study the stellar mass and mass ratio dependencies)  
re-populate the galaxies with the ``correct'' masses according to the HOD. 

{\bf (3)} Substructure: Instead of using a halo-halo merger rate with some ``delay'' 
applied to model subhalos in the HOD, we can attempt to follow subhalos directly after the halo-halo 
merger, and define the galaxy-galaxy merger when the subhalos 
are fully merged/destroyed. We compare the \citet{hopkins:merger.rates} 
rates to those obtained 
tracking the halo+subhalo populations in cosmological simulations from
\citet{stewart:merger.rates} (populating subhalos according to the same HOD). 
We also compare 
with the results using the different subhalo-based methodology 
described in \citet{hopkins:groups.qso} (essentially, beginning from the subhalo 
mass function constructed from cosmological simulations and evolving this 
forward in short time intervals after populating it, at each time, according to 
the HOD constraints). We compare two different 
constructions of the subhalo mass functions: that from 
cosmological dark-matter only simulations in 
\citet{kravtsov:subhalo.mfs} \citep[see also][]{zentner:substructure.sam.hod} 
and that from the extended Press-Schechter formalism coupled to 
basic prescriptions for subhalo dynamical evolution, 
described in \citet{vandenbosch:subhalo.mf}. 

{\bf (4)} Merger Timescales: Another commonly applied method in 
determining galaxy-galaxy merger rates as distinct from halo-halo mergers is 
to assign a simple delay to the galaxy-galaxy merger after the halo-halo merger, 
based on e.g.\ the dynamical friction time for the subhalo+galaxy orbit to decay, as 
calibrated in high-resolution N-body simulations that resolve the galaxy and 
follow the entire system self-consistently. 
The \citet{hopkins:merger.rates} model adopts the traditional dynamical friction time, 
using the calibration from numerical simulations in 
\citet{boylankolchin:merger.time.calibration}. We compare with the model re-run using 
the characteristic timescale for pair-pair gravitational 
capture in group environments, calibrated to simulations following 
\citet{mamon:groups.review}. We again repeat, considering 
capture in angular momentum space rather than gravitationally, 
following \citet{binneytremaine}, as well as the gravitational cross-section 
timescale (similar to the group capture timescale) for capture between passages in e.g.\ 
loose group or field environments, calibrated from simulations 
in \citet{krivitsky.kontorovich}. 

{\bf (5)} Halo Occupation: We re-run the \citet{hopkins:merger.rates} 
model using a different set of halo occupation constraints 
to determine the galaxy mass-halo mass relations. 
The default \citet{hopkins:merger.rates} employs the fits to these 
distributions from \citet{conroy:hod.vs.z}, which adopt a monotonic 
rank-ordering approach to assign galaxies to halos in a manner 
constructed to match observed mass functions 
and clustering 
\citep[see the compilation in Fig.~1 therein and][]{bell:mfs,
drory:2005.stellar.mfs,borch:mfs,fontana:highz.mfs,
panter:2007.sdss.stellar.mf,perezgonzalez:mf.compilation}. 
Gas masses are added to galaxies according to the observed 
correlation between galaxy gas fractions and stellar mass, 
from observations spanning $z=0-3$ \citep{belldejong:disk.sfh,mcgaugh:tf,
calura:sdss.gas.fracs,shapley:z1.abundances,erb:lbg.gasmasses,
puech:tf.evol,mannucci:z3.gal.gfs.tf,cresci:dynamics.highz.disks,
forsterschreiber:z2.sf.gal.spectroscopy,erb:outflow.inflow.masses,
mannucci:z3.gal.gfs.tf} 
compiled in \citet{hopkins:merger.rates} (\S~2.2.2 therein). 
We compare to the results using the same monotonic rank-ordering 
HOD methodology, but with galaxy stellar mass function constraints 
from \citet{perezgonzalez:mf.compilation} or 
\citet{fontana:highz.mfs} (each determined up to redshift $z\sim4$). We also compare 
to the results selecting the fit to $M_{\ast}(M_{\rm halo})$ 
and its scatter for central and satellite galaxies from the observed SDSS 
clustering at $z=0$ \citep{wang:sdss.hod}; in this case, we simply adopt 
the $z=0$ fit at all redshifts -- we do not allow for evolution. 

We will discuss each of these aspects in detail in what follows. However, 
it should already be clear in Figure~\ref{fig:mgr.rate.vs.model} that these 
choices, while not negligible, introduce only factor $\sim2$ uncertainties 
in merger rates. Order-of-magnitude effects, it seems, must depend either on 
some particular combination of effects above, or relate to the baryonic 
physics of galaxy formation, such that the galaxy-halo mappings predicted 
are significantly different from those empirically determined by HOD 
models.

\subsection{A Priori Models Compared}
\label{sec:empirical:vs.sams}

We now consider various {\em a priori} models for galaxy formation. 
In these cases, there are of course similar choices to those discussed 
above, for quantities such as the halo-halo merger rate, merger timescales 
and/or subhalo tracking, and cosmological parameters. However, the halo 
occupation distribution is not determined from observations, but is predicted as a 
consequence of attempting to model galaxy formation in an a priori 
manner based on some set of physical assumptions. 
That is not to say this is completely unconstrained, however; the models 
(especially semi-analytic models) are generally adjusted so as to reproduce 
various observations such as the stellar mass function and large-scale 
galaxy clustering. As a result, at least certain portions of the model 
should be similar to the HOD models. 

The specific models we will consider throughout the paper are briefly as follows: 
\\

{{\bf (a) \citet{somerville:new.sam}:}} A semi-analytic model updating those in 
\citet{somerville99:sam,somerville:sam}. 
Halo growth is tracked using the extended 
Press-Schechter formalism (using a slightly modified 
version of the method of \citet{somerville:merger.trees}, as described in 
\citet{somerville:new.sam}, to match the results of $N$-body simulations), 
with merging of sub-halos in virialized halos followed with appropriate 
dynamical friction times (including mass loss and tidal destruction). 
Cooling is calculated in the typical manner, distinguishing between 
``cold'' and ``hot'' mode regimes 
\citep[gas accretes onto central galaxies in 
a dynamical time in halos with $M_{\rm halo}\lesssim10^{12}\,\msun$, 
but forms quasi-static pressure-supported halos at 
larger masses, as motivated by cosmological simulations; see e.g.][]{keres:hot.halos}. 
Disks form assuming conservation of specific angular momentum as described in 
\citet{somerville:disk.size.evol}, and star formation 
follows \citet{kennicutt98} with supernova feedback. Black holes and bulges 
grow in mergers, with feedback associated with bright quasars and 
low-luminosity AGN, the latter of which suppresses new cooling in 
the ``hot mode'' of massive halos. 
This model adopts the prescription for bulge formation calibrated 
to simulations in \citet{hopkins:disk.survival}, given the gas fractions and mass 
ratios of mergers. 
Satellite galaxies are assumed to lose their hot gas reservoirs to 
their new parent halos (and cease to accrete new gas) upon halo-halo merger in the EPS tree, but 
do not lose the recycled gas from star formation internally in their pre-existing cold 
disks. 
\\

{\bf (b) \citet{khochfarsilk:new.sam.dry.mergers} 
\citep[see also][]{weinzirl:b.t.dist}:} This semi-analytic model differs from 
that of \citet{somerville:new.sam} primarily in that there is no 
``quasar mode'' of feedback, and the prescriptions for the low-luminosity 
``radio mode'' of AGN feedback and stellar wind feedback differ 
somewhat in detail. Galaxy structural properties are followed using 
energetic arguments and scalings motivated by simulations 
\citep{khochfar:size.evolution.model}. 
Bulge formation is followed with a more simplified recipe common in most 
semi-analytic models: the secondary stellar disk is destroyed (added to the bulge), 
but the primary disk is only violently relaxed in a major merger, and starbursts 
in mergers are tracked following \citet{cox:massratio.starbursts}. The differences 
stemming from the bulge formation model are discussed in detail in 
\citet{hopkins:disk.survival.cosmo}. 
Satellite galaxies retain their cold and hot gas reservoirs until the  
merger with the primary central galaxy (but do not accrete new gas). 
\\

{\bf (c) \citet{bower:sam}:} This iteration of the Durham Galform SAM 
revises previous versions 
\citep{laceycole:halo.merger.rates.93,benson:sam,baugh:sam} by 
adopting halo merger trees determined in N-body simulations, specifically 
the Millenium simulation \citep{springel:millenium}, and tracking 
substructure within those halos, as well as adding a 
mode of low-luminosity AGN feedback (a ``radio mode'') 
which suppresses new cooling in ``hot mode'' halos. Cooling is followed 
as in the models above, and 
bulge formation with the prescription later adopted by \citet{khochfarsilk:new.sam.dry.mergers}, 
with the addition of allowed bulge formation in a strong disk  
instability mode that dominates at high redshift. 
This mode sets in when the disk becomes self-gravitating, and 
in consequence the disk is assumed to largely collapse to a bulge. 
Satellite galaxies are assumed to lose all of their hot gas to the new parent halo 
upon first linkage of the parent halos (often several $R_{\rm vir}$), 
and gas which galaxies ``expel'' owing to the assumed 
stellar wind mode of feedback is considered part of the halo gas and treated 
similarly. 
\\

{\bf (d) \citet{font:durham.sam.update}:} The most recent iteration of the 
Durham SAM . 
The model improves on that of \citet{bower:sam} by including a 
physically-motivated prescription for the removal of hot gas from 
subhalos after they are first linked to the main halo group 
(as a function of e.g.\ their position in the primary halo and 
the background gas density; motivated by ram-pressure stripping models). 
It is otherwise identical. 
We will show in \S~\ref{sec:compare.hods:satellites} 
that this has a large effect on the predicted merger rates. 
\\

{\bf (e) \citet{croton:sam}:} Another semi-analytic model based on 
the output of the Millenium simulation and broadly similar to 
that in \citet{bower:sam}, but with important differences in the 
prescriptions for stellar feedback and cooling 
in both central and satellite 
galaxies (see \S~\ref{sec:compare.hods:satellites}). 
Accretion in the ``cold mode'' is delayed by a galaxy dynamical time. 
Satellite halos are instantly stripped as in \citet{bower:sam}; 
the mass expelled by star formation is contained in a separate reservoir to 
be recycled, but this too is lost to the parent halo for satellite galaxies. 
The merger trees are constructed from the same simulation, but with 
different algorithms and cuts. 
The model also includes a much weaker disk instability mode of 
bulge formation, relative to that in \citet{bower:sam}. 
The implementation of AGN feedback is qualitatively similar 
between the two models, but quantitatively the \citet{croton:sam} 
mode is ``stronger'' and yields more radio power. 
\\

{\bf (f) \citet{delucia:sam}:} The most recent iteration of the Munich SAM, 
this is a somewhat revised version of the \citet{croton:sam} SAM. The 
primary modifications include the adoption of a different stellar IMF 
and revised dust attenuation calculation, and the removal of a dynamical 
time delay in ``cold mode'' accretion (making 
``quenching'' more rapid). Also, in this version, dynamical friction times 
assigned to galaxies before mergers are increased relative to the 
canonical dynamical friction time by a factor $\sim2$, with the goal of 
suppressing mergers at the massive end of the galaxy mass function 
(since the models tend to have difficulty reproducing this). 
\\

{\bf (g) \citet{maller:sph.merger.rates}:} The results of direct cosmological 
hydrodynamic simulations, incorporating cooling and star formation self-consistently, 
but without feedback of any kind from stars or BHs. Galaxy-galaxy mergers 
are directly identified, and their mass ratio quantified at the time just before 
merger. Hydrodynamic simulations with very large dynamic ranges in mass and redshift 
remain prohibitively expensive, so the comparison is restricted to intermediate 
and high-mass galaxies at redshifts $z\lesssim2$. As the authors note, without 
feedback, the mass function at $z=0$ is over-produced (and the break is not 
as strong as observed). They adopt a mean correction for this; 
essentially, re-normalizing their stellar masses by a systematic factor of $\sim3$. 
\\
\, \\

\begin{figure*}
    \centering
    \scaleup
    \plotter{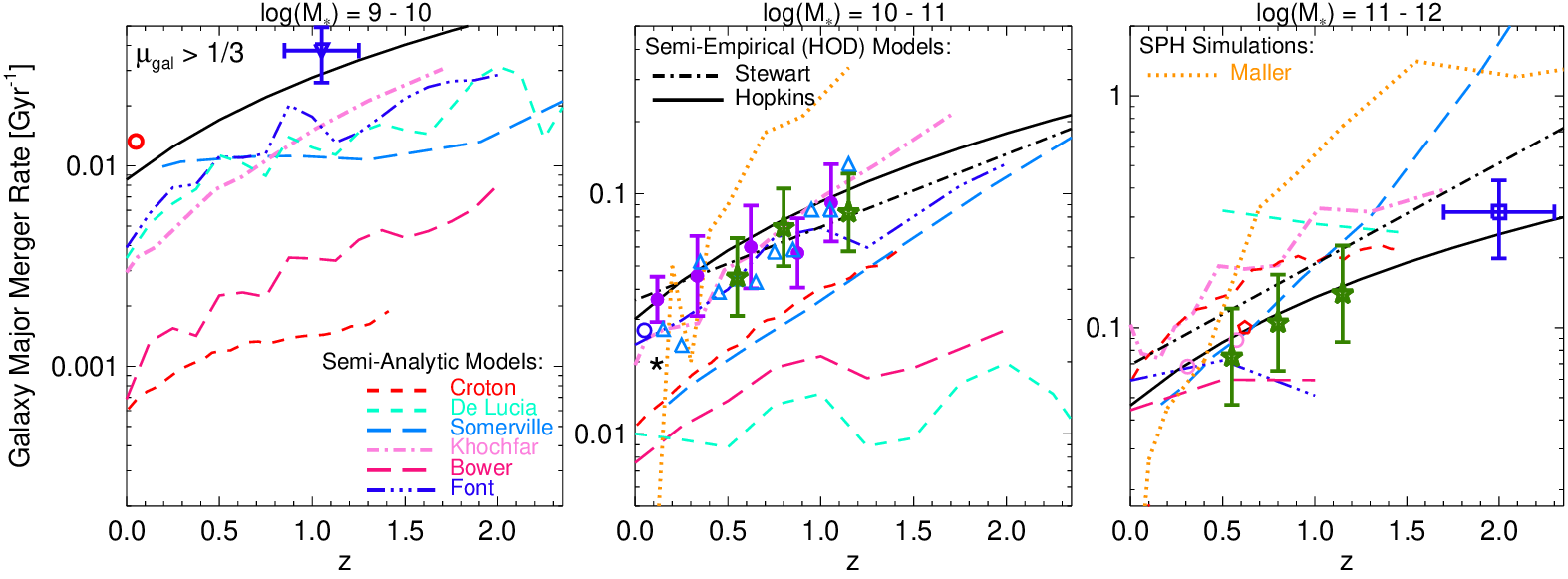}
    \caption{
    As Figure~\ref{fig:mgr.rate.vs.model}, but comparing the 
    HOD-based models with different predictions from 
    {\em a priori} galaxy formation models, in different galaxy stellar 
    mass intervals ($\log{(M_{\ast}/\msun)}$, as labeled). 
    We compare the standard semi-empirical HOD 
    models from \citet{hopkins:merger.rates} and \citet{stewart:merger.rates} 
    (described in \S~\ref{sec:empirical:robustness}) to 
    the direct predictions of cosmological hydrodynamic simulations 
    \citep[with cooling and star formation, but without feedback from either 
    stars or AGN;][]{maller:sph.merger.rates} and the 
    predictions of various semi-analytic galaxy formation models 
    \citep[][described in \S~\ref{sec:empirical:vs.sams}]{croton:sam,bower:sam,
    delucia:sam,khochfarsilk:new.sam.dry.mergers,somerville:new.sam,
    font:durham.sam.update}. We compare these predictions with 
    observations in each mass interval, based on close pair counts 
    ($r<20\,h^{-1}\,{\rm kpc}$) with calibration of merger timescales specific for 
    each observed separation and selection criteria from 
    large suites of high-resolution galaxy merger simulations \citep{lotz:merger.selection}. 
    The observations are compiled in \citet{hopkins:merger.rates} from
    \citet[][blue triangles]{kartaltepe:pair.fractions}, 
    \citet[][blue inverted triangles]{conselice:mgr.pairs.updated}, 
    \citet[][pink circles]{lin:merger.fraction,lin:mergers.by.type}, 
    \citet[][blue circle]{xu:merger.mf}, 
    \citet[][black asterisk]{depropris:merger.fraction}, 
    \citet[][blue squares]{bluck:highz.merger.fraction}, 
    \citet[][green stars]{bundy:merger.fraction.new}, and 
    \citet[][red pentagons]{bell:dry.mergers,bell:merger.fraction}.
    \label{fig:mgr.rate.vs.sams}}
\end{figure*}

Figure~\ref{fig:mgr.rate.vs.sams} compares the predicted merger rates 
from these models. 
We compare these with observational determinations of the galaxy-galaxy merger 
rate, from close pairs. 
Specifically, from the fraction of {\em major} ($\mu_{\rm gal}>1/3$) 
pairs with small projected separations $r_{p}<20\,h^{-1}\,{\rm kpc}$ 
(often with the additional requirement of a
line-of-sight velocity separation $<500\,{\rm km\,s^{-1}}$), 
and stellar masses in a specific range (labeled). 
For each mass bin, the pair fractions 
as a function of redshift can be empirically converted to a merger rate 
using the merger timescales at each radius. \citet{lotz:merger.selection} 
specifically calibrate these timescales for the same projected separation and 
velocity selection from a detailed study of a large suite of hydrodynamic 
merger simulations (including a 
range of galaxy masses, orbital parameters, gas fractions and star formation 
rates) using mock images obtained by applying realistic radiative transfer models, 
with the identical observational 
criteria to classify mock observations of the galaxies at all times 
and sightlines during their evolution. For this specific pair selection 
criterion, they find a median 
merger timescale of $t_{\rm merger}\approx 0.35\,$Gyr, with relatively 
small scatter and very little dependence on simulation parameters ($\pm0.15\,$Gyr).\footnote{
The merger timescale from simulations at this radius is somewhat shorter than 
the time obtained assuming dynamical friction and circular orbits in e.g.\ an 
isothermal sphere, as has commonly been done 
\citep[this is assumed in e.g.\ both][]{patton:merger.fraction,
kitzbichler:mgr.rate.pair.calibration}. This 
owes to two effects: first, angular momentum loss at these radii is {\em not} 
dominated by dynamical friction (at least in the 
traditional sense of a small mass moving through a smooth, isotropic 
velocity dispersion background), but rather by exchange in strong 
resonances between the 
baryonic components that act much more efficiently. Second, by these radii, even initially 
circular orbits have become highly radial, leading to shorter merger times. 
Because of these effects, the remaining merger time at this scale depends 
only weakly on initial conditions or orbital parameters --
essentially, these processes have erased most of the ``memory'' of 
the original orbital configuration. 
}
We use this median $t_{\rm merger}$ to convert the observations to a 
merger rate. Completeness corrections are discussed in the various papers; 
we also adopt the standard correction from \citet{patton:mgr.rate.vs.rmag}, 
calibrated to high-resolution simulations, for the fraction of systems on early 
or non-merging passages (to prevent double-counting systems on multiple 
passages); but this is relatively small \citep[$20-40\%$; see also][]{lotz:merger.selection}. 
For more detailed discussion of these observations and related constraints 
from other methods and observations, we refer to \citet{hopkins:merger.rates}.

We are not attempting to comprehensively discuss
the predicted merger rates here: those rates are 
presented in other papers in much greater 
detail.\footnote{For predicted merger rates from the 
\citet{somerville:new.sam} model, see \citet{hopkins:disk.survival.cosmo,
hopkins:merger.rates}. For those from \citet{khochfarsilk:new.sam.dry.mergers}, 
see \citet{weinzirl:b.t.dist}. From \citet{bower:sam}, 
see \citet{mateus:mgr.rate.sam.obs.compare} 
and \citet{parry:sam.merger.vs.morph}. 
From \citet{delucia:sam}, see \citet{guo:gal.growth.channels}, 
\citet{kitzbichler:mgr.rate.pair.calibration}, and \citet{bertone:stellar.wind.recycling.sam}. 
The \citet{maller:sph.merger.rates} rates are discussed therein. 
The model predictions from \citet{delucia:sam,bower:sam,font:durham.sam.update} 
can also be obtained online from the Millenium public database, 
at \MPAurl\ or \Durhamurl 
(quantities shown here from these two models are obtained from the 
same database).} 
However, we do wish to examine what can affect the predicted 
merger rate. It is clear in Figure~\ref{fig:mgr.rate.vs.sams} that the 
variation in predictions between different {\em a priori} models 
is significantly larger than that within the  
semi-empirical models. This motivates our further examination of 
the differences and similarities in these models. 

For clarity, we have limited the number of semi-analytic models 
shown in Figure~\ref{fig:mgr.rate.vs.sams}. However, we have compared with a 
number of other such models and find similar results 
\citep[including e.g.][]{benson:sam,baugh:sam,
kang:sam,cattaneo:sam,monaco:sam,bertone:stellar.wind.recycling.sam}. 
The models considered span the representative behavior 
in these other semi-analytic models as well.

\section{The Dark Side}
\label{sec:compare.dark}

First, we examine the differences in the dark matter side of 
the merger rate, i.e.\ effects on the predicted merger rates 
that are independent of the detailed baryonic physics of galaxy formation.

\subsection{Cosmological Parameters}
\label{sec:compare.dark:cosmology}

As shown in Figure~\ref{fig:mgr.rate.vs.model}, the assumed cosmology 
makes little difference for merger rates, so long as models are normalized 
to produce similar stellar mass functions at each redshift. 
Present constraints on the cosmological parameters are also quite 
strong \citep{komatsu:wmap5}; there is little freedom to change the 
cosmology in a significant fashion. At high redshifts, the differences as a 
consequence of e.g.\ the value of $\sigma_{8}$ become larger, 
but so do the other sources of uncertainty below -- at no point do we 
find that the uncertainties in cosmological parameters dominate the 
variations in predicted merger rates. Moreover, 
most of the models considered here adopt an identical cosmology 
(for example, all of the semi-analytic models in Figure~\ref{fig:mgr.rate.vs.model} 
have been run or have available versions adopting the ``concordance'' 
cosmological parameters defined in \S~\ref{sec:intro}), and exhibit nearly identical variation between 
models. 

Comparing the halo mass function and halo accretion histories with 
different cosmological parameters \citep[see e.g.][]{neistein:natural.downsizing}, 
the dominant effect is the predicted halo mass function shifting to higher masses 
with larger $\sigma_{8}$. However, if we re-normalize the model to 
enforce the {\em same} observed galaxy stellar and baryonic mass functions 
and clustering, then these differences are largely normalized out, at 
least for variation in $\sigma_{8}$ in the relatively small observationally 
allowed range (here $0.7-0.9$ considered; much larger differences 
may significantly change dynamical friction times and other higher-order effects). 
\citet{elahi:substructure.mf.vs.powerspectrum.ns} show that the quantity of greatest 
importance for our conclusions, the normalized substructure mass function or 
(equivalently) dimensionless merger rate (mergers per halo per Hubble time 
per unit mass ratio) is almost completely independent of 
cosmological parameters including e.g.\ the power spectrum shape $n_{s}\sim1$ and 
amplitude over the range of variations here (not until 
one goes to much larger effective $n_{s}\sim3$ does one see 
this function change shape).

\subsection{Halo-Halo Merger Rates}
\label{sec:compare.dark:halos}

\subsubsection{Comparison of Different Simulations and EPS Calculations}
\label{sec:compare.dark:halos:compare}

\begin{figure*}
    \centering
    \scaleup
    \plotter{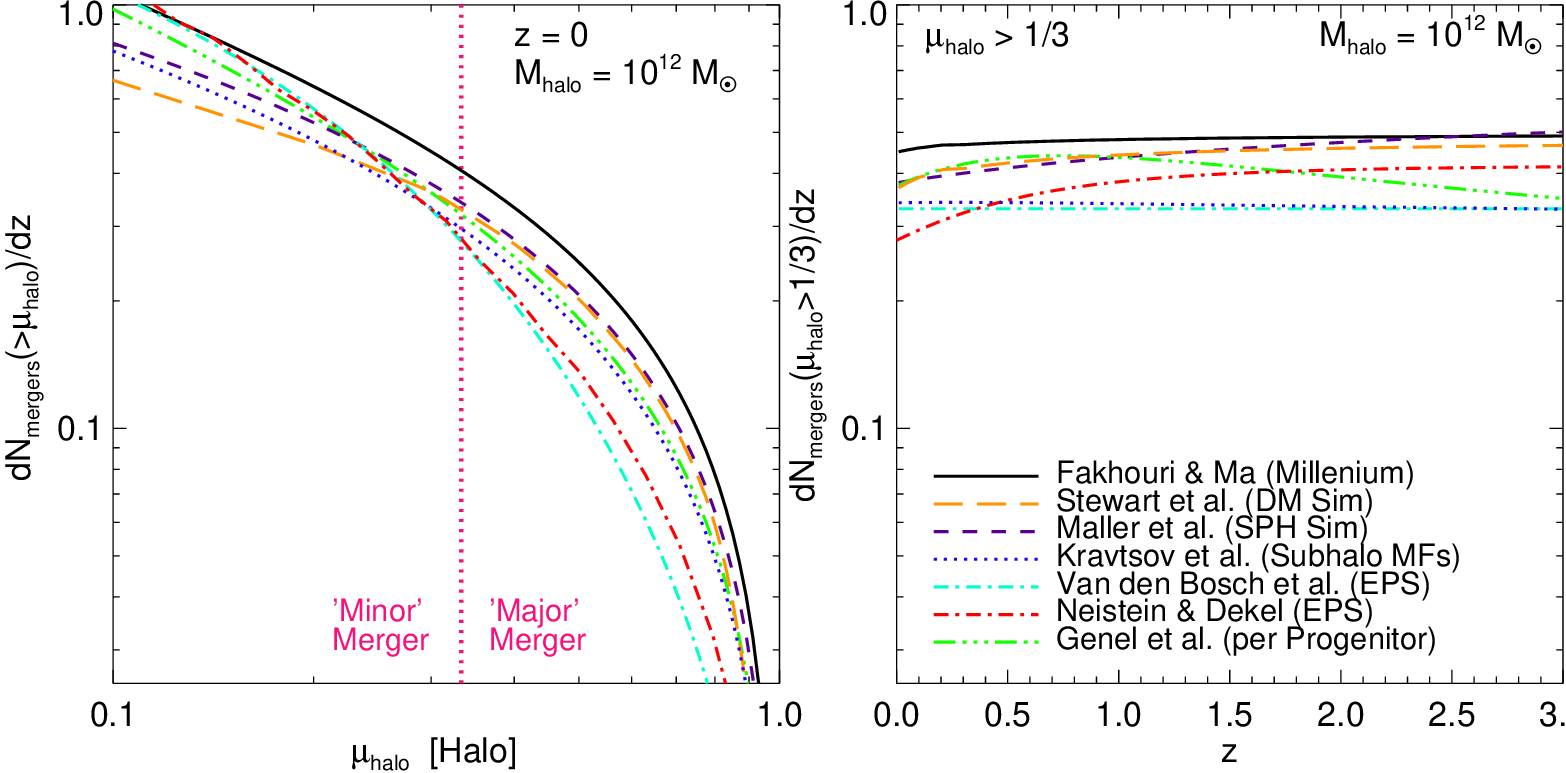}
    \caption{{\em Left:} DM halo-halo merger rates (mergers per 
    halo per unit redshift) as a function of {\em halo} 
    mass ratio (at $z=0$), from different simulations and methodologies shown. 
    \citet{fakhouri:halo.merger.rates} and \citet{genel:merger.rates.perprogenitor} 
    extract merger rates from the Millenium and Millenium-II DM-only simulations, with different 
    methodologies. \citet{stewart:merger.rates} determine rates from independent 
    DM simulations. The \citet{kravtsov:subhalo.mfs} rates
    are based on the integral constraint from subhalo mass functions, rather than 
    an instantaneous merger rate. 
    Rates from \citet{maller:sph.merger.rates} are from 
    hydrodynamic cosmological simulations (using galaxies as halo tracers). 
    Rates from \citet{vandenbosch:subhalo.mf} and 
    \citet{neistein:eps.merger.tree.calibration} 
    are constructed with the EPS formalism. 
    {\em Right:} Corresponding halo major merger rate versus redshift. 
    Most different simulated merger rates agree well 
    (within $\sim20-50\%$ for $\mu>0.1$), provided definitions are appropriately 
    accounted for. Results are shown for $M_{\rm halo}=10^{12}\,\msun$ 
    (at the given redshift), but depend very weakly on halo mass.
    \label{fig:model.halorates}}
\end{figure*}

Even given the same cosmological parameters, different 
methods used to determine the halo-halo merger rate can yield different answers. 
In Figure~\ref{fig:model.halorates}, we plot the 
dependence of the halo-halo merger rate on halo mass ratio 
$\mu_{\rm halo}\equiv M_{\rm halo,\,2}/M_{\rm halo,\,1}<1$ 
and redshift, 
from different cosmological simulations and analytic calculations. 
Specifically, we plot the number of mergers per unit redshift above some $\mu_{\rm halo}$ 
at $z=0$, 
and the evolution with redshift for major ($\mu_{\rm halo}>1/3$) mergers 
(evaluated in a narrow mass interval around $10^{12}\,\msun$ halos, 
in terms of post-merger mass, although all 
these models find that in these units, the mass dependence 
is weak). 

We consider several different determinations of the halo-halo merger rate. 
\citet{fakhouri:halo.merger.rates} calculate the merger rate 
in the Millenium simulation in the usual fashion (determining the 
number of mergers per descendant halo in terms of the 
halo friends-of-friends group mass and mass ratio $\mu_{\rm halo}<1$). 
These are updated with the results of the higher-resolution 
Millenium-II simulation in \citet{fakhouri:millenium.2.merger.rates}, but 
the results are for our purposes identical.
\citet{genel:merger.rates.perprogenitor} analyze the same simulation, 
but determine the merger rate using another methodology: defining 
halo masses in a different manner and quoting the merger 
rate per progenitor halo; they discuss in detail how this leads to subtle 
but potentially important variations in the merger rate fits. However, converted to the 
same units (following the methodology in their Appendix), the results are 
similar. \citet{stewart:merger.rates} determine the halo-halo merger 
rate from an independent dark matter simulation, using a slightly revised 
methodology and different time-step criteria. 
We also compare these simulation results with an analytic approximation: 
using the methodology in \citet{vandenbosch:subhalo.mf} 
to predict the merger rate analytically from a modified extended Press-Schechter 
formalism (note that these authors consider only the main branch mergers; 
because we are considering instantaneous merger rates here, this 
makes no difference, but it will lead to a different integrated 
number of mergers if not accounted for).  
\citet{neistein:eps.merger.tree.calibration} also present 
Press-Schechter trees modified to better match numerical simulations. 
Yet another methodology is described in \citet{hopkins:groups.qso}; 
those authors take the subhalo mass function from 
cosmological dark matter simulations, and 
infer the halo-halo merger rate required to maintain 
this mass function allowing for dynamical evolution of the subhalos 
(i.e.\ assume steady-state subhalo mass functions over a 
narrow redshift interval, given certain 
initial conditions). 
We specifically adopt their estimate using the subhalo mass functions 
presented in \citet{kravtsov:subhalo.mfs} 
and \citet{zentner:substructure.sam.hod}. 
In that paper, however, they consider several other simulation results as 
well, and find similar results in all cases \citep{shaw:cluster.subhalo.statistics,
springel:cluster.subhalos,
tormen:cluster.subhalos,delucia:subhalos,gao:subhalo.mf,nurmi:subhalo.mf}.

The mergers of halo barycenters are the real mergers of interest, and 
it is possible that the presence of baryons could alter halo mass functions and 
merger rates at some level. These and other baryonic effects 
can only be followed in cosmological hydrodynamic simulations. Of course, 
such models may not reproduce the correct observed galaxy baryonic masses. However, 
we can still use such simulations as tracers of galaxy-galaxy mergers; using the 
information about the position and mergers of galaxies, but replacing the galaxy masses 
according to the observed HOD (essentially using galaxies in the simulation to 
`tag' the relevant halos merging).\footnote{We do this by simply replacing the 
rank-ordered galaxy masses with those obtained by abundance matching.}
This also allows for the possibility that the 
baryonic material changes the halo-halo merger rate, by e.g.\ contributing to 
halo self-gravity, although this is typically estimated to be a 
$\sim10-20\%$ effect for the major-merger regime of interest 
\citep[see e.g.][]{weinberg:baryons.and.substructure}. 
Figure~\ref{fig:model.halorates} shows the 
results of this procedure given the cosmological smoothed particle 
hydrodynamics simulations in \citet{maller:sph.merger.rates}. 

Overall, there is quite good agreement between these different calculations -- 
halo-halo merger rates as determined in cosmological simulations are 
generally converged, at least for intermediate masses and redshifts $z\lesssim2-3$, 
at the $\sim50\%$ level or better. The scatter in different simulation-based determinations grows 
to a factor $\sim2$ at high/low masses. 

Furthermore, many of these differences owe to choices of definition that, treated 
properly, will not enter into the galaxy-galaxy merger rate. For example, 
\citet{genel:merger.rates.perprogenitor} demonstrate in detail how changing the 
halo mass definition and definition of merger rate in terms of mergers 
per progenitor (and allowing for $\mu_{\rm halo}>1$; i.e.\ mergers into larger halos) leads 
to differences in the halo-halo merger rate relative to the fits presented in 
\citet{fakhouri:halo.merger.rates} \&\ \citet{fakhouri:millenium.2.merger.rates} 
from the same simulation, and the differences 
can be significant (factor $\sim2$). However, the merger trees in terms 
of e.g.\ halo {\em barycenters} in both 
cases are almost identical. 
In terms of galaxy-galaxy merger rates, the precise halo mass definition, 
for example, should make no difference, so long as the halo occupation statistics 
are appropriately renormalized to yield the same galaxy mass function and galaxy clustering 
versus luminosity (halo mass is simply a label to use in tagging galaxies 
to halos). Many of the differences in Figure~\ref{fig:model.halorates}, therefore, 
are actually reduced in terms of their effect on galaxy-galaxy merger rates, 
so long as sufficient care is taken in correcting for the definitions used. 

Considerable effort has also gone into tuning EPS-predicted merger rates to 
better correspond to the results of N-body simulations 
\citep[see e.g.][]{zentner:eps.methodology.review,
neistein:eps.merger.tree.calibration}. 
With these improvements, many current-generation 
EPS trees differ from the simulation results they are tuned to by as little 
as $\sim10-20\%$. For detailed comparison of EPS and simulation 
merger rates, we refer to \citet{genel:merger.rates.perprogenitor}. 
Clearly, this is not the dominant source of uncertainty in 
merger rates, as the differences between various counts of mergers from the same 
simulation differ by larger amounts, depending on their exact methodology.

\subsubsection{Some Caveats in Adopting Halo-Halo Merger Rates}
\label{sec:compare.dark:halos:caveats}

We do not mean to imply that choices of definition are entirely negligible. 
Depending on the application, a halo merger rate 
defined in terms of e.g.\ progenitor galaxies or descendant galaxies is more 
appropriate (the former being an estimate of the probability that a given halo 
{\em will have} a merger; the latter being an estimate of the probability that 
a given halo {\em did have} a merger). And there are means of constructing 
halo catalogs that could give quite different halo-halo merger 
rates, despite being ultimately a 
reflection of the same merger tree. 
The application of any halo merger rate requires careful accounting for 
consequences of such definitions.

\begin{figure}
    \centering
    \scaleup
    \plotter{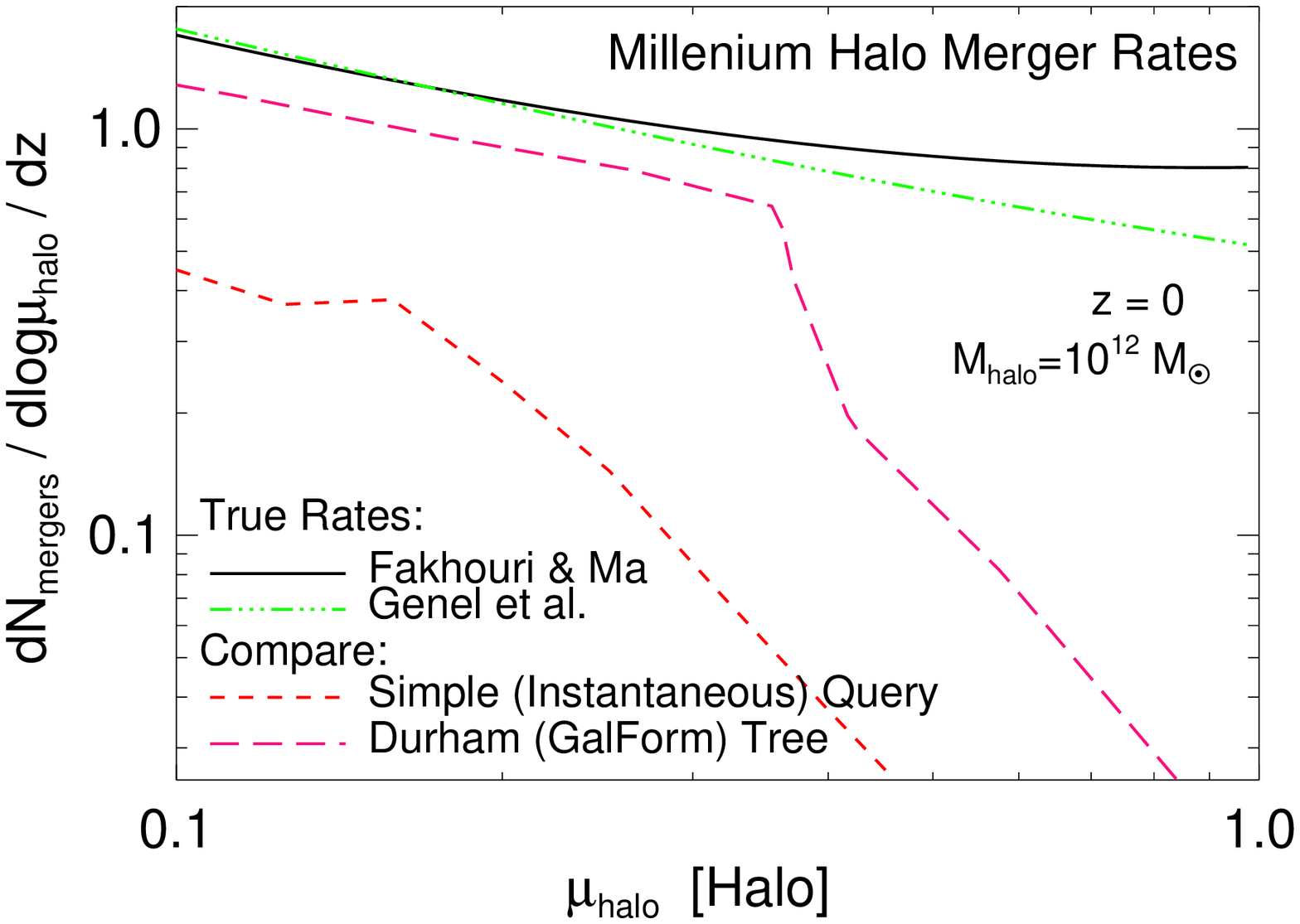}
    \caption{As Figure~\ref{fig:model.halorates}, showing the differential 
    number of halo-halo mergers per log mass ratio (at $z=0$) from the 
    Millenium DM-only simulation. The \citet{fakhouri:halo.merger.rates} 
    and \citet{genel:merger.rates.perprogenitor} rates are intended to 
    represent the halo-halo merger rate. We compare with the instantaneous 
    rate taken directly from the Durham semi-analytic model halo trees 
    \citep{harker:marked.correlation.function}, or from a simple query of the 
    simulation database of the 
    Munich halo trees using the instantaneous mass ratios just before 
    the secondary is ``lost'' in the tree. 
    Despite this being the same set of trees, these ``instantaneous'' 
    mass ratio definitions lead to an apparent large suppression of major halo-halo mergers.  
    This is an artificial effect of 
    the definitions used; we show the differences to illustrate 
    the importance of accounting for different halo mass ratio definitions 
    in constructing the halo-merger rate. 
    \label{fig:model.halorates.durham}}
\end{figure}

\begin{figure}
    \centering
    \scaleup
    \plotter{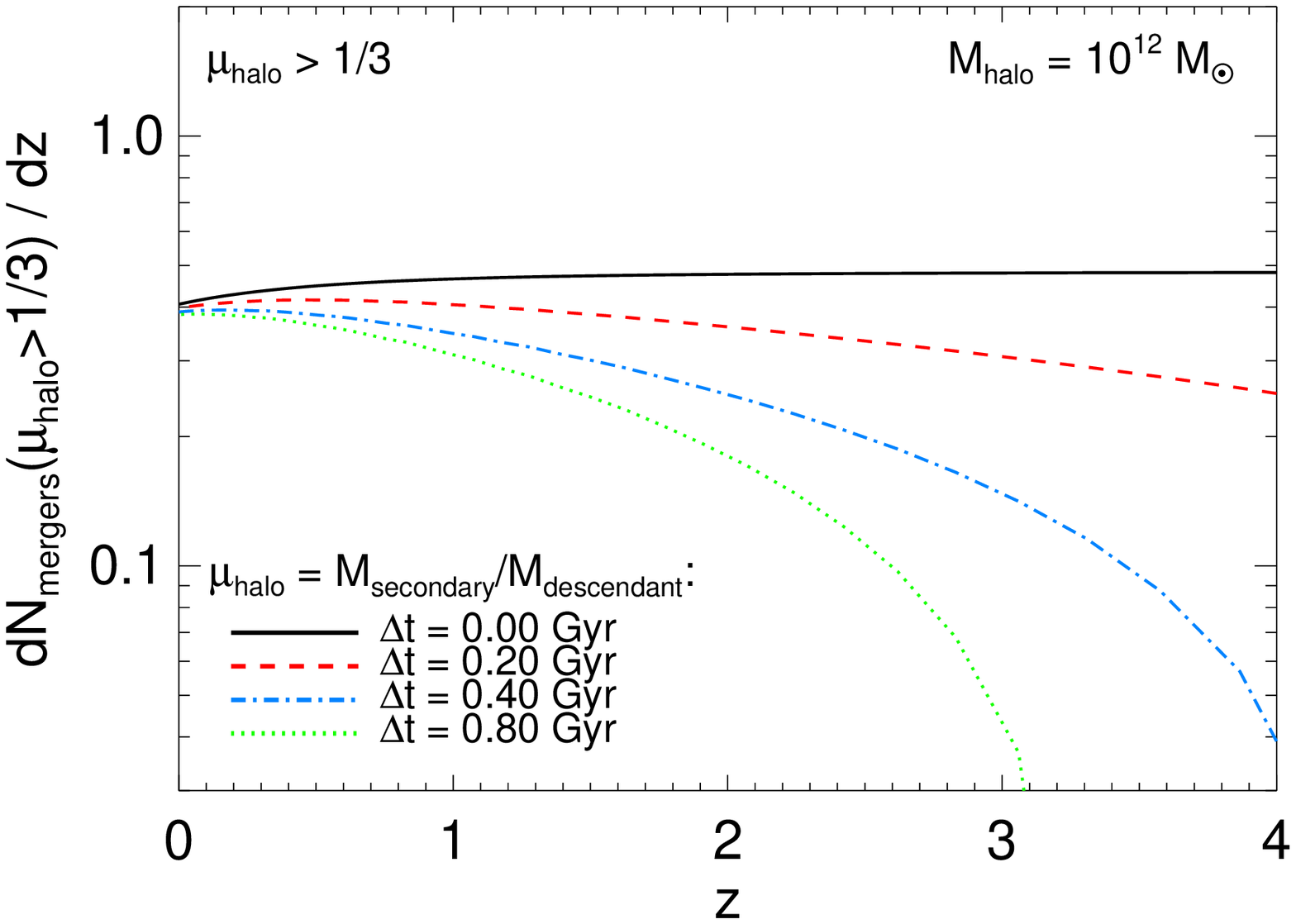}
    \caption{As Figure~\ref{fig:model.halorates}, comparing the 
    halo merger rate as a function of redshift from the \citet{fakhouri:halo.merger.rates} 
    trees. We compare the merger rate using the minimum available time spacing 
    ($\Delta t=0.00\,$Gyr, solid; in fact the spacing is finite but the halos are tracked on-the-fly 
    in the simulation, so it is small for our purposes), to the rate inferred from 
    comparing snapshots only with the given time spacing, if the mass ratio is defined 
    by comparison of subsequent snapshots. Large timestepping means major 
    mergers appear minor or are lost. If a quantity depends super-linearly on halo 
    mass (as e.g.\ galaxy mass does at low masses), then the apparent suppression 
    will be enhanced. 
    \label{fig:model.halorates.timestepping}}
\end{figure}

Figure~\ref{fig:model.halorates.durham} illustrates this. We compare 
the halo-halo merger rates constructed from the Millenium simulation by 
\citet{fakhouri:halo.merger.rates,fakhouri:millenium.2.merger.rates} and \citet{genel:merger.rates.perprogenitor} 
to those constructed from the same simulation for use in the 
Durham semi-analytic models \citep{bower:sam,font:durham.sam.update}, 
according to the method described in \citet{harker:marked.correlation.function}. 
There is an abrupt cutoff in the number of major halo-halo mergers in 
the latter tree. The differences are analyzed in detail in the Appendix 
of \citet{fakhouri:halo.merger.rates}, but they largely owe to two post-processing 
cuts made after the original halo+subhalo tree construction: subhalos are 
removed from their parent friends-of-friends (FOF) 
identified halo if their centers are outside twice the half-mass 
radius, or if they have retained at least $75\%$ of their mass at the time they were 
last an independent halo. Subhalos (and their progenitors) that are removed from groups in this way 
are discarded from the trees.

The intention of these cuts is to reduce the number of spurious 
linkings from the FOF group finder, and both cuts above probably do so. However, 
they also suppress the number of halo-halo major mergers in the tree for three 
reasons. First, the discarded subhalos change the shape of the halo mass function -- 
some high-mass halos lose a non-negligible fraction of their mass and, on the exponential 
tail of the mass function, this can have large effects (since small changes in mass 
correspond to large changes in number densities). 
Second, the secondary (sub)halo in a major merger
is (fractionally) 
more robust to stripping than a small secondary in a minor merger, so major mergers 
are preferentially affected by the mass retention threshold. 
And third, in an equal mass 
merger, by definition the halos first overlap 
when the secondary center is at $\sim2\,R_{\rm vir}$ from 
the primary (in fact given the FOF algorithm used for linking, 
the distance can be $\sim3-5\,R_{\rm vir}$), 
so many of its subhalos will be instantaneously split off (being outside 
twice the half-mass radius at this instant) and thus 
not contribute to the secondary halo mass at the moment of merger 
(so the instantaneous mass ratio suddenly ``dips'' just before the merger). 

This is not necessarily problematic for galaxy merger rates. 
The trees so constructed are designed for 
tracking the galaxy population in a specific semi-analytic model; they are {\em not} 
designed to represent the global halo-halo merger rate. 
The effects above mainly pertain to definitions; 
they mostly apply to the early stages of subhalo-halo mergers, 
not necessarily the final, relaxed stages when a galaxy merger will occur.
As such they only indirectly affect galaxies, 
for example, via the small changes to the total halo mass introduced 
and corresponding gas mass assumed available for cooling. 
So the apparent difference in halo-halo merger rates translates much less directly 
in terms of galaxy-galaxy merger rates (basically, the baryonic physics act as a 
buffer between the halo-halo merger rate and galaxy-galaxy merger rate). But 
clearly, simply adopting the halo trees without accounting for how this processing 
is different from what is typically done (where e.g.\ halo mass ratio is the maximum pre-interaction  
halo mass ratio, for example) could lead to an apparent deficit of mergers (where 
the same mergers are, in fact, present). 

This is related to another illustration in 
Figure~\ref{fig:model.halorates.durham}, showing the result of adopting 
what might seem to be the most obvious definition of halo-halo mergers given 
a merger tree. Using a pair of closely spaced snapshots, and a descendant halo in 
the later snapshot, we simply take all the halos+subhalos in the earlier snapshot and define the mass 
ratio as the ratio of their instantaneous mass in that snapshot (the last where they 
are independent) to that of the main (most massive) progenitor. Specifically, we 
apply this algorithm to the standard 
Millenium simulation MPA merger trees. Mergers are again dramatically suppressed;
this was noted in \citet{bundy:mf.evol.vs.mergers}.
The issue is that the definition of ``halo-halo merger'' is not trivial: in the merger tree used 
to compute this particular example, halos ``survive'' in the tree so long 
as they can still be recognized as distinct subhalos/substructure/density peaks. 
In effect, this query is measuring the merger 
rate in terms of the instantaneous mass of a secondary subhalo just before it is no longer 
identifiable as a distinct subhalo/structure in the simulation (typically well after the 
two original halos have overlapped and become part of the same FOF group). 
By this time the smaller halo will have been 
stripped significantly, and with infinite resolution this particular way of 
defining mergers would make all mergers have a secondary mass $\rightarrow0$. 
The correct definition of secondary mass is clearly some maximum mass or mass before 
interaction/joining into the same FOF group.\footnote{These issues 
are described in detail at \KBurl\  
(K.\ Bundy). There, the authors both outline the definitions 
that lead to this artificial merger rate suppression and present a modified 
query of the Millenium database (courtesy of V.\ Springel \&\ S.\ White)
that defines halo-halo mergers in terms of the 
more proper pre-accretion secondary mass, that does not suffer from this problem.}

Another, somewhat related caveat, regards the time-stepping of simulation 
outputs. We illustrate one of the possible 
dangers in Figure~\ref{fig:model.halorates.timestepping}. 
Obviously, for a process to be well-resolved in time, 
simulation or model outputs must be spaced with some $\Delta t $ which 
is much less than the characteristic evolutionary timescale of interest. 
For galaxy halos, this is the Hubble time at each $z$. Halos grow with a characteristic 
doubling time that is of order the Hubble time (only weakly dependent on mass); 
also, at each redshift, the time between mergers is of order the Hubble time; 
and the timescale from dynamical 
friction for a major merger to complete 
is a factor of $\sim1/10-1/5$ of the Hubble time. 
In short, $\Delta t \ll t_{H}(z)$ is required, or in terms of redshift, $\Delta z \ll 1$ 
at all $z$ where resolved dynamics are desired. 

Without such resolution, mergers will be entirely missed -- for example, 
a system with initial mass $m_{1}=m$ could experience a merger with 
mass $m_{2}=3\,m$, then the product (mass $=4\,m$) merge with a 
mass $m_{3}=12\,m$ system. The central system has experienced two consecutive 
major mergers ($\mu=1/3$) in this case. However, if the timestepping is large, 
the outputs of the simulation will only indicate that at some initial time, 
halos of mass $m_{1}=m$, $m_{2}=3\,m$, and $m_{3}=12\,m$ were separate, 
then at a later time they were all together, and the most common assumptions 
applied in re-constructing merger histories will assume each merged onto 
the ``primary'' ($m_{3}$) separately, giving two minor mergers 
(a 1:12 and 1:4 merger, as opposed to the correct 1:3 and 1:3 merger). 
Considering the merger history along all branches of a given halo tree, 
the resulting uncertainty in merger rates can grow quickly if the 
time-stepping used is too large. 

If one adopts a ``merger timescale'' to follow the galaxies in their last stages 
of infall after the subhalo is unresolved, then information is obviously lost without 
narrow timesteps -- the merger time will necessarily be based on the last 
resolved dynamics, at large radii, which tends to lead to a significant 
over-estimate of the remaining ``merger time'' (see 
the discussion in \S~\ref{sec:compare.dark:subhalos:caveats}). 
This and the above can be significant concerns at high redshifts in semi-analytic 
models that interpolate between individual snapshots of the dark matter 
background at various times.

Moreover, if merger rates or mass ratios are defined in terms of 
the ``descendant'' mass -- i.e.\ if the mass ratio is defined by 
the mass of a progenitor $m_{i}$ (in one output) to the mass of the 
final halo $m_{f}$ (in the next output), as is common in many analyses, 
or if the halo mass bin for which the merger is ``counted'' (in e.g.\ asking 
the merger rate at a given mass) is $m_{f}$ -- 
then there can be a significant bias without narrow time-stepping because 
the halo will grow by a non-trivial amount in mass between the 
two timesteps (not just from the merger itself). 

Figure~\ref{fig:model.halorates.timestepping} illustrates this. We 
extract the halo-halo merger rate from the Millenium simulation using a 
simple query -- for every halo in a given timestep, we ask whether it had a 
progenitor (other than its primary/main-branch progenitor) in the 
previous timestep with mass $>1/4$ the final mass (corresponding to a 
$>1/3$ merger). 
Mass here, to avoid the other problems above, is 
defined as the maximum pre-accretion mass. But we vary the length of that timestep. 
If the timestep is large, then  
at redshifts $z>2$, the apparent merger rate obtained by such a method 
plummets, even though it is well-known that the halo-halo merger rate continues 
to rise indefinitely with redshift. This is because the time-spacing of simulation 
outputs becomes large ($\Delta z\ll 1$ is no longer satisfied) and halos 
grow non-trivially between timesteps, so major mergers may represent 
only a small fraction of the mass of the ``final'' halo after a large time 
(in units of the Hubble time) passes.

\breaker
\subsection{Substructure: Tracking Subhalos and Their Dynamics}
\label{sec:compare.dark:subhalos}

\subsubsection{The Subhalo Merger Rate}
\label{sec:compare.dark:subhalos:compare}

\begin{figure*}
    \centering
    \scaleup
    \plotter{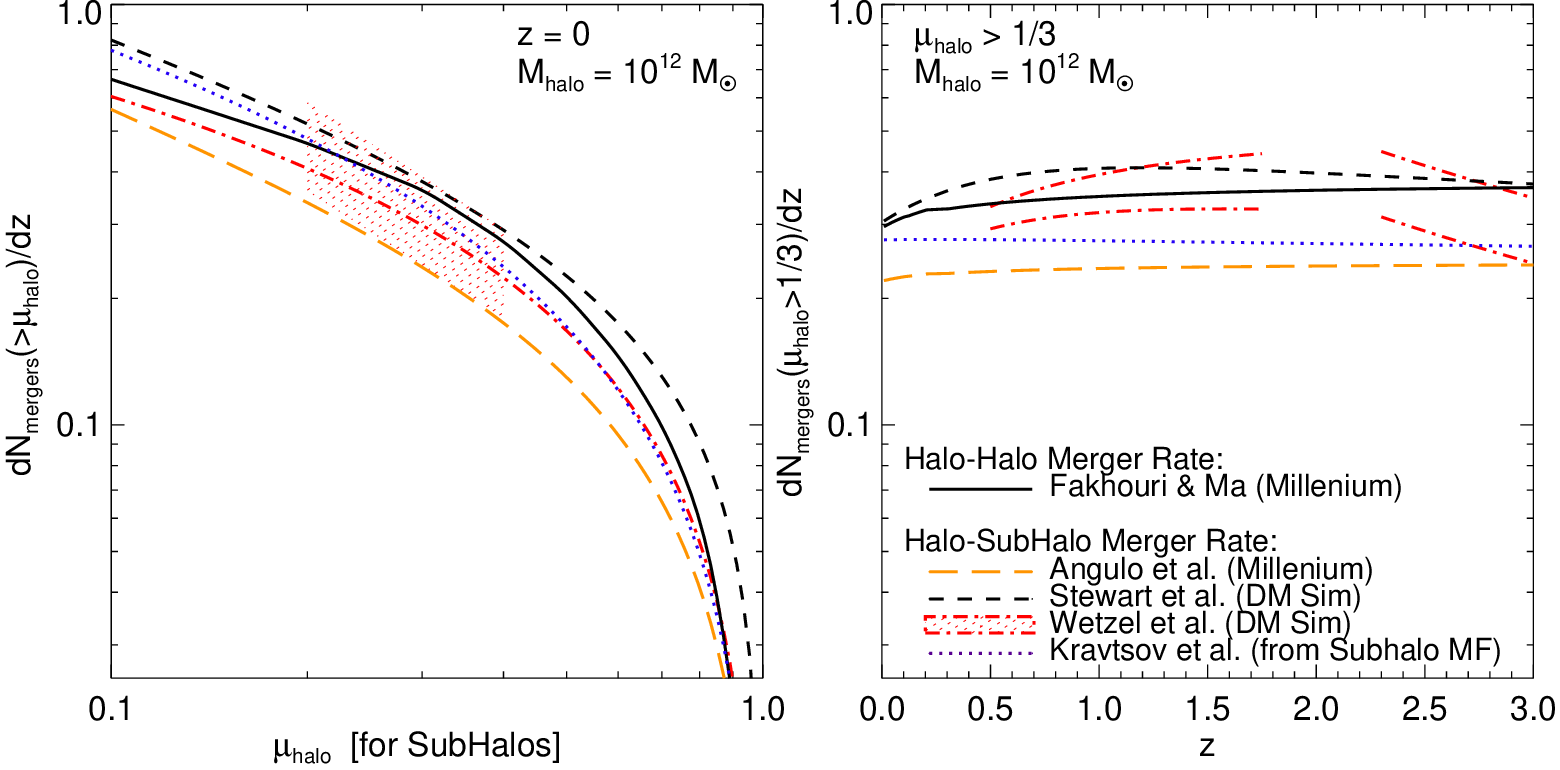}
    \caption{As Figure~\ref{fig:model.halorates}, but comparing 
    halo-halo merger rates \citep[from][]{fakhouri:halo.merger.rates} 
    to subhalo-halo merger rates. 
    Subhalo mergers are defined when subhalos are fully merged 
    or destroyed with $\mu_{\rm halo}=M_{\rm subhalo}/M_{\rm halo}$; 
    where $M_{\rm halo}$ is the instantaneous total primary 
    halo mass, and $M_{\rm subhalo}=M_{\rm infall}$ 
    is the subhalo's infall/maximum pre-accretion (pre-stripping) halo mass. 
    We compare with the results from the subhalo trees in the Millenium 
    simulation \citep[the same simulation as][]{fakhouri:halo.merger.rates}, taken from 
    \citet{angulo:millenium.substructure.survival,kitzbichler:mgr.rate.pair.calibration}; 
    with the subhalo trees constructed 
    from the independent DM simulation in \citet{stewart:merger.rates}; 
    from the subhalo mass functions determined in \citet{kravtsov:subhalo.mfs} with the 
    methodology outlined in \citet{hopkins:groups.qso}; and with the 
    subhalo mergers defined in another different manner in \citet{wetzel:mgr.rate.subhalos} 
    For the \citet{wetzel:mgr.rate.subhalos} results, 
    shaded range ({\em left}) shows typical statistical and systematic uncertainties; 
    double lines ({\em right}) bracket this range, they are broken as a function of 
    redshift because of the different volume simulations used to probe different redshifts. 
    Despite the fact that subhalos can survive a significant time after halo-halo 
    merger, the rates are similar within a factor $\sim2$ to the halo-halo merger 
    rates; different subhalo identification methodologies also yield similar results 
    to within the same factor. 
    \label{fig:model.subhalorates}}
\end{figure*}

Of course, a halo-halo merger is not a galaxy-galaxy merger; there 
will be some finite time where the galaxy is part of a satellite/substructure 
system before the final galaxy-galaxy coalescence. To follow this, 
simulations have made great improvements in tracking not just primary halos 
but also substructure in those halos, identifying subhalos on the fly 
\citep[see e.g.][]{springel:cluster.subhalos}. Figure~\ref{fig:model.subhalorates} 
plots the merger rates of such subhalos, measured from cosmological 
dark matter simulations: the subhalo ``merges'' when it is effectively 
destroyed in the simulation (it is mixed/stripped to the point where no 
subhalo can be identified by the algorithm, at least above the simulation 
mass resolution limit; we discuss some aspects of these limits below).\footnote{Implicit 
in Figure~\ref{fig:model.subhalorates} is the assumption that all subhalos merge 
with the central one, not with each other. This is not strictly true, but 
simulations find it effects merger rates at only the $\sim10-20\%$ level, 
much less than the differences between different models shown \citep{wetzel:mgr.rate.subhalos,
angulo:millenium.substructure.survival}.}
The subhalo is continuously losing mass to stripping after the 
initial halo-halo merger (and by definition its mass is effectively zero 
when it ``merges''), so the mass ratio is most appropriately defined in terms of the 
ratio of the {\em maximum} secondary halo mass (maximum mass 
it had when it was its own halo before any onset of stripping) to the 
instantaneous primary mass (technically primary minus subhalo mass, since we define 
an equal mass merger as 1:1, or $\mu_{\rm halo}<1$). 

We show the results of this determination from several different sources: 
the subhalo trees from the Millenium simulation (the ``MPAhalo'' catalog); 
the comparable but independent dark matter simulations in 
\citet{stewart:merger.rates}; and from a third set of simulations 
presented in \citet{wetzel:mgr.rate.subhalos}. We also consider 
subhalo tracking using a slightly different methodology, beginning with e.g.\ the 
subhalo mass function (rather than the differential subhalo merger rate) 
inside a radius $< R_{\rm vir}$, as determined from 
cosmological simulations in \citet{kravtsov:subhalo.mfs}, and evolving 
these subhalos forward from this already small radius according to the 
results of high-resolution galaxy-galaxy merger simulations used to calibrate 
e.g.\ dynamical friction times 
\citep[see][]{hopkins:groups.qso,hopkins:groups.ell}. 

Despite the fact that these are all independent simulations, with 
somewhat different algorithms used to identify and track subhalos, 
the agreement is reasonably good: differences in the major subhalo merger 
rate are at the factor $<2$ level. Moreover, the predicted subhalo major merger 
rate is quite similar to the predicted halo-halo major merger rate; depending 
on the calculation, subhalo major mergers appear to be between a systematic factor 
of $\sim2$ lower than, or equal to, the halo-halo major merger rate.

\subsubsection{The Halo Merger Rate with a ``Merger Delay''}
\label{sec:compare.dark:subhalos:delay}

\begin{figure}
    \centering
    \scaleup
    \plotter{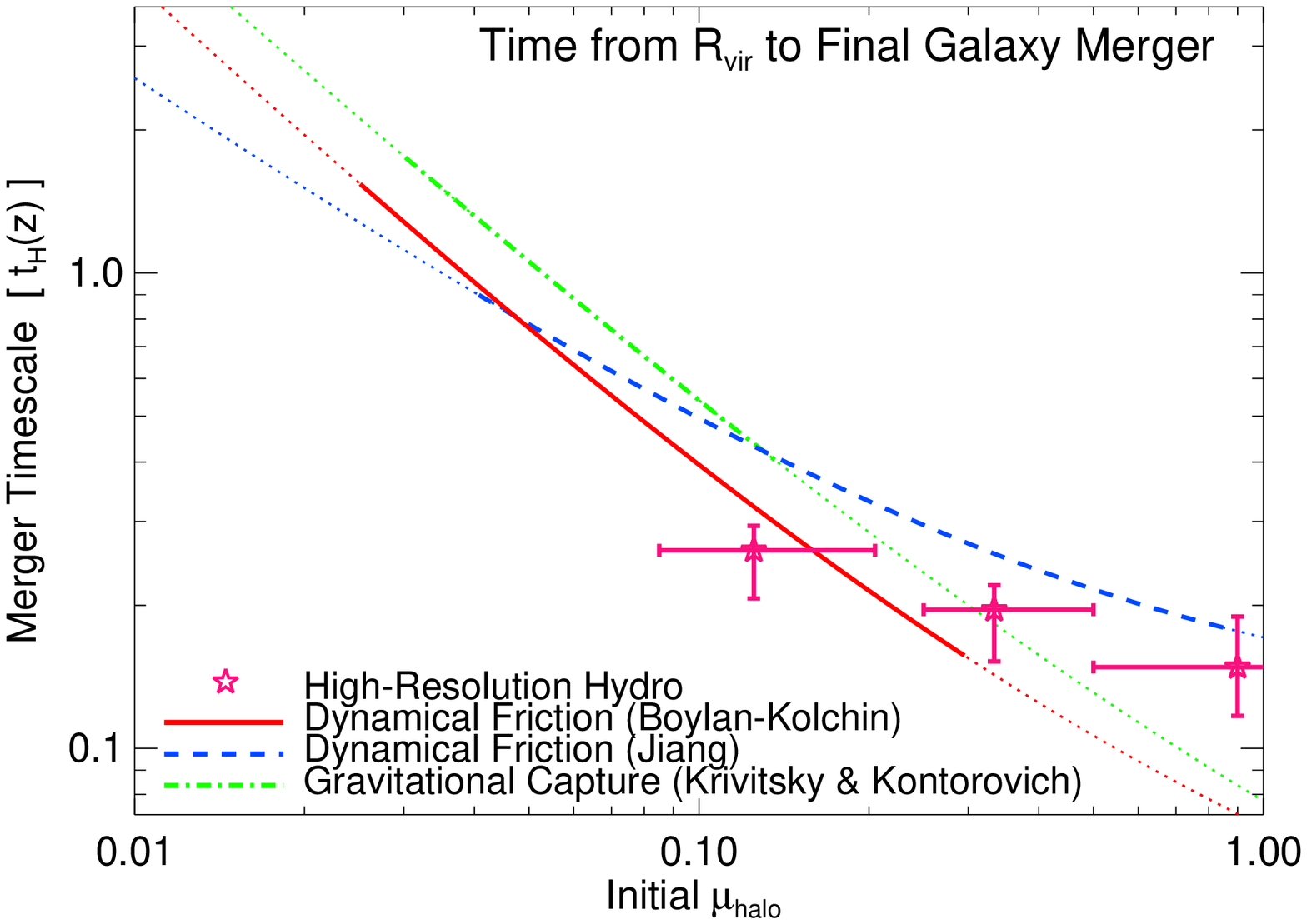}
    \caption{Comparison of different calibrations of the ``merger timescale'' 
    (average time between halo-halo merger defined at crossing of $R_{\rm vir}$ 
    and central galaxy-galaxy merger) in units of the Hubble time as a function of 
    initial mass ratio. 
    Each comes from high-resolution N-body simulations: 
    dissipationless binary galaxy+halo merger simulations surveying orbital parameters 
    from \citet{boylankolchin:merger.time.calibration} (thick line corresponds to the range 
    explicitly simulated; thin dotted line to extrapolation beyond this range), 
    hydrodynamic binary galaxy+halo mergers \citep{cox:kinematics,younger:minor.mergers}; 
    lower-resolution hydrodynamic cosmological simulations \citep{jiang:dynfric.calibration}, 
    and dissipationless gravitational/group capture cross sections for 
    galaxies in groups \citep{krivitsky.kontorovich}. All yield a similar factor $\approx2$ scatter 
    in timescales owing to cosmological distributions of orbital parameters 
    (lines are for median circularity and other orbital parameters in cosmological simulations). 
    Different estimators agree well where calibrated. 
    Major-merger timescales are $\ll t_{\rm Hubble}$; as a result, subhalo 
    major merger rates are not very different 
    from halo major merger rates. 
    \label{fig:tmerger.compare}}
\end{figure}

Another common approach, especially given low-resolution 
cosmological simulations or analytic (e.g.\ extended Press-Schechter) approaches 
where subhalo tracking is not possible, is to assign some ``merger timescale.'' 
This is a delay between the initial halo-halo merger and the final galaxy-galaxy merger, 
approximating the time for the decay of the secondary orbit and final resonant tidal interaction 
and coalescence of the two galaxies. Typically some variant of the standard 
Chandrasekhar dynamical friction time is employed, but this does not have to be 
the case: in fact, the angular momentum loss and 
orbital decay in the final stages of the merger are governed by resonant tidal processes, 
not dynamical friction against the smooth background. Because of these uncertainties, 
a number of authors have attempted to calibrate these timescales using 
high resolution simulations of galaxy-galaxy mergers and halo-subhalo or group 
encounters: we compare 
such calibrations in Figure~\ref{fig:tmerger.compare}. 

We consider: 
{\bf (a)} The directly extracted time required for the final merger 
in a galaxy-galaxy hydrodynamic (SPH) merger simulation (of disk+bulge+halo systems), 
including star formation, multi-component galaxies, and gas physics 
\citep{cox:kinematics,younger:minor.mergers,
hopkins:disk.survival,lotz:merger.selection}. 
We date the final merger by the coalescence of the two galactic nuclei. 
Error bars show the variation in timescales corresponding to sampling 
the full (isotropic) range of relative initial disk inclinations and orbital parameters 
(prograde mergers being a factor $\sim2$ more rapid to coalesce than retrograde 
mergers, owing to the enhanced tidal responses dissipating angular momentum). 
{\bf (b)} Calibration of a revised dynamical 
friction formula from gas-free (dissipationless) bulge+halo merger simulations in 
\citet{boylankolchin:merger.time.calibration}. These authors also survey a 
wide set of orbital parameters: we show the result for median and $\pm1\,\sigma$ 
range of orbital parameters measured in cosmological simulations 
\citep{benson:cosmo.orbits,khochfar:cosmo.orbits}. We show the full fit, but note that 
the simulations used span $\mu_{\rm halo}=\mu_{\rm gal}=0.025-0.3$, over which range they agree 
with the SPH simulations. 
{\bf (c)} A similar fit, for cosmological SPH simulations 
\citet{jiang:dynfric.calibration} \&\ \citet{jiang:baryon.fx.on.mgr.times}, calibrated from 
$\mu_{\rm halo}=0.04-1.0$. The cosmological nature of the simulations allows full 
sampling of representative orbits, with self-consistent galaxies; however the lower 
resolution suppresses resonant effects in major mergers, contributing to a slightly 
longer merger time. 
{\bf (d)} The characteristic timescale for pair-pair gravitational 
capture in loose group or field environments, calibrated from 
simulations in \citet{krivitsky.kontorovich}
\citep[see also][]{mamon:groups.review}. 
Although dynamical friction can be a reasonable approximation to these 
scalings, such capture cross-sections are technically more appropriate 
for collisions in e.g.\ small groups or field environments, or major 
mergers, where the idea of a long-lived, 
slow inspiral is not appropriate. An alternative calibration in angular-momentum 
space gives similar results \citep[see][]{binneytremaine}.

These timescales are discussed in greater detail in 
\citet{hopkins:merger.rates} and \citet{hopkins:groups.qso}. For our purposes 
they all agree reasonably well over the range 
where they are calibrated -- to a factor better than $\sim2$.
Where they are extrapolated, of course, they should be regarded 
with some caution. 

\begin{figure*}
    \centering
    \scaleup
    \plotter{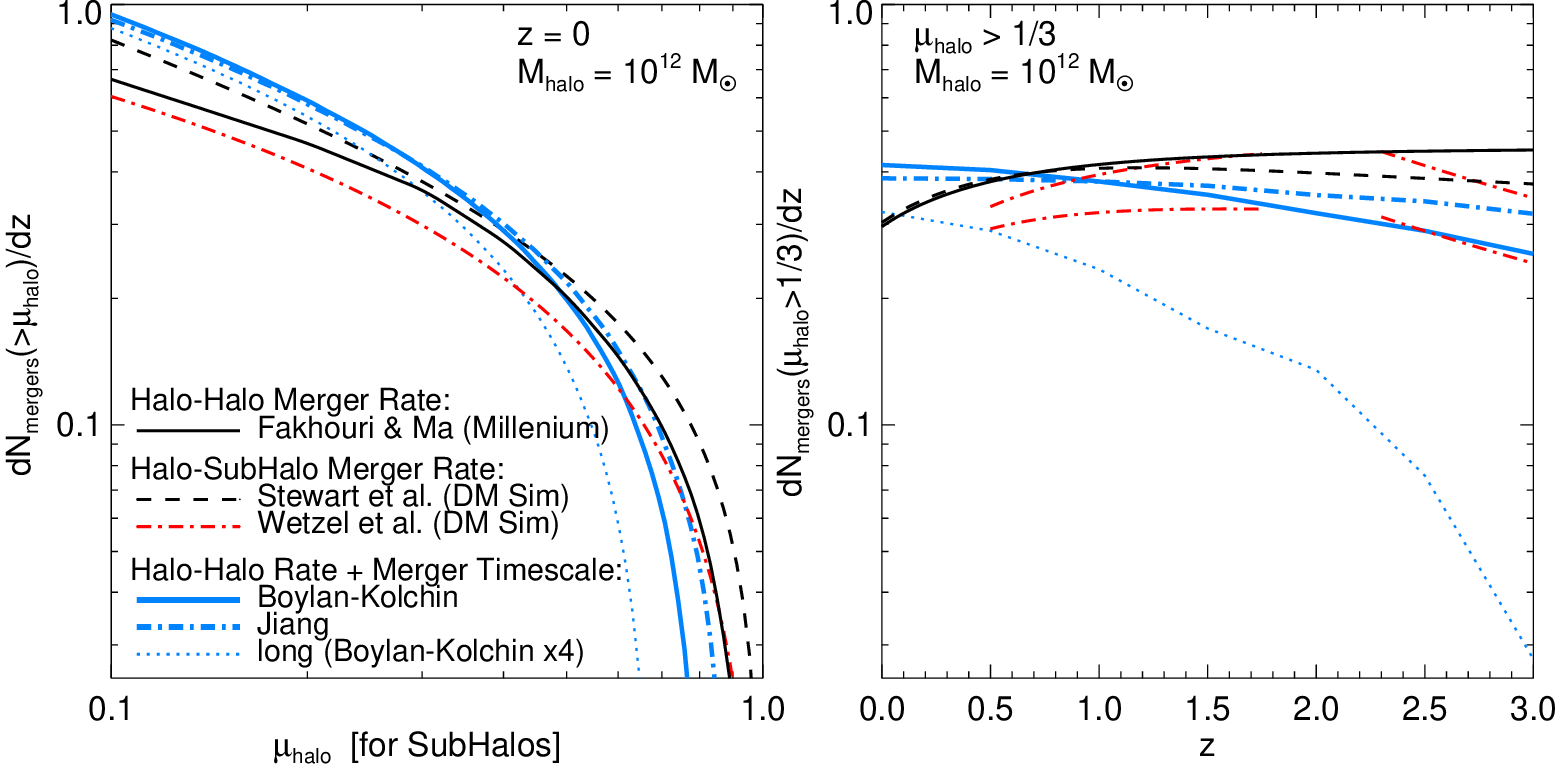}
    \caption{Comparison of halo and subhalo 
    merger rates from Figure~\ref{fig:model.subhalorates}, to 
    the merger rate determined by taking an ``initial'' halo-halo merger 
    rate \citep[that from][]{fakhouri:halo.merger.rates} and combining  
    with the merger timescales from Figure~\ref{fig:tmerger.compare} 
    (i.e.\ including a ``delay'' between the time of halo-halo merger 
    in the simulation and the final recording of the merger or assumed 
    subhalo-halo merger, given by the different merger timescale 
    fits in Figure~\ref{fig:tmerger.compare}). 
    The timescales of \citet{boylankolchin:merger.time.calibration} and 
    \citet{jiang:dynfric.calibration} bracket the major-merger timescales therein. 
    The difference is relatively small.
    We also compare the result given by artificially increasing the merger time 
    by a large factor ($=4$) relative to the calibration of 
    \citet{boylankolchin:merger.time.calibration}; in this case the merger timescale 
    becomes so long ($\sim t_{\rm Hubble}$) that high-redshift mergers 
    are substantially suppressed. 
    \label{fig:model.subhalorates.tdf}}
\end{figure*}

Figure~\ref{fig:model.subhalorates.tdf} shows the results of combining these 
merger timescales with halo-halo merger rates. In detail, each halo-halo merger 
is assigned a ``merger time'' based on the curves in Figure~\ref{fig:tmerger.compare}, 
appropriate for the redshift of the initial halo-halo merger and the halo-halo mass 
ratio at that moment, assuming the 
secondary begins at $R_{\rm vir}$ (which is what the fitted merger times are calibrated to use). 
The final merger is then delayed by this interval, during which the primary halo continues to 
grow and accrete. The mass ratio of the final merger is defined as the ratio of the 
maximum (pre-accretion) secondary halo to the primary halo mass at this time of final 
merger (as is done for subhalo merger rates). We consider both the fitted merger times 
from \citet{boylankolchin:merger.time.calibration} and 
\citet{jiang:dynfric.calibration}, which bracket the 
other calculations in Figure~\ref{fig:tmerger.compare}. 

To see how merger rates would be affected by very long merger timescales, 
we have also considered the \citet{boylankolchin:merger.time.calibration} 
time multiplied by a constant factor of $4$ (for major mergers, 
roughly, the \citet{jiang:dynfric.calibration} time 
multiplied by a factor of $\sim2$, or $\sim0.5\,t_{\rm Hubble}(z)$). This 
represents a fairly extreme case. 
We compare to the subhalo and halo-halo merger 
rates, as a function of mass ratio and redshift. 

The merger rates calculated with this approach agree to within a factor $\sim2$ with those 
determined from a direct tracking of subhalos in simulations: it appears that evolution 
on the timescales 
adopted for ``typical'' orbital parameters in simulations, shown in Figure~\ref{fig:tmerger.compare}, 
is a good approximation to the full dynamics. The redshift evolution is slightly different: 
high redshift mergers (when halo growth is more rapid) have their mass ratios somewhat more 
suppressed, but in terms of mergers per unit redshift the evolution is still weak and within the 
range of various subhalo calculations.\footnote{If satellite-central 
mergers were simply a delayed version of halo-halo mergers, keeping mass ratio fixed 
in infall, then the merger rate distribution versus mass ratio would have the same shape 
as that of halo-halo mergers but with a redshift lag -- and because merger rates 
increase with redshift, it would be higher than the immediate halo-halo merger rate.
But differential mass evolution in this time leads to fewer major mergers. 
Together these drive the effects seen in Figure~\ref{fig:model.subhalorates.tdf} 
\citep[see][]{wetzel:mgr.rate.subhalos}.}
The differences between each 
different merger time calibration are small (for major mergers): it is only when we artificially increase 
the merger timescale by a factor of several that a significant difference appears, in the sense that 
high-redshift mergers are strongly suppressed. 
This emphasizes that, although existing 
calibrations from high-resolution simulations agree well and yield little difference 
when used properly, adopting a significantly longer merger 
time can substantially suppress the merger rate, especially at high redshifts.

\subsubsection{Caveats: Subhalo 
Identification and The Application of Dynamical Friction to Orphans}
\label{sec:compare.dark:subhalos:caveats}

Despite the reassuring agreement seen above, it should be noted that merger 
rates are sensitive at a non-trivial level to issues of subhalo identification: 
in particular the choice of algorithm and criteria used to 
determine subhalo ``membership,'' and the time, mass, and force resolution 
of the simulation. These effects have been examined in a number of simulations  
\citep[see e.g.][and references therein]{klypin:subhalos.vs.resolution,springel:cluster.subhalos,
shaw:subhalo.masses,wetzel:mgr.rate.subhalos,wetzel:subhalo.disruption,
giocoli:subhalo.mf.universality}. 
We note here a couple of possible pathological regimes. 

First, it is possible in some particular circumstances that subhalos could survive ``too long'' 
in the simulations shown above, depending on the definition of the time of merger. This can 
happen, for example, because the contribution of 
baryonic galaxies to speeding up the merger is neglected. 
Including baryons keeps subhalos more tightly 
bound and changes the mass distribution, shortening the dynamical friction time 
\citep[see][]{weinberg:baryons.and.substructure}. Moreover, 
the baryonic components actually control the final merger, allowing for resonant 
interactions once the secondary is within a radius 
that contains a mass less than or of order its own mass (at which point the 
Chandrasekhar dynamical 
friction approximation is no longer valid). 
This makes the merger time within $\sim20-50\,$kpc shorter by a factor 
$\approx2$, relative to that estimated without baryonic galaxies 
\citep[see e.g.][]{lotz:merger.selection,
lotz:mgr.timescale.vs.massratio.morphology,lotz:gasfraction.vs.mergertime}, and can even 
lead to the merger and/or destruction of the baryonic galaxies 
{\em before} the subhalo (their bound dark matter) is significantly gravitationally disrupted 
or stripped \citep[see e.g.][]{donghia:resonant.stripping.dwarfs}. 
The lack of very high resolution and accurate {\em ab initio} formation of 
realistic disks (in e.g.\ their scale heights and sizes) in full cosmological 
simulations
contributes to the fact that the major merger timescales in the 
calibration of 
\citet{jiang:dynfric.calibration} are significantly longer than those in 
high-resolution hydrodynamic simulations of individual mergers of 
disk galaxies. 
And in the opposite, infinite-resolution limit of dark-matter simulations, a subhalo 
might never be completely disrupted (there always being a few tightly 
bound particles), even when the contained galaxies would have long ago merged. 

Correcting for these effects is difficult. Subhalos could be identified as 
``merged'' if they have lost some large fraction \citep[$\gtrsim90-99\%$; 
for calibrations see][]{wetzel:subhalo.disruption} of their 
mass (as opposed to simply reaching some resolution limit), but in low-resolution 
simulations this may prematurely ``destroy'' the subhalo. One could attempt 
to account for the resonant effects of baryons by merging any systems that 
have a passage within a close radius $r$ to another galaxy, such that the enclosed 
mass $M(<r)$ is comparable to the 
subhalo/secondary mass (really the secondary galaxy mass plus tightly bound dark matter 
within $\sim$a couple $R_{e}$) -- within any such radius the resonant and baryonic interactions 
will be efficient and 
dominate over the dark matter dynamics and simple dynamical friction is not applicable. 

A more common problem, especially in cosmological simulations (which by 
necessity have less than ideal resolution), is that subhalos are lost below 
the resolution limit ``too early.'' Identifying a 
subhalo as a distinct bound substructure is demanding in terms of 
resolution, and at some point subhalos will fall below these 
limits as they are stripped. 
Lack of inclusion of baryons (which would keep the subhalo more tightly 
bound), lower force resolution, or conservative subhalo 
membership criteria will all accelerate this. Subhalos in the Millenium simulation, 
for example, typically fall below the resolution limit within $\sim0.5-1\,R_{\rm vir}$ 
(at low masses in particular). \citet{wang:sdss.hod} show that 
if one simply took only the still-resolved subhalos in this simulation 
at any given time, it would not be possible to reproduce the clustering/halo 
occupation of observed galaxies on small scales 
($\lesssim100\,$kpc), or that of subhalos in higher force-resolution 
simulations. Similar conclusions are reached in \citet{wetzel:subhalo.disruption}.

\begin{figure}
    \centering
    \scaleup
    \plotterr{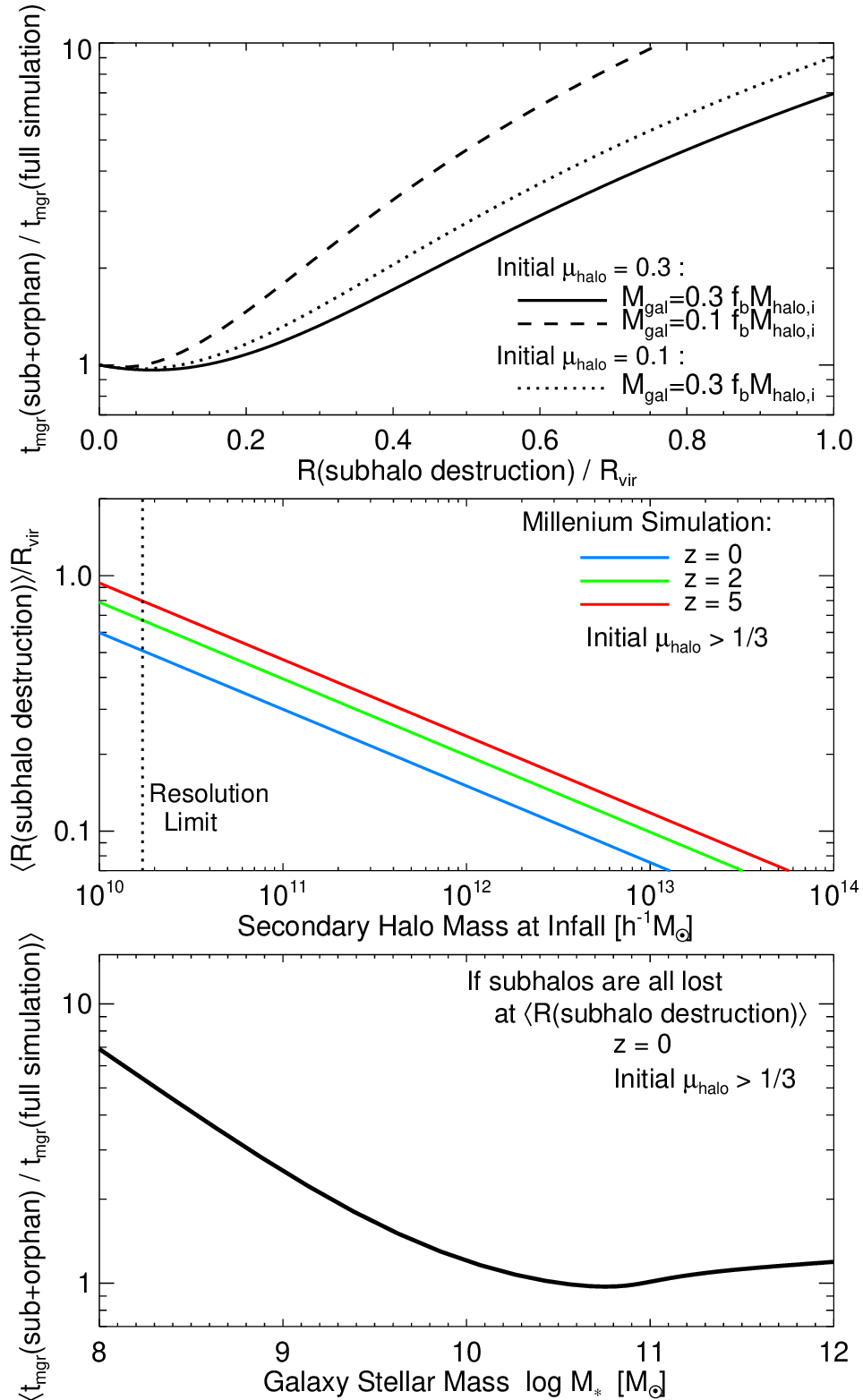}
    \caption{Caveats of using a subhalo+merger time (``orphan'')
    calculation of the total merger timescale.     
    {\em Top:} Ratio of the ``subhalo plus orphan'' timescale $t_{\rm mgr}(\rm sub+orphan)$
    to the true merger timescale $t_{\rm mgr}(\rm full\ simulation)$ 
    from $R_{\rm vir}$ to final merger. 
    The true timescale $t_{\rm mgr}(\rm full\ simulation)$ is given by the 
    calibrations in Figure~\ref{fig:tmerger.compare} -- simulations that follow 
    galaxies fully from $R_{\rm vir}$ to $R\rightarrow0$. 
    The subhalo+orphan timescale comes from following a subhalo self-consistently 
    to some $R$(subhalo destruction) where the subhalo is lost (falls 
    below some resolution limit) -- a timescale given by the same calibrations -- 
    then re-applying the merger time formula based on the instantaneous 
    $R$(subhalo destruction) and remaining galaxy+subhalo mass. 
    The ratio of this timescale to the true merger time is shown as a function 
    of $R$(subhalo destruction), the initial halo-halo mass ratio, 
    and galaxy mass. If $R$(subhalo destruction)/$R_{\rm vir}\gtrsim0.2$, 
    the total merger time can be significantly over-estimated. 
    {\em Middle:} Average $R$(subhalo destruction) relative to $R_{\rm vir}$, where 
    subhalos in the Millenium simulation fall below the 
    mass resolution limit, as a function of 
    the secondary/satellite halo infall mass and redshift (S.\ Genel; private 
    communication). Small satellites, which are close to the simulation resolution 
    limit, naturally fall below the resolution limits at relatively large $R$ in any simulation. 
    {\em Bottom:} Effect on the ``typical'' inferred merger time for $1:3$
    mergers, assuming subhalos are lost at the radii of the $z=0$ curve at {\em middle} 
    and given the methodology shown at {\em top}, as a function of galaxy mass 
    (using the $M_{\rm gal}-M_{\rm halo}$ relation from \citealt{wang:sdss.hod}).
    If subhalos are only marginally resolved,  
    caution should be used in applying merger times based on instantaneous 
    subhalo properties; better calibration is needed. 
    \label{fig:tmerger.orphan.issues}}
\end{figure}

To compensate for 
this, many models built on halo+subhalo trees invoke a population of 
``orphan'' galaxies. This is the same principle as the ``merger timescale'' 
above: the models follow subhalos in the cosmological 
simulation as far as possible, and then assign a ``merger time'' to the galaxy 
based on the last recorded subhalo properties in the last timestep where 
the subhalo could be identified. 
Usually, the assigned timescale follows the 
Chandrasekhar dynamical friction formula,
\begin{align}
\label{eqn:tdf}
t_{\rm merger} &\approx
1.17\,\frac{V_{\rm vir}\,r_{\rm sat}^{2}}{G\,m_{\rm sat}\,\ln{(1+M_{\rm vir}/m_{\rm sat})}}\\
\nonumber &=(1.05\langle\epsilon\rangle^{0.47}+0.6)\,
\frac{R_{\rm vir}}{V_{\rm vir}}\,
\frac{M_{\rm vir}/m_{\rm sat}}
{\ln{(1+M_{\rm vir}/m_{\rm sat})}}\,
\left \langle
\frac{R_{c}}{R_{\rm vir}}
\right \rangle^{1/2}
\end{align}
from \citet{binneytremaine}, with the second equality coming from 
the calibration in \citet{jiang:dynfric.calibration} assuming 
initial $r_{\rm sat}=R_{\rm vir}$ 
(their Equation~8; $\epsilon=j/j_{c}$ is the orbital eccentricity 
with $\langle \epsilon \rangle=0.5$, and 
$R_{c}/R_{\rm vir}$ is related to the initial pericentric passage distance, 
both taken from the distributions they measure). 
Fits to high-resolution simulations with somewhat more freedom in the scalings give 
\begin{equation}
t_{\rm merger}=0.56\,{\Bigl(}\frac{R_{\rm vir}}{V_{\rm vir}}{\Bigr)}\,
\frac{(M_{\rm vir}/m_{\rm sat})^{1.3}}{\ln{(1+M_{\rm vir}/m_{\rm sat})}}\,
{\Bigl (}\frac{R_{\rm sat}}{R_{\rm vir}}{\Bigr )}
\label{eqn:tdf.bk}
\end{equation}
from \citet{boylankolchin:merger.time.calibration} (again adopting a 
the median initial circularity $j/j_{c}=0.5$), 
where $m_{\rm sat}$ and $r_{\rm sat}$ are the satellite mass and location 
at the last time its subhalo could be identified, and 
$V_{\rm vir}$ and $M_{\rm vir}$ are the virial velocity and mass of the 
primary (minus the subhalo/secondary itself). 
Similar scalings are obtained from analytic considerations 
(at least in the small $m_{\rm sat}/M_{\rm vir}$ limit), accounting 
for subhalo mass loss and orbital parameter evolution \citep[see][]{taylor:substructure.evolution}.

These particular formulations, discussed above 
(see Figure~\ref{fig:tmerger.compare}), 
have been tested in simulations and shown to be a 
reasonable approximation to the {\em total} merger 
time from initial halo-halo contact at $r_{\rm sat}=1-2\,R_{\rm vir}$, 
with $m_{\rm sat}$ defined as the maximum (pre-stripping) mass of the secondary 
(this is the input when used to calculate the merger time for a halo-halo merger, 
as in \S~\ref{sec:compare.dark:subhalos:delay}). 
However, these formulations have {\em not} been calibrated for application to a satellite at 
a later time, after the subhalo has self-consistently evolved and is heavily stripped. 
Assume, for example, that a subhalo with initial (pre-stripping) 
mass $m_{s}$ self-consistently decays in orbit 
from an initial radius $R_{i}\sim R_{\rm vir}$
according to the calibration from \citet{boylankolchin:merger.time.calibration} 
(Equation~\ref{eqn:tdf.bk} above), but at some point -- when the subhalo is 
at a radius $r^{\prime}_{s}\equiv R[{\rm subhalo\ destruction}]$ and 
time $t=t_{\rm Eqn.~\ref{eqn:tdf.bk}}(R_{i}\rightarrow r^{\prime}_{s}\,|\,m=m_{s})$
-- can no longer be resolved. At this point the subhalo mass is $m^{\prime}_{s}$. 
Now assume that the remaining merger time is then assigned according to 
the same Equation~\ref{eqn:tdf.bk}, with these revised input values, 
i.e.\ a remaining $t=t_{\rm Eqn.~\ref{eqn:tdf.bk}}(r^{\prime}_{s}\rightarrow0\,|\,m=m^{\prime}_{s})$. 

The {\em total} merger time is then simply the time from $R_{\rm vir}$ to $r^{\prime}_{s}$ 
plus the assigned remaining time. But we also know what the total merger 
time should be from the initial $R_{\rm vir}$ radius to $r=0$ (i.e.\ without breaking into 
sub-steps in this manner) -- this is in fact what the 
formulae used above are explicitly calibrated from high-resolution simulations 
to reproduce (with the inputs of initial $R_{\rm vir}$ and $m_{s}$). Comparing 
the two, we obtain the ratio of the total merger time estimated from this 
two-stage breakdown, to the total merger time 
from $R_{\rm vir}$ to $r=0$ directly from the simulations. 
This is 
\begin{align}
\nonumber &\frac{t_{\rm mgr}({\rm sub+orphan})}{t_{\rm mgr}({\rm full\ simulation})} \\
\nonumber &=\frac{t_{\rm Eqn.~\ref{eqn:tdf.bk}}(R_{\rm vir}\rightarrow r^{\prime}_{s}\,|\,m=m_{s}) + 
t_{\rm Eqn.~\ref{eqn:tdf.bk}}(r^{\prime}_{s}\rightarrow 0\,|\,m=m^{\prime}_{s})}
{t_{\rm Eqn.~\ref{eqn:tdf.bk}}(R_{\rm vir}\rightarrow 0\,|\,m=m_{s})} \\ 
&= 1 + 
{\Bigl \{ }\frac{m_{s}^{1.3}\,\ln{[1 + M_{\rm vir}/m_{s} ]}}{m^{\prime 1.3}_{s}\,\ln{[1 + M_{\rm vir}/m^{\prime}_{s} ]}}
\,{\Bigl(}\frac{r^{\prime}_{s}}{R_{\rm vir}}{\Bigr)}-1
{ \Bigr \}}\,{\Bigl(}\frac{r^{\prime}_{s}}{R_{\rm vir}}{\Bigr)}
\end{align}
But $m_{s}/M_{\rm vir}$ is just the initial halo-halo mass ratio 
($\mu_{s}$). The final mass $m^{\prime}_{s}$, evaluated 
instantaneously when the subhalo is nearly or entirely stripped, 
will by definition just be the galaxy mass $M_{\rm gal} = \eta\,f_{b}\,M_{\rm halo}$, 
where $f_{b}$ is the Universal baryon fraction, $M_{\rm halo}$ is the {\em pre-accretion} 
subhalo mass, and $\eta$ is some star formation efficiency ($\eta=0.3$ around $\sim L_{\ast}$ 
where star formation is most efficient and drops rapidly at lower or higher 
masses). If we include the mass of the subhalo at the last time it was resolved, we 
add to this roughly the simulation resolution limit; but for intermediate to high 
mass galaxies and and the resolution limits of simulations of interest, this is 
similar ($\sim10^{9}-10^{10}\,\msun$). For typical 
$\sim L_{\ast}$ galaxies and the parameters of e.g.\ the Millenium simulation, 
with an initial 1:3 halo-halo merger, this yields 
$t({\rm sub+orphan})/t({\rm full\ simulation})\approx 1+ (6.5\,x-1)\,x$, 
where $x=r^{\prime}_{s}/R_{\rm vir}$ is the fraction of the virial radius 
where the subhalo is no longer resolved. 

Figure~\ref{fig:tmerger.orphan.issues} illustrates this as a function 
of initial mass ratio, galaxy mass/resolution threshold, and the radius where 
the subhalo becomes unresolved. In short, if subhalos can be self-consistently followed to 
small radii $<0.1-0.2\,R_{\rm vir}$, where the remaining merger time is short, there 
is no issue. However, if subhalos fall below the resolution limit at large radii, 
then the merger time can be over-estimated by a large factor. 
We compare the typical radii in the Millenium simulation where subhalos 
of a given initial (pre-accretion) mass become unresolved (after being 
accreted into a larger halo), at several redshifts (S.\ Genel, 
private communication). Unsurprisingly, when the subhalo 
initial mass is close to the resolution limit, subhalos 
are lost at $\sim R_{\rm vir}$. Given the $M_{\rm gal}(M_{\rm halo})$ relation 
and these median radii as a function of mass, the 
Figure illustrates how this affects the expected merger time for satellites of a given 
mass: at high masses, subhalos are massive and can be followed self-consistently, 
at low masses however, subhalos are lost quickly and the application 
of Equations~\ref{eqn:tdf}-\ref{eqn:tdf.bk} to the ``orphan'' tends 
to over-estimate merger timescales. This will have a large effect on 
predicted galaxy mass assembly at high redshifts. 

The issue is that fitting formulae such as Equations~\ref{eqn:tdf} \&\ \ref{eqn:tdf.bk}
are not necessarily calibrated for these multi-tiered 
applications. This is not to say that dynamical friction is fundamentally invalid -- 
orbits of small satellites can be well-approximated by application of an 
instantaneous dynamical friction drag force and tidal stripping
\citep{quinn86:dynfric.on.sats,hernquist:sat.orbital.decay,
velazquezwhite:disk.heating,
font:sats.on.large.orbits,villaloboshelmi:minor.mergers,benson:heating.model}. 
(Although for major mergers, the underpinning assumptions of dynamical friction 
are not strictly valid). But this is not as simple as a single timescale, and 
implicitly assumes a high-resolution limit -- they are not necessarily designed to account for a 
resolution ``floor'' self-consistently. 
Moreover, many estimates of the merger time average out or ignore 
the detailed orbital information; often, cases where the subhalo is quickly dissociated/stripped
correspond to highly radial orbits, and simply re-applying an 
orbit-averaged formula is not ideal. If a system has just had a radial passage, for example, 
it will most likely be ``caught'' in a snapshot at apocenter -- applying the dynamical friction time 
from this point, assuming the orbit is circular (as in most models) will in fact over-estimate 
the remaining merger time by a factor of $\approx 6$.

\breaker
\section{The Light Side: How Galaxies Occupy Halos}
\label{sec:compare.hods}

The issues discussed thus far pertain to all models, whether 
they are halo occupation, semi-analytic, or simulation-based. 
Although we have seen that some of the aspects of the treatment of 
dark matter mergers can be important, we know from our 
comparison in \S~\ref{sec:empirical} and the discussion above 
that most of the order-of-magnitude differences in merger rates must primarily 
owe to the treatment of baryons. 
Once we know how halos/subhalos behave, then the other necessary ingredient to 
predict galaxy merger rates is some prescription for how galaxies populate those 
halos (in other words, the ``halo occupation'' statistics). We therefore investigate how 
differences in these halo occupation 
aspects of the models give rise to different merger rates.

\subsection{Halo Occupation Statistics of {Central} Galaxies}
\label{sec:compare.hods:central}

\subsubsection{At Low-Redshift}
\label{sec:compare.hods:central:local}

\begin{figure*}
    \centering
    \scaleup
    \plotter{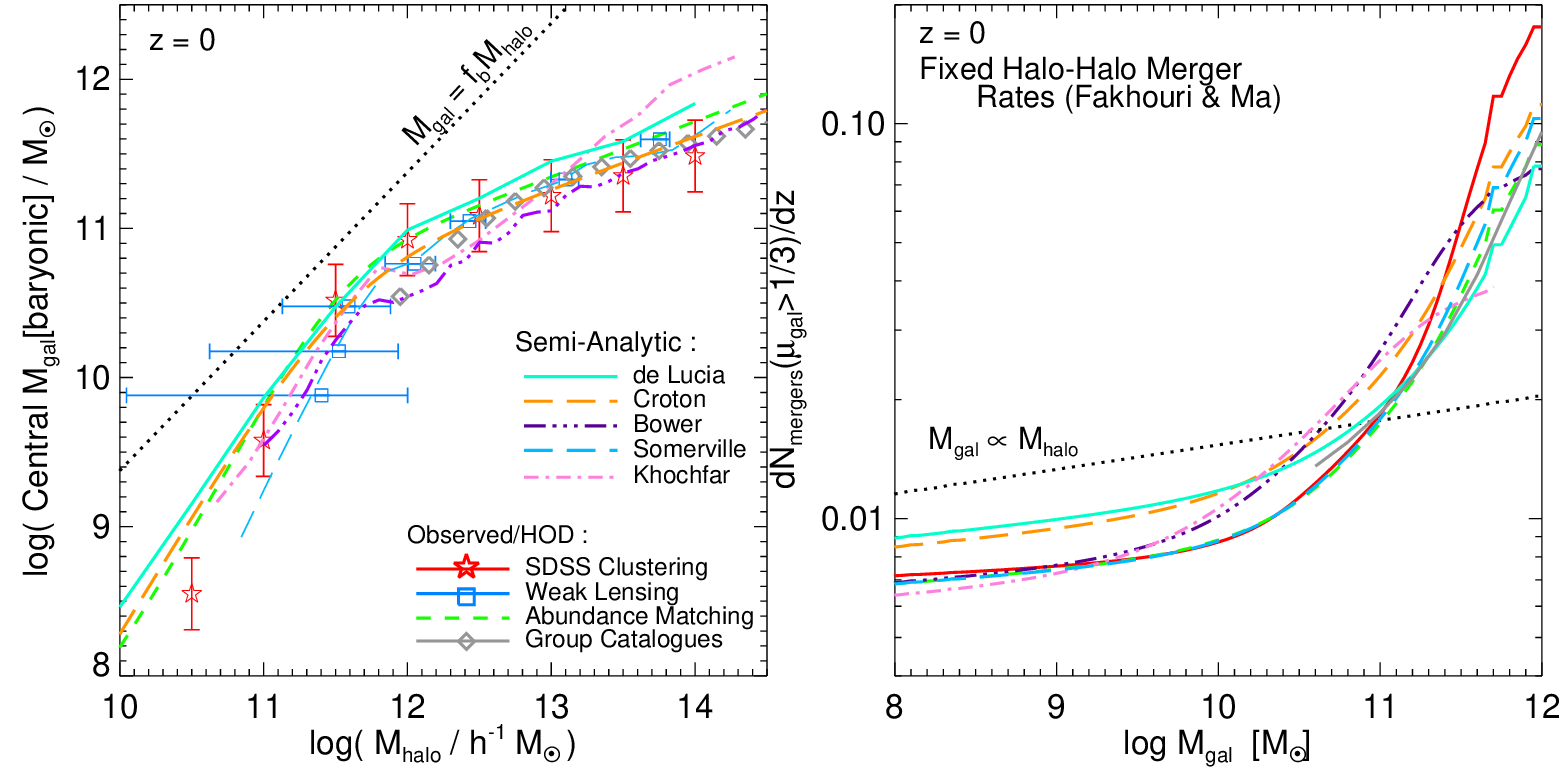}
    \caption{{\em Left:} Halo occupation statistics (mean galaxy baryonic 
    mass as a function of host halo mass) from 
    observations of weak lensing \citep{mandelbaum:mhalo}, 
    fits to halo occupation models from clustering 
    data in \citet{wang:sdss.hod}, using the monotonic abundance matching method 
    in \citet{conroy:monotonic.hod}, or from group catalogs with satellite 
    abundances and kinematics \citep{yang:clf.update.bycolor}. 
    We compare with different semi-analytic 
    model predictions, and the simplest efficient star formation model 
    ($M_{\rm gal}=f_{b}\,M_{\rm halo}$; similar to hydrodynamic simulations without 
    feedback).
    {\em Right:} Corresponding merger rates (at $z=0$) as a function of $M_{\rm gal}$ 
    (in $\pm0.5$\,dex bins of $M_{\rm gal}$), combining these HODs with 
    the dark matter merger rates from \citet{fakhouri:halo.merger.rates}. We hold 
    all other properties fixed. For now, we consider just the {\em central} galaxy HOD, 
    and assume satellites obey the same relation (where $M_{\rm halo}$ is the maximum 
    pre-accretion mass for satellites). Different observational constraints 
    yield small (factor $\sim1.5$ near $\sim L_{\ast}$) differences at $z=0$; 
    SAMs (being constrained to reproduce the 
    stellar mass function) agree within factors $\sim1.5-2$. However, 
    simply taking halo-halo merger rates or $M_{\rm gal}=f_{b}\,M_{\rm halo}$, 
    the merger rate predicted is different by factors of $\sim3-10$. 
    \label{fig:model.HODs}}
\end{figure*}

With some complete {\em a priori} knowledge of subhalo evolution and mergers, 
the galaxy-galaxy merger rate, as a function of galaxy mass and 
mass ratio, can be determined knowing the distribution of galaxy masses in each halo. 
To lowest order, this is given by the function $M_{\rm gal}(M_{\rm halo},\,z)$ and its 
scatter. Of course, this could depend on other variables such as environment, 
but as discussed below, observations indicate that such additional parameters have 
much smaller systematic effects than the relevant uncertainties 
in $M_{\rm gal}(M_{\rm halo},\,z)$. 

For simplicity, and to isolate the important physics, we first consider only 
{\em central} galaxies (i.e.\ those where $M_{\rm halo}$ is the parent halo, 
not satellites in subhalos), at $z=0$. 
Figure~\ref{fig:model.HODs} shows the median 
$M_{\rm gal}(M_{\rm halo})$, as a function of $M_{\rm halo}$, at $z=0$ 
\citep[we could also compare the scatter in this quantity, but 
all the models we consider predict a similar, relatively small $\lesssim 0.15-0.2\,$dex 
scatter; see e.g.][]{yang:clf.update.bycolor,
moster:stellar.vs.halo.mass.to.z1,
behroozi:mgal.mhalo.uncertainties,
wetzel:subhalo.disruption,more:2009.halo.mass.cen.mass.relation,
guo:2010.hod.constraints}. 
Of course, since every satellite was a central galaxy in some halo of mass 
$M_{\rm halo}$ before being accreted, the halo occupation function and 
$M_{\rm gal}(M_{\rm halo})$ function of satellites should be identical 
(barring strong satellite-specific physics and/or strong redshift 
evolution in the $M_{\rm gal}(M_{\rm halo})$ function since 
the time the satellite was accreted), 
so long as $M_{\rm halo}$ for satellites is defined as the maximum (pre-accretion) 
mass of the satellite subhalo (i.e.\ maximum mass when the satellite was central 
in its own halo). We will free this assumption and discuss 
the implications in \S~\ref{sec:compare.hods:satellites} below. 
It is worth noting that the vast majority of the galaxy number density and mass 
density at all galaxy masses is in central galaxies: this is the 
quantity that is in practice
most strongly constrained by a model's fit to the galaxy stellar mass 
function or the clustering amplitude 
(large-scale bias) of galaxies as a function of mass/luminosity. 

Given some function $M_{\rm gal}(M_{\rm halo})$, the galaxy-galaxy merger rate is simply 
this function convolved with the population of halo-halo mergers, 
with the appropriate accounting for delays before the final subhalo merger/destruction 
or the ``merger timescale'' from e.g.\ dynamical friction. 
Since we aim to isolate the specific effects of 
differences in the distribution $M_{\rm gal}(M_{\rm halo})$, we show in Figure~\ref{fig:model.HODs} 
the resulting merger rate after convolving each 
presented halo occupation function with the {\em same} halo merger rate 
\citep[in this case that from][]{fakhouri:halo.merger.rates}. A different choice 
of halo merger rates or subhalo treatment will systematically 
change the predicted rates along the lines shown in \S~\ref{sec:compare.dark}, 
but will not change how the galaxy-galaxy 
merger rate systematically depends on HOD quantities. 

We compare the predicted $M_{\rm gal}(M_{\rm halo})$ distributions from different 
theoretical models with several observational estimates. 
First, fitting to the correlation function of galaxies with a given baryonic or 
stellar mass\footnote{Because it is closer to the relevant parameter in 
e.g.\ merger simulations, and because it better highlights the 
relevant systematic effects, we will show our comparisons here usually in terms of 
the galaxy baryonic mass $M_{\rm gal}$. However, our qualitative statements in 
this section apply to stellar mass $M_{\ast}$ as well. 
Where necessary, we will convert between observed stellar and baryonic masses 
using the fits to the observed $M_{\rm gas}/M_{\ast}$ relation of galaxies as a 
function of redshift from $z=0-3$ given in \citet{stewart:disk.survival.vs.mergerrates}; 
this is consistent with a large number of direct observations over this redshift 
range \citep{belldejong:disk.sfh,mcgaugh:tf,
calura:sdss.gas.fracs,shapley:z1.abundances,erb:lbg.gasmasses,
puech:2010.baryonic.tf.z06,mannucci:z3.gal.gfs.tf,cresci:dynamics.highz.disks,
forsterschreiber:z2.sf.gal.spectroscopy,erb:outflow.inflow.masses}.} 
yields a characteristic host halo mass for each galaxy mass bin -- we 
show the results of this exercise performed on SDSS galaxies at 
$z=0$ from \citet{wang:sdss.hod}.
We also show the empirical 
``abundance matching'' or ``rank ordering'' results: it has been shown that good fits 
to halo occupation statistics (e.g.\ group counts, correlation functions as a function of 
galaxy mass and redshift, etc.) over a range of 
redshifts $z=0-4$ are obtained by simply rank-ordering all galaxies 
and halos+subhalos in a given volume and assigning one to another in a monotonic 
one-to-one manner \citep[see e.g.][]{conroy:monotonic.hod,valeostriker:monotonic.hod}. 
Here we perform this exercise using the redshift-dependent 
stellar mass functions from \citet{fontana:highz.mfs} (showing the result at $z=0$); 
but as discussed in \S~\ref{sec:empirical:robustness} other choices of 
stellar mass function make little difference. 
Recent studies \citep{guo:2009.structural.props.central.gals.vs.env,
moster:stellar.vs.halo.mass.to.z1,tinker:quenching.by.mergers.preferred,
behroozi:mgal.mhalo.uncertainties} obtain similar results 
(they are statistically indistinguishable from the other observational 
results shown in Figure~\ref{fig:model.HODs}). 
We compare these results to independent estimates of the 
$M_{\rm gal}(M_{\rm halo})$ distribution, 
where halo masses are determined from weak lensing studies in 
\citet{mandelbaum:mhalo}. 
Other independent measurements give nearly identical constraints: these include 
estimates of halo masses in low-mass spirals as a function of mass 
from rotation curve fitting \citep[see e.g.][]{persic88,persic90,persic96,persic96:data,
belldejong:disk.sfh,belldejong:tf,borriello01,borriello03,shankar06,avilareese:baryonic.tf} 
and estimates of the halo masses of massive groups and clusters from e.g.\ 
X-ray gas and group kinematics, as a function of 
central galaxy mass in massive brightest 
group or brightest cluster galaxies \citep[BGGs/BCGs; see][and references 
therein]{eke:groups,yang:obs.clf,brough:group.dynamics,
vandenbosch:concordance.hod}. 

In semi-empirical or halo occupation-based models of galaxy-galaxy merger 
rates, the required $M_{\rm gal}(M_{\rm halo})$ distribution is simply adopted 
from these or similar empirical constraints, and enforced on all the galaxies 
in the model. Such models are therefore different only insofar as different empirical 
constraints yield different estimates of $M_{\rm gal}(M_{\rm halo})$. 
We find that the effects of these differences are relatively small 
(factor $<2$ near $\sim L_{\ast}$, at $z\lesssim2$); the level of variation 
within such models owing to observational uncertainties is discussed in 
detail in \citet{hopkins:merger.rates} and also \citet{stewart:merger.rates}, and 
summarized in \S~\ref{sec:empirical:robustness}. Figure~\ref{fig:model.HODs} 
also clearly illustrates that the different observational estimates of the 
$M_{\rm gal}(M_{\rm halo})$ distribution yield similar merger rates; 
the uncertainties are largest at very low 
mass (where observations are difficult and knowing the true gas fractions of 
galaxies, which are gas-dominated at these masses, is critical) and 
very high mass (where statistics are few and, given the stellar and 
halo mass function shapes, the merger 
rate will be most sensitive to small changes in the HOD). 

Now compare these results with those from the 
$M_{\rm gal}(M_{\rm halo})$ distribution predicted by various {\em a priori} models. 
The semi-analytic models, for the most part, match the observational constraints 
quite well. This should not be entirely surprising: a major constraint used in 
building those models is that they reproduce the observed $z=0$ galaxy 
luminosity and stellar mass functions. Such a result (an important 
achievement in and of itself, we stress) implicitly guarantees a reasonable match 
to the observations in Figure~\ref{fig:model.HODs}. Of the specific choices 
shown, the \citet{bower:sam} model appears to deviate from the observations 
to a slightly greater extent, around $\sim L_{\ast}$. The manner of this discrepancy 
reflects a common tendency for models to not reproduce quite as sharp a break 
in the stellar mass function as is observed. In a very narrow stellar mass bin, this 
could lead to non-trivial differences in the predicted merger rate, but averaged over a 
reasonable baseline ($\sim0.5-1$\,dex in stellar mass), the 
{\em shape} of the predicted HOD is nevertheless in sufficiently good agreement with 
the observational constraints that it yields merger rates converged at the $\sim50\%$ 
level. It seems clear that, insofar as they succeed in reproducing the observed 
stellar mass functions, the differences between merger rates in semi-analytic 
models owing to differences in the predicted $M_{\rm gal}(M_{\rm halo})$ distribution of 
{\em central} galaxies at low redshifts 
are no larger than the differences that result when one attempts to directly adopt 
the $M_{\rm gal}(M_{\rm halo})$ distribution from observational constraints. 

As discussed at length in \citet{hopkins:merger.rates}, 
the HOD shape is non-trivial, and this leads to a significant galaxy-mass 
dependence of the galaxy-galaxy major merger rate, even though the 
halo-halo merger rate depends only weakly on halo mass. 
If the galaxy formation efficiency were to be independent of halo mass, 
so that $M_{\rm gal}\propto M_{\rm halo}$, then 
galaxy-galaxy mergers would be a direct, trivial 
reflection of halo-halo mergers (regardless of the absolute value of that 
constant of proportionality). However, at low masses, the dependence of 
$M_{\rm gal}$ on $M_{\rm halo}$ is steep, $M_{\rm gal}\propto M_{\rm halo}^{1.5-2}$. 
So a ``typical'' 1:3 halo-halo merger is a 1:9 galaxy-galaxy merger, and 
the galaxy-galaxy major merger rate is correspondingly suppressed 
(relative to the halo-halo merger rate). At high masses, 
the dependence is shallow, $M_{\rm gal}\propto M_{\rm halo}^{0.3-0.5}$, 
so a typical 1:9 halo-halo merger is a 1:3 galaxy-galaxy merger, 
and the galaxy-galaxy merger rate is enhanced. 

This is important for model merger rates in hydrodynamic galaxy formation 
simulations: it is well-established that in cosmological hydrodynamic simulations 
(at least those without detailed prescriptions for both stellar and AGN feedback), 
galaxy formation is efficient, with $M_{\rm gal} \sim f_{b}\,M_{\rm halo}$ 
\citep[see e.g.][and references therein]{springel:lcdm.sfh,
maller:sph.merger.rates,naab:etg.formation}. 
Although considerable progress is being made 
\citep[see e.g.][]{governato:disk.formation,governato:disk.rebuilding,
sijacki:radio,scannapieco:fb.disk.sims,dimatteo:cosmo.bhs,croft:cosmo.morph.density}, 
no cosmological hydrodynamic simulation 
has yet been successful at reproducing the $z=0$ stellar mass function and HOD; 
and simulations without feedback tend to produce something much closer to 
efficient cooling/galaxy formation at all masses, with $M_{\rm gal}\propto M_{\rm halo}$ 
(or at least closer to this limit than to the $M_{\rm gal}(M_{\rm halo})$ distribution 
observed). 
As a consequence, such a simulation will predict a galaxy 
merger history that simply 
reflects the halo growth history: fewer major (and more minor) mergers, with a 
factor of several bias in the predicted merger rate.
This effect is especially important at masses $\gtrsim L_{\ast}$; a cosmological 
hydrodynamic simulation that does not correctly reproduce galaxy 
``quenching'' and match the observed bright end of the galaxy stellar 
mass function can under-predict the number of major mergers 
by factors of $\sim5-10$.

In fact, these effects have been seen in such simulations 
\citep{maller:sph.merger.rates,naab:etg.formation}. 
Our comparison here demonstrates that indeed this is expected in a 
simulation of this nature 
and reflects halo growth. However, if we take these same simulations, 
and simply re-populate all of the galaxies according to the observed HOD 
constraints (i.e.\ take the simulation but replace the galaxies just before merger 
with ones of the ``correct'' mass according to the observational constraints), 
then we obtain the results from Figure~\ref{fig:mgr.rate.vs.model}. 
These results agree to within a factor $\sim2$ with the 
semi-empirical models and observed merger rates.

\subsubsection{High-Redshift Results}
\label{sec:compare.hods:central:highz}

\begin{figure*}
    \centering
    \scaleup
    \plotter{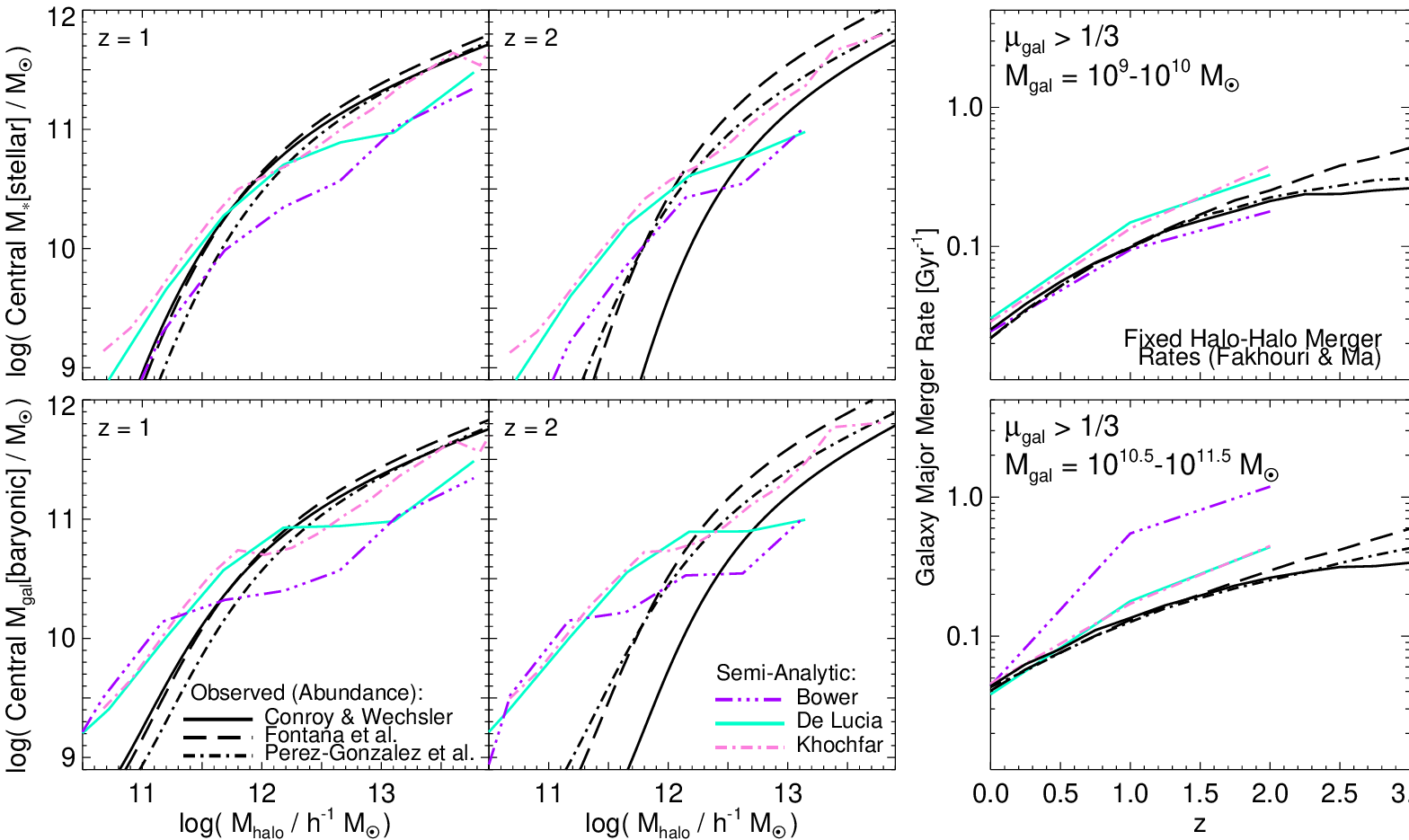}
    \caption{As Figure~\ref{fig:model.HODs}, but comparing empirical HOD constraints 
    and {\em a priori} model predictions at different redshifts. 
    {\em Top Left:} Stellar HOD (median central stellar mass versus $M_{\rm halo}$) 
    at $z=1$ ({\em left}) and $z=2$ ({\em right}). We compare the empirical 
    HOD from abundance-matching, using stellar mass functions/HOD fits 
    from \citet{conroy:hod.vs.z,
    fontana:highz.mfs} and \citet{perezgonzalez:mf.compilation}. 
    We contrast predictions from the \citet{bower:sam}, \citet{delucia:sam}, 
    and \citet{khochfarsilk:new.sam.dry.mergers} SAMs. 
    {\em Bottom Left:} Same, in terms of baryonic mass. 
    {\em Right:} Resulting galaxy-galaxy major merger rate as a function of 
    redshift, in a given baryonic mass bin, from the different HOD curves.
    In general, where the {\em shape} of the HOD is similar 
    (regardless of normalization), the merger rates are in agreement. 
    \label{fig:model.HODs.z}}
\end{figure*}

Thus far we have only considered results at $z=0$. 
Figure~\ref{fig:model.HODs.z} compares the $M_{\rm gal}(M_{\rm halo})$ distribution 
as a function of redshift, from different 
observational estimates and semi-analytic models. The agreement is reasonable around 
$\sim L_{\ast}$, but less so at higher and lower masses. This reflects well-known aspects 
of the comparison between such models and observed galaxy stellar mass functions -- 
the SAMs tend to predict more low-mass and fewer high-mass galaxies at high redshifts 
\citep{fontana:highz.mfs,ilbert:cosmos.morph.mfs,
fontanot:downsizing.vs.sams,
marchesini:highz.stellar.mfs}. Of course, 
the uncertainties in SAMs grow at high redshift, but so do observational uncertainties. 
Low-mass galaxies are subject to completeness concerns and selection effects; 
high-mass galaxies may have non-trivial corrections from extended intra-group light 
and/or uncertainties in their photometrically inferred stellar masses and photometric 
redshifts \citep{maraston:ssps,maraston:ssp.effects}, 
and there may be evolution in the stellar initial mass function 
\citep{hopkinsbeacom:sfh,vandokkum:imf.evol,dave:imf.evol}. 

The growing differences and uncertainties in $M_{\rm gal}(M_{\rm halo})$ 
highlight the need for further study and observational constraints. Nevertheless, 
the resulting merger rates, integrated over a reasonably broad stellar mass 
interval, are still similar to within a factor $\sim2$. 
This owes to the fact that, at a given stellar mass 
$M_{\ast}$, the merger rates do not depend on the absolute 
value of $M_{\rm gal}(M_{\rm halo})$, but rather the {\em shape} of this function.

\subsection{Halo Occupation Statistics of {Satellite} Galaxies}
\label{sec:compare.hods:satellites}

We now generalize our previous comparison to allow for {\em different} 
$M_{\rm gal}(M_{\rm halo})$ distributions in 
satellite or central galaxies. In other words, whereas before, a satellite galaxy's properties 
were set by its maximum pre-accretion halo mass 
(i.e.\ its state when it was last a central galaxy), we now free 
that assumption.

\subsubsection{The Central-Satellite Difference and its Effect on Merger Rates}
\label{sec:compare.hods:satellites:overview}

\begin{figure*}
    \centering
    \scaleup
    \plotter{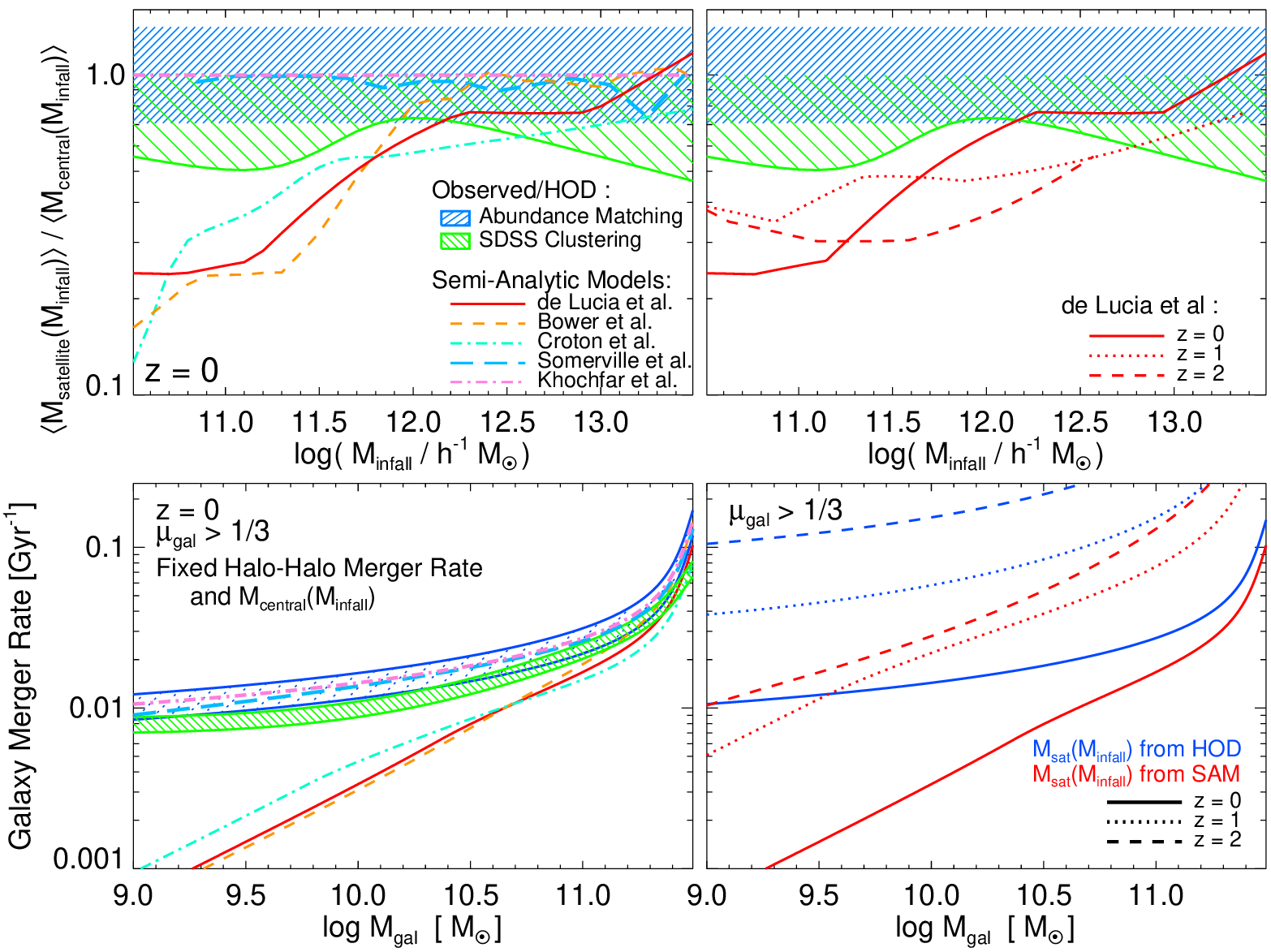}
    \caption{Possible effects of allowing the {\em satellite} HOD to differ from the 
    {\em central} HOD by a significant amount. 
    {\em Top Left:} Average satellite versus average central galaxy mass 
    (total baryonic mass, not stellar mass)
    at fixed maximum or infall halo mass (for the centrals, just $M_{\rm halo}$, 
    for the satellites, their maximum pre-accretion mass -- i.e.\ halo mass when 
    they were themselves last a central galaxy). A model where 
    satellites were totally un-stripped would have this $=1$ exactly. 
    Blue shaded range shows the assumed range of systematic offsets in 
    the abundance-matching HOD models matched to abundance and clustering 
    versus {\em stellar} mass following \citet{conroy:hod.vs.z}; 
    the same $\fgas-M_{\ast}$ relation 
    is applied to all systems (i.e. no gas stripping allowed). 
    By assumption, the typical HOD model has no offset. 
    Green range show the same, but fitting both $M_{\ast}-M_{\rm halo}$ 
    relations for satellites and centrals separately, to local SDSS clustering 
    measurements from \citet{wang:sdss.hod}. 
    We compare the predictions from various SAMs (labeled), at $z=0$.
    Since we are interested in satellites ``about to merge,'' we 
    restrict the SAM `satellites' to those in the `orphan' stage with a 
    remaining merger time $<t_{\rm Hubble}(z)/2$ (by assumption, 
    this gives the same result in the HOD models, but is 
    not directly constrained by observations); differences 
    are much smaller when all satellites are included (or when 
    stellar, not baryonic, mass is plotted). 
    {\em Top Right:} Same, comparing the prediction from 
    \citet{delucia:sam} at three redshifts. 
    {\em Bottom Left:} Resulting ($z=0$) galaxy-galaxy major merger 
    rate (mergers per galaxy per Gyr, for galaxies with mass $>M_{\rm gal}$), 
    from convolving halo-halo merger rates with the above HOD for 
    both centrals and satellites. 
    In each, the halo-halo merger rate is held fixed 
    \citep{fakhouri:halo.merger.rates}, 
    as is the central galaxy $M_{\rm gal}(M_{\rm halo})$ relation 
    \citep[from abundance matching following][]{conroy:hod.vs.z}; 
    differences stem from the predicted difference in satellite-central 
    properties at fixed infall mass. 
    Linestyles correspond to the models above. 
    {\em Bottom Right:} Same, comparing just the abundance-matching 
    prediction (blue lines) and \citet{delucia:sam} model at 
    $z=0,\,1,\,2$. 
    In low mass, gas-rich galaxies, the difference between 
    models with all gas retained (the HOD case above;  
    an upper limit) and satellite gas stripping can translate to 
    significant differences in predicted {\em baryonic} 
    major merger rates. 
    \label{fig:model.HODs.sat}}
\end{figure*}

Figure~\ref{fig:model.HODs.sat} shows the relevant comparison. 
Specifically, we compare the median $M_{\rm gal}(M_{\rm halo})$ relation, for both central and 
satellite galaxies. 
As always, we define $M_{\rm halo}$ for satellites as the 
maximum $M_{\rm halo}$ before their accretion, i.e.\ their ``infall'' mass $M_{\rm infall}$. 
We show two representative models: first, the HOD constraints from 
\citet{conroy:monotonic.hod}, inferred from the observational constraints, in which 
the two $M_{\rm gal}(M_{\rm halo})$ relations (that for satellites and centrals) 
are nearly {\em identical} \citep[with this pre-accretion definition of $M_{\rm halo}$; 
see][]{gao:subhalo.mf,nagai:radial.galaxy.dist.in.clusters}. Indeed, 
that the two distributions can be nearly the same is one of the major 
findings of such empirical HOD models. 
To be specific, the observational input in this method typically 
constrains the stellar mass $M_{\ast,\,\rm sat}/M_{\ast,\,\rm cen}$, 
not the baryonic mass $M_{\rm gal,\, sat}/M_{\rm gal,\, cen}$ -- we {\em assume} 
here similar $f_{\rm gas}(M_{\ast})$ for the {\em cold}, bound, disk gas component. 
However we will show below (Figure~\ref{fig:sat.overquenching}) that observations 
show no difference in $f_{\rm gas}(M_{\ast})$ for the 
specific (major-merger candidate) systems of interest here. 
Further exploration of these distinctions is important, however 
(e.g.\ Neistein et al., in prep) -- for now, this should more conservatively be 
considered an assumption in the semi-empirical models.

Observationally, the large-scale clustering of galaxies (and, to first order, the 
stellar mass function) constrain primarily central galaxies. However, 
the small-scale galaxy correlation function (the ``one halo'' term) probes 
satellite populations, and independent constraints are available from 
e.g.\ group counting statistics and satellite kinematics. 
As noted in \S~\ref{sec:compare.hods:central}, 
most halo occupation models find good agreement with these observations 
by assuming that satellites/subhalos are described by the {\em same} 
$M_{\rm halo}$ distribution as central galaxies, so long as the subhalo mass $M_{\rm halo}$ used to 
assign the subhalo galaxies (satellites) their masses is the maximum 
pre-accretion mass, as opposed to their 
instantaneous (stripped) subhalo mass \citep[see e.g.][]{conroy:monotonic.hod,zheng:hod.evolution,
conroy:mhalo-mgal.evol,valeostriker:monotonic.hod,conroy:hod.vs.z,perezgonzalez:hod.ell.evol,
stewart:merger.rates,hopkins:merger.rates}. This includes all satellites in the model, 
even those about to merge (at very small scales $\lesssim20-100\,$kpc), 
and results in excellent agreement with the observed number of pairs on those 
scales as a function of galaxy mass and redshift \citep{conroy:monotonic.hod,bell:merger.fraction,
stewart:merger.rates}. 

The top panel of Figure~\ref{fig:model.HODs.sat} shows this 
result: $M_{\rm gal}(M_{\rm halo})$ 
is identical for both central and satellite galaxies, defined in these terms. 
But although this is a typical assumption in halo occupation models, the 
observational constraints may not necessarily require identical satellite and 
central properties at fixed infall mass. 
Therefore, in terms of the ratio $M_{\rm satellite}/M_{\rm central}$ -- the typical satellite versus 
central galaxy mass at fixed halo infall mass, the top panel shows the allowed 
scatter or systematic offset that could be tolerated within observational errors, given 
the constraints presented in \citet{conroy:monotonic.hod}, \citet{conroy:hod.vs.z} 
and \citet{stewart:merger.rates}. The authors 
find that less than $\sim 0.2$\,dex in 
scatter or systematic offset of satellite populations from central populations is 
acceptable, before agreement with the observed clustering on small scales breaks down 
\citep[see also][]{moster:stellar.vs.halo.mass.to.z1,
guo:2009.structural.props.central.gals.vs.env,
tinker:quenching.by.mergers.preferred,
pasquali:2009.satellite.age.metal.vs.models,weinmann:sam.sats.whats.needed.to.keep.blue}.

The conclusion is not strongly redshift-dependent at least for redshifts 
$z=0-4$ where clustering and stellar mass function measurements are available 
\citep[and observations of other populations even at $z\sim6$ suggest 
it remains true there as well;][]{white:2008.qso.mhalo.scatter.z6,
shankar:2009.smbh.demographics.review,
shen:2009.smallscale.qso.clustering.highz}. 
This has also been found in a number of other studies 
\citep{zheng:hod,zheng:hod.evolution,yang:clf.update.bycolor,
tinker:2009.drg.clustering.hod,more:2009.halo.mass.cen.mass.relation,
moster:stellar.vs.halo.mass.to.z1}. 

For example, 
\citet{wang:sdss.hod} arrive at similar conclusions with independent data sets, simulations, 
and different fitting methodologies. They find reasonable agreement (especially 
at low masses, which is the regime of particular interest here) assuming that 
centrals and satellites obey an identical $M_{\rm halo}$ distribution. 
They do obtain a marginal 
improvement for high-mass galaxies if the 
preferred satellite $M_{\rm gal}(M_{\rm halo})$ is offset from the central 
$M_{\rm gal}(M_{\rm halo})$ by a small amount, shown in 
Figure~\ref{fig:model.HODs.sat}; satellites might be 
systematically less massive at the 
same halo infall mass by a factor $\sim 1.2$ around $\sim L_{\ast}$ 
and $\sim2$ at $L\gg L_{\ast}$. But differences 
in the fitted {\em scatter} actually ``cancel out'' much of the predicted 
offset, and the statistical difference at low masses is 
small.\footnote{Note that the simulations 
used in \citet{wang:sdss.hod} require the inclusion of an ``orphan'' (unresolved subhalo 
but un-merged) population 
in order to match the observed small-scale clustering: repeating the analysis 
with subhalo populations in some other simulations \citep[e.g.][]{stewart:merger.rates} 
that require a smaller orphan population yields a smaller satellite-central 
offset.}

Interestingly, cosmological hydrodynamic simulations predict a nearly identical 
$M_{\rm gal}(M_{\rm halo})$ for satellite and central galaxies, in agreement with the 
HOD model fits (including the assumption that the $f_{\rm gas}(M_{\ast})$ relation 
is unchanged for satellites and centrals, and therefore that there is no difference 
either in $M_{\ast}(M_{\rm halo})$ or $M_{\rm gal}(M_{\rm halo})$). 
The reasons for this are discussed below 
(\S~\ref{sec:compare.hods:satellites:resolution}). 

We compare these empirical HOD constraints with the predictions of the 
semi-analytic model from \citet{delucia:sam} (the results are similar 
for other SAMs shown, we show this as a representative case).
Since we are interested in effects on mergers, we specifically isolate populations 
in the ``orphan'' stage, i.e.\ for which 
the ``merger clock'' in the model has been initialized and for which the remaining 
time until merger is less than the Hubble time.\footnote{This is not 
exactly the same as the satellite population in the halo occupation models 
constrained as described above. However, in the HOD models, the 
satellite mass as a function of infall mass is 
by definition preserved at smaller radii, and comparing the systems constrained by 
observed clustering at small radii $<R_{\rm vir}$ (let alone $\ll 100\,$kpc), which 
are near-merger, 
they are almost all in the orphan population in these particular SAMs. Regardless, 
the selection from the SAMs does not qualitatively change our conclusions. }
In this model, and in many other SAMs, there is actually a significant 
difference between the two $M_{\rm gal}(M_{\rm halo})$ relations (especially 
at low masses), owing to 
satellite-specific physics that lower $M_{\rm gal}(M_{\rm halo})$ for 
satellites relative to its value at the time of infall/accretion.
Semi-analytic models tend to predict that low-mass satellites (especially 
those about to merge -- the population of interest) have substantially lower baryonic 
masses than central galaxies with the same halo mass at the time of 
the satellite accretion (``infall mass''). 
The baryonic mass difference is about a factor 
of $\sim2-3$ at the lowest masses, where disks are 
gas-dominated.\footnote{Note that this distinction does not necessarily 
appear in the stellar mass as well (S.~White, private communication), 
but we focus on baryonic masses here since it more appropriately focuses 
on the relevant dynamical major merger events 
(but see \S~\ref{sec:definitions:mass.ratio} below).}
We consider the 
difference in the predicted distribution ($M_{\rm gal}(M_{\rm halo})$) 
for satellites and centrals at higher 
redshifts ($z=1$ and $z=2$) as well, and find that the gap between 
centrals and satellites, in the models, extends to higher 
masses at higher redshifts. 

Given this, we can estimate how large an effect the different satellite-versus-central 
distributions will have on merger rates. As above, 
we can compare the galaxy-galaxy merger rate on an even footing by 
adopting a fixed halo-halo merger rate \citep[here that in][]{fakhouri:halo.merger.rates}, 
and populating each halo according to the $M_{\rm gal}(M_{\rm halo})$ relation. 
Unlike in the previous section, however, we do not assume that both galaxies 
obey the same $M_{\rm gal}(M_{\rm halo})$ relation. Rather, the primary 
and secondary subhalos 
separately follow the relations for central and satellite galaxies, shown here. 

Recall that differences between semi-analytic and semi-empirical 
estimates of the $M_{\rm gal}(M_{\rm halo})$ distribution for {\em central} 
galaxies have relatively little effect on merger rates. Thus, the variations in 
merger rates reflected in Figure~\ref{fig:model.HODs.sat} almost entirely 
owe to the {\em difference} (or lack thereof) between 
the $M_{\rm gal}(M_{\rm halo})$ of satellites and that of centrals. This is 
in contrast to the HOD-based model in which the $M_{\rm gal}(M_{\rm halo})$ is 
nearly the same for centrals and satellites, leading to the same curves 
predicted in \S~\ref{sec:compare.hods:central}. In the SAM, 
the median $M_{\rm gal}(M_{\rm halo})$ can be significantly lower 
for satellites at the same infall/pre-accretion mass (especially 
at low masses). Thus even for a merger of exactly equal-mass halos 
(which, on average, must have equal-mass galaxies just before the halo-halo 
merger, i.e.\ at infall), the median 
expectation in the SAM is that the ``secondary'' (whichever happens to 
have slightly less initial mass) will be only some fraction of the mass of the 
primary by the time the actual galaxy-galaxy merger occurs, if it 
were initially gas-rich. The number 
of major mergers can therefore be substantially suppressed. 

It is clear how a larger 
discrepancy between $M_{\rm gal}(M_{\rm halo})$ of satellites 
versus centrals at low masses leads to a stronger suppression of the merger rate. 
Likewise, at high redshifts, if the gap between central and 
satellite masses grows, the major merger rate will be further suppressed. 
At high masses, though, and in 
regimes where galaxies are gas-poor, both SAM and HOD constraints find a similar result 
where the $M_{\rm gal}(M_{\rm halo})$ distributions of satellites and centrals 
are nearly identical -- as a result, their predicted merger rates are 
not dramatically different. 

The results from the \citet{bower:sam} and \citet{croton:sam} 
semi-analytic models are qualitatively similar. 
In contrast, the semi-analytic models of \citet{somerville:new.sam} and 
\citet{khochfarsilk:new.sam.dry.mergers} predict little 
discrepancy between the $M_{\rm gal}(M_{\rm halo})$ relations of satellites 
and centrals. This is because they do not include any mechanism for the 
satellite galaxies to lose a significant mass in gas before their final merger; 
we discuss this further below.

\subsubsection{Effective Mass Loss in Satellites: What Drives These Differences?}
\label{sec:compare.hods:satellites:massloss}

Satellite-specific physics in the semi-analytic models are clearly important for the predicted 
merger rates at low masses. Why do some models predict different 
$M_{\rm gal}(M_{\rm halo})$ distributions for satellites versus central galaxies?

\begin{figure*}
    \centering
    \scaleup
    \plotter{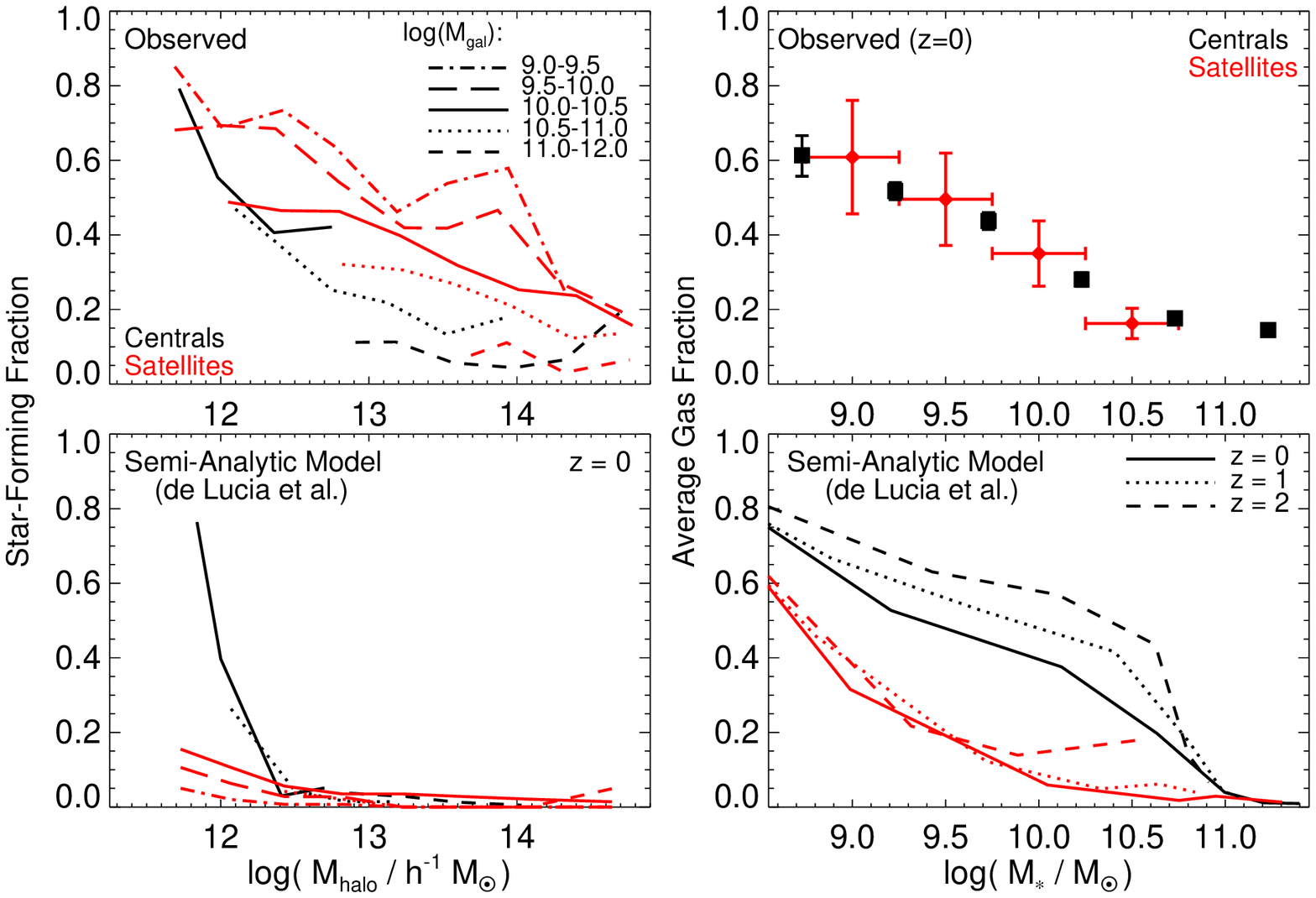}
    \caption{Illustrations of the effect seen in Figure~\ref{fig:model.HODs.sat}, where 
    low-mass satellites in many SAMs have suppressed baryonic masses. 
    {\em Left:} Fraction of satellite and central 
    galaxies that are star-forming (at $z=0$), as a bivariate 
    function of galaxy stellar and halo mass. 
    {\em Top Left:} Observed distributions from \citet{kimm:passive.sats.vs.centrals}. 
    Low-mass satellites have large star-forming fractions, similar to central galaxies. 
    {\em Bottom Left:} Prediction from the \citet{delucia:sam} model. Satellites 
    tend to be uniformly quenched.
    {\em Right:} Average gas fractions (and scatter in gas fractions) versus 
    galaxy mass, for satellites and centrals. 
    {\em Top Right:} Observed distributions at $z=0$ for 
    field centrals \citep{kannappan:gfs}, and satellites in 
    ``normal'' groups \citep[{\em not} massive clusters, where 
    satellite properties are significantly different;][]{belldejong:disk.sfh}. 
    Barring extreme (rare) environments or very minor mass ratios, 
    satellites are not strongly gas-depleted. 
    {\em Bottom Right:} Prediction from the SAM for 
    satellites and centrals. Model satellites are strongly gas-depleted. 
    These are general trends in a number of semi-analytic models.
    \label{fig:sat.overquenching}}
\end{figure*}

In general, the predicted differences in $M_{\rm gal}(M_{\rm halo})$ between 
satellites and centrals relate to a well-known tendency for semi-analytic models 
to predict a satellite population that is ``over-quenched.'' 
Satellites in these models tend to be too red, too gas-poor, too low-mass, 
and insufficiently star-forming relative to observed satellite galaxies. 
Figure~\ref{fig:sat.overquenching} briefly illustrates 
this effect: we show first the observed fraction of passively evolving (``quenched'') 
central and satellite galaxies, as a function of galaxy stellar mass and halo mass. 
We compare these observations to the predicted passive 
fractions in semi-analytic models.
 
The observational results are taken from \citet{kimm:passive.sats.vs.centrals}, 
from their analysis of the \citet{yang:group.finder,yang:clf.update.bycolor} SDSS ($z=0$) 
galaxy catalogs, with the passive population identified by the observed 
bimodal distribution of specific star formation rates at each mass. 
The observations show that, 
at fixed stellar and halo mass, the fraction
of satellite galaxies that have ``quenched'' 
is not much different from that of central galaxies. At low stellar masses, 
in particular, {\em most} satellites are still star-forming (even those satellites in 
high-mass halos). Similar 
results have been found by other analyses and in a wide range of different 
surveys, over a range of 
redshifts from $z=0-1$ \citep[see e.g.][]{gerke:blue.frac.evol,
haines:passive.frac.dept.on.mgal.not.mhalo,
weinmann:obs.hod,weinmann:group.cat.vs.sam,
weinmann:2009.no.satellite.specific.physics.in.env,
wang:sat.shutdown.slower.vs.sam,yang:clf.update.bycolor,
guo:2009.structural.props.central.gals.vs.env,
vandenbosch:2008.only.strangulation.on.sats}. 
This agrees well with the expectation from semi-empirical models and 
HOD constraints (independently determined via the observed small-scale 
clustering of galaxies), that (at least massive) satellites should have properties 
that reflect their halo when it was last a central, but otherwise differ 
little from central galaxies. 

This is also reflected in galaxy gas content, shown in 
Figure~\ref{fig:sat.overquenching}. We show the observed 
gas fraction distribution of central and satellite galaxies 
(separately), as a function of stellar mass. We specifically show 
compilations of observed disk gas fractions as a function of 
their stellar mass from \citet{belldejong:disk.sfh}, \citet{mcgaugh:tf}, and 
\citet{kannappan:gfs}. 
These observations span both satellite and central galaxies. 
The sample of \citet{kannappan:gfs}, being simply mass-selected, is 
dominated by central galaxies at all masses. But from the other 
samples we can separate both central (chiefly massive field) galaxies 
and satellite galaxies, in particular the Ursa Major Cluster sample 
of \citet{tully:ursa.major.cluster.disk.sample}. 
This is ideal for our comparison, as it represents a rich, but not 
extremely massive (and thus unusual) group with a group velocity 
dispersion of $\sim150\,{\rm km\,s^{-1}}$ -- this is typical of 
the host halos of central/primary galaxies with $M_{\rm gal}\sim10^{10}-10^{11}\,\msun$. 
We stress that we are {\em not} trying to compare with galaxies in e.g.\ massive 
clusters, where the gas-rich galaxies tend to be 
much less massive than their cluster host -- stripping 
and other processes may be efficient in such galaxies, but those would not represent major 
mergers in {\em any} model. 
This is also seen in many of the color or SFR-based studies above; 
for example, \citet{weinmann:2009.no.satellite.specific.physics.in.env,
guo:2009.structural.props.central.gals.vs.env,
wang:2009.isolated.red.dwarfs,
vandenbosch:2008.only.strangulation.on.sats,
kang:2008.satellite.stripping.efficiency} find that 
only in the most extreme (rare) environments could there be significant gas 
depletion in satellites -- over a wide range of satellite and halo masses, 
they find that there is at most very minor depletion (for 
major-merger candidates) and, in particular, that the amount of depletion 
does {not} increase or decrease with halo mass in any significant 
manner. 
This is true even in the central 
regions of groups, i.e.\ systems ``about to merge,'' for which the predicted 
satellite-central gas mass difference in some models is maximized 
(see references above). 
The important result observed is, for galaxies with similar $M_{\ast}$ in 
similar-mass halos, which have some possibility of being major merger 
pairs, there is not much difference in gas fractions between 
central and satellite systems.

We compare both of these diagnostics -- the ``quenched'' fraction and gas 
mass fractions in central and satellite galaxies -- to the predictions of the 
semi-analytic models. 
The differences are immediately apparent. The SAMs predict 
that essentially all satellites are quenched, especially those in 
halos above the critical ``quenching mass'' of $\sim10^{12}\,\msun$. 
The average predicted gas fraction of satellites at low mass is much less 
than that of equivalent central galaxies, in contrast to the observations. 

For clarity, we 
show in Figure~\ref{fig:sat.overquenching} just the predictions from the 
\citet{delucia:sam} model, but \citet{kimm:passive.sats.vs.centrals} 
and others have demonstrated that this 
satellite ``overquenching'' tendency is quite general in recent SAMs, 
including e.g.\ those of 
\citet[][]{croton:sam}, 
\citet[][]{monaco:sam}, 
\citet[][]{bower:sam},
\citet[][]{kang:sam}, \citet{cattaneo:sam}, and \citet{bertone:stellar.wind.recycling.sam}. 

As noted above, we are re-stating a well-known result; the 
same physical deficiencies are reflected in a number of 
other measures: in semi-analytic models, satellite 
mass functions tend to be under-massive, the fraction of blue satellites is 
lower than that observed, and the observed small-scale clustering of star-forming or blue 
galaxies tends to be poorly reproduced 
\citep[see references above and][]{cooper:z1.color.density,
cooper:color.density.evol,park:redfrac,blanton:smallscale.env,
li:clustering,coil:z1.clustering.update}. 

What, in these models, drives the ``overquenching'' of satellites? 
In most, once a system becomes a satellite, 
its halo gas reservoir is immediately added to that of its ``parent'' halo, and (as a 
consequence) no new cooling occurs onto the secondary. 
But if this removal of un-cooled, pristine halo gas were the 
only effect, the consequences would not be so pronounced. In all of these 
models, strong stellar feedback is assumed to affect star-forming 
galaxies -- generally some mass loss of order several times the star 
formation rate is ejected from the disk, per unit stars formed. Given such 
strong mass ejection, in order to maintain observed (high) star formation 
rates in disk galaxies, the ejected gas must be recycled rapidly -- either 
processed through the halo or some (for numerical reasons) designated 
``reservoir'' and returned quickly to the galaxy. But when a system is a 
``satellite'' in these models (even if the mass ratio is close to 1:1), the 
mass is simply lost to the parent halo. The gas mass is therefore 
quickly depleted and the satellite quenches as it has ejected most of its 
gas.\footnote{This is despite the fact that many of the models do not 
nominally include an 
additional explicit 
``ram-pressure stripping prescription'' for the cold disk gas.}

This can be a very large effect. A system with $\sim50\%$ gas will lose $\sim1/2$ 
its baryonic mass in just a short time after accretion. Thus an initial 1:2 major 
merger will become a 1:4 minor merger (or a more minor merger, 
if the parent ``steals'' this 
ejected mass). As a consequence, there will be almost no 
very {\em gas-rich} major mergers. 

This problem is further exacerbated in many of these models by the assumption 
of instantaneous recycling from stellar evolution. When some stars form 
in the new satellite, the mass loss of $\sim50\%$ that will occur over their 
lifetime is immediately turned into gas, which can then suffer the 
consequences above. A more gradual mass loss over actual stellar lifetimes 
would, obviously, yield somewhat less rapid an evolution (especially 
for major mergers, where the merger timescale might be only a Gyr). 

Since the consequences of this gas loss will clearly be stronger 
for systems with higher gas content, low-mass galaxies 
are preferentially affected. This gives rise to the 
mass-dependent trend in the discrepancy between $M_{\rm gal}(M_{\rm halo})$ 
predicted for satellite and central galaxies by such models 
(shown in Figure~\ref{fig:model.HODs.sat}), where low-mass 
satellites have their masses more suppressed relative to centrals. Likewise, 
since at high redshifts all galaxies are more gas-rich, 
the discrepancy grows in the models with redshift. This, in turn, 
suppresses predicted merger rates at low masses and high redshifts. 

Differences between different semi-analytic models largely depend on 
exactly what prescription is used for stellar feedback (controlling how efficiently 
the system self-quenches) as well as e.g.\ when and where this occurs 
and how it is connected to the lifetimes/stripping/identification of subhalos/satellites. 
For example, in some models based on cosmological dark matter merger trees 
\citep[e.g.][]{delucia:sam,bower:sam}, the 
system is labeled a ``satellite'' when the halos 
are first linked by the friends-of-friends (FOF) halo finder algorithm, 
which occurs for major pairs at anywhere from $\sim2-5\,R_{\rm vir}$. 
In such a case, these effects can ``strip'' the secondary even before 
it enters the primary halo. In contrast, simulations and observations 
find that, if any stripping occurs, it often only does so in the central, most 
dense and high-pressure regions of massive halos 
\citep{weinmann:obs.hod,weinmann:group.cat.vs.sam,
wang:sat.shutdown.slower.vs.sam,
yang:2009.subhalo.disruption,
weinmann:2009.no.satellite.specific.physics.in.env,
vandenbosch:2008.only.strangulation.on.sats,
kang:2008.satellite.stripping.efficiency,
mccarthy:ram.pressure.revised.model}. 
In fact, cosmological hydrodynamic simulations show that 
satellite galaxies tend to keep growing and accreting {\em new} material 
(both onto their subhalo and into the galaxy) until they are in the 
final stages of orbiting a relaxed massive system near its center 
\citep{simha:sat.growth.after.acc}. 
Observations likewise see no significant difference 
between satellites in the outer regions of even massive groups, 
relative to centrals in similar-mass subhalos. 
The tendency of satellites to continue accreting in simulations 
is stronger at higher redshifts, and observational constraints from 
$z=0-2$ require that the gas-rich, star-forming satellite fraction 
remain large over this entire interval \citep{tinker:quenching.by.mergers.preferred}, 
enhancing the discrepancy with these models at high redshift. 

Unsurprisingly, at high masses, where gas fractions tend to 
be small ($\sim10-20\%$), these effects are minimized. Thus the 
$M_{\rm gal}(M_{\rm halo})$ distributions for satellites and centrals 
are similar, and there is little difference in predicted 
merger rates at these masses.

\subsubsection{Towards Resolution: What Physics for Satellites?}
\label{sec:compare.hods:satellites:resolution}

It is important 
to recall that the vast majority of mergers by both number and 
mass density are in the field (essentially single 
halo-halo mergers) or small/loose group environments (analogous to e.g.\ the 
Local Group). The observations indicate that 
satellite-specific physics appear to be important 
only for systems with the smallest mass ratios $\mu_{\rm gal}\lesssim 0.01-0.05$ in 
these environments \citep[specifically see e.g.][]{haines:passive.frac.dept.on.mgal.not.mhalo,
weinmann:2009.no.satellite.specific.physics.in.env,
kimm:passive.sats.vs.centrals}, and then only in the final 
stages and/or close passages, {\em not} just at the moment of 
crossing $R_{\rm vir}$ \citep[references above and][]{
yang:2009.subhalo.disruption,
vandenbosch:2008.only.strangulation.on.sats,
kang:2008.satellite.stripping.efficiency}. 
Even if stripping were extremely efficient in dense environments 
(e.g.\ for clusters, or for dwarf galaxies in massive halos 
such as the Local Group), such that it suppressed all such major 
mergers, it would have almost no effect on the {\em overall} merger rate. 

In contrast, much of the intuition 
used to design the prescriptions above 
is built up from satellite-specific studies 
based on massive clusters with dense, high-pressure ICM gas. 
But such systems contain only a small 
fraction of the bulge mass density of the Universe \citep[$\lesssim 5\%$; 
see e.g.][]{carollo98,kormendy.kennicutt:pseudobulge.review,
allen:bulge-disk,ball:bivariate.lfs,
driver:bulge.mfs,gadotti:sdss.pseudobulge.properties}. 

Indeed, with the possible exception of dwarf galaxies 
in massive halos \citep{donghia:resonant.stripping.dwarfs,
wang:2009.isolated.red.dwarfs}, the observations favor a ``gentle'' treatment 
of satellites. Considering a volume-limited sample of satellites 
(i.e.\ one not dominated by extreme environments) that might become 
major mergers (i.e.\ are not so different in mass from their centrals), the 
observed specific star formation rates, colors, and gas fractions can 
be well-fitted (to first order) by a model in which there is {\em no} 
removal of the hot or cold satellite gas (including the ``ejecta'' or recycled 
material from stellar winds and supernovae). The only constraint is that 
the satellite halo stops growing, so e.g.\ the total baryonic mass supply 
is ``frozen in'' at the value it had when the subhalo was accreted. 

Such a simple model leads to a color 
and passive/quenched galaxy distribution as a bivariate function of 
galaxy mass and halo mass similar to that of centrals (with just slightly 
more ``quenching'' in low mass satellites), in better agreement with 
observations \citep{weinmann:obs.hod,weinmann:group.cat.vs.sam,
vandenbosch:2008.only.strangulation.on.sats,
kang:2008.satellite.stripping.efficiency,
kimm:passive.sats.vs.centrals}, and similar gas 
fractions (as the subhalo retains ejected stellar wind feedback), as 
observed \citep{belldejong:disk.sfh,mcgaugh:tf,kannappan:gfs}. 
This also naturally leads to the observed result that 
satellite quenched fractions are a much stronger function of the 
satellite stellar mass, and weaker function of e.g.\ the local density or 
halo mass, than currently predicted \citep{weinmann:group.cat.vs.sam,
haines:passive.frac.dept.on.mgal.not.mhalo}, at both $z=0$ and 
high redshifts $z\sim1-1.5$ \citep{gerke:blue.frac.evol}. 
And the observed weak dependence of red satellite fraction on distance from 
halo center \citep{weinmann:obs.hod} is similarly obtained. The 
observationally favored gradual decline of satellite star formation rates 
relative to those that such systems had when they were central 
galaxies \citep[see e.g.][]{wang:sat.shutdown.slower.vs.sam} 
arises naturally in an effectively ``closed-box'' model, 
given the observed star formation rate-gas density scaling laws 
\citep{kennicutt98}, as systems deplete their original baryon 
reserve without new halo growth. 

High-resolution hydrodynamic simulations 
appear to give consistent results. Gas-rich systems are in the ``cold accretion'' 
regime: their halos are dense, with a cooling time shorter than the local dynamical 
time, so gas outflows may be efficiently halted by the dense ISM  
and more gas in new accretion is brought into the galaxy in cold filaments on 
the dynamical time. Especially in this regime, these cold flows dominate 
accretion -- the idea of spherical accretion from a quasi-static halo 
accounts for only a small amount of residual cooling in the most massive 
halos (essentially, this is the old ``cooling flow'' problem in massive clusters; 
but it contributes negligibly to the global cold gas budget or star formation rate density). 
These cold filaments are sufficiently dense to survive 
disruption by the diffuse parent halo gas when systems become satellites; the 
simulations find that even relatively small satellites continue to hold onto 
their halo gas. 

In fact, the simulated satellites continue to accrete new material, 
and grow by cooling, well into their parent halo virial radius and 
for $\sim$Gyr after become satellites \citep[see e.g.][]{simha:sat.growth.after.acc}. 
In this regime, satellite growth in the simulations is indistinguishable from 
central galaxy growth, for the same halo/infall mass. 
At high redshift, the simulations suggest that even the ``frozen in'' models may quench 
satellites too severely; subhalos in the simulations continue to accrete new 
halo gas at a rate equal to the central galaxy well into the halo. In effect, it is as 
if the system does not ``know'' it is a satellite until at very small radii from the 
primary center, just before the final galaxy-galaxy merger \citep{simha:sat.growth.after.acc,
keres:cooling.revised,keres:fb.constraints.from.cosmo.sims}. 

Recall also that we are ultimately most interested in the case of major mergers: in these cases, 
it is clear that the distinction of ``satellite'' versus ``central'' is somewhat semantic. 
Physically, the two are mutually gravitating (one is not 
moving in the halo or potential of the other; rather the two mutually coalesce in a 
single halo/potential) and (by definition) have similar halo 
properties (and therefore similar halo gas masses, virial temperatures, etc.). 
Simple physical considerations (and simulations) imply that there should be 
no significant difference in the cooling/mass loss of one system relative 
to the other (let alone ram-pressure stripping of one and not the other). 

Of course, it is easier to point out these differences than to design 
physical prescriptions that effectively model them -- let alone 
prescriptions that are robust across the entire regime of interest 
(from dwarf galaxies and/or small satellites in massive clusters to 
equal-mass pairs in field or Local Group-analogue environments). And as 
noted above, it is not enough to simply retain pristine halo gas associated 
with the satellite -- ejected gas from stellar feedback should be recycled in 
some manner appropriate to how it is implemented in the different models. 

Considerable progress is being made in this area.
For example, \citet{font:durham.sam.update} updates the \citet{bower:sam} model, 
with the primary change being a 
more detailed satellite cooling model. Specifically, the models are otherwise identical, 
but instead of removing subhalo gas instantly \citep[as in][]{bower:sam}, 
the revised model strips the gas from the assumed extended subhalo gas halo 
according to the model in \citet{mccarthy:ram.pressure.revised.model}, as a function of e.g.\ the local 
gas density (itself a function of radius from the primary, for assumed isothermal 
sphere gas profiles). There are still some sources of uncertainty in e.g.\ the exact 
wind treatment and efficiency of subhalo gas stripping (and the prescription adopted 
may not be applicable if most accretion comes via cold flows), but the agreement with 
the star-forming fractions of satellite galaxies is substantially improved.

\begin{figure*}
    \centering
    \scaleup
    \plotter{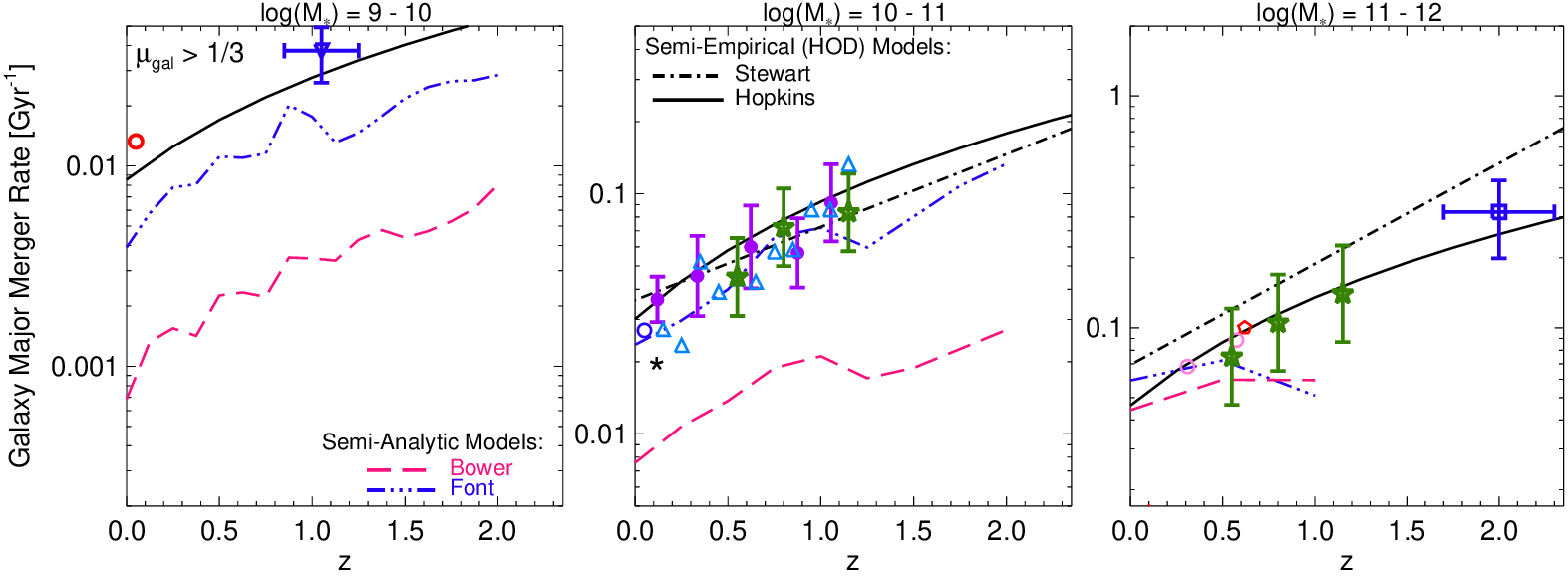}
    \caption{Comparison of the predicted major galaxy merger 
    rates (as Figure~\ref{fig:mgr.rate.vs.sams})
    in two similar semi-analytic 
    models, \citet{bower:sam} and \citet{font:durham.sam.update}.
    The SAMs are otherwise identical, but \citet{font:durham.sam.update} implement a 
    more physical, less severe satellite stripping/quenching model, 
    improving the agreement with e.g.\ the observations in 
    Figures~\ref{fig:model.HODs.sat}-\ref{fig:sat.overquenching}. 
    We compare the predictions to estimates directly adopting the observational 
    HOD constraints, and to observed merger rates. 
    Rates at high $M_{\rm gal}$, where cooling/gas fractions are negligible, 
    are unchanged. Rates at low masses are significantly higher 
    in the revised model, as expected from Figure~\ref{fig:model.HODs.sat}, 
    and agree more closely with direct HOD models. 
    \label{fig:font.vs.bower}}
\end{figure*}

Figure~\ref{fig:font.vs.bower} compares the merger rates predicted by 
this revised model to those from the previous \citet{bower:sam} iteration of the model 
(with strong satellite quenching). We also compare the rates determined 
from HOD-based approaches, and observational constraints. 
At high masses where systems are gas-poor and not star-forming, 
all the models (and observations) agree well. At low masses, where galaxies 
are gas-rich, however, there is an order-of-magnitude difference in the 
predicted merger rates. The revised \citet{font:durham.sam.update} yields merger rates more 
similar to the HOD-based models, as expected, since the primary adjustment to the model 
is intended to bring it into better agreement with the observed HOD statistics used as 
input in the HOD-based models. This yields significantly improved agreement with 
observational estimates of the merger rates. There are clearly remaining differences; 
however, they are at the factor $\sim2-3$ level, typical as we have shown of 
the other differences between the models. 

This is reinforced by comparison of Figures~\ref{fig:mgr.rate.vs.sams} \&\ \ref{fig:model.HODs.sat}. 
Besides the \citet{font:durham.sam.update} model discussed above, the two 
semi-analytic models that predict merger rates in best agreement with 
the halo occupation models, over the entire galaxy mass range, 
are those from \citet{khochfarsilk:new.sam.dry.mergers} and \citet{somerville:new.sam}. 
These models do not include any mechanism for cold gas (including that affected by 
stellar feedback) to be removed from the satellite before a merger; 
the \citet{somerville:new.sam} model allows for some stripping of hot gas, but 
only inside of the parent virial radius where the timescale for a major merger 
is short and so there is little effect on the satellite mass. 
In Figure~\ref{fig:mgr.rate.vs.sams}, we showed that these models are precisely 
those which predict little or no discrepancy between the $M_{\rm gal}(M_{\rm halo})$ 
relation of satellites and centrals. This does not mean that there are not other, 
perhaps very significant, differences between the models (nor does it necessarily 
mean that the same answer is being obtained for the correct reasons); but it does 
imply that, controlling for differences in predicted satellite properties removes a 
significant factor driving different predictions for merger rates.

\breaker
\section{The Other Side: Definitions of Mass and Effects of Mergers on Morphology}
\label{sec:definitions}

\subsection{The Definition of Mass Ratio}
\label{sec:definitions:mass.ratio}

\begin{figure*}
    \centering
    \scaleup
    \plotter{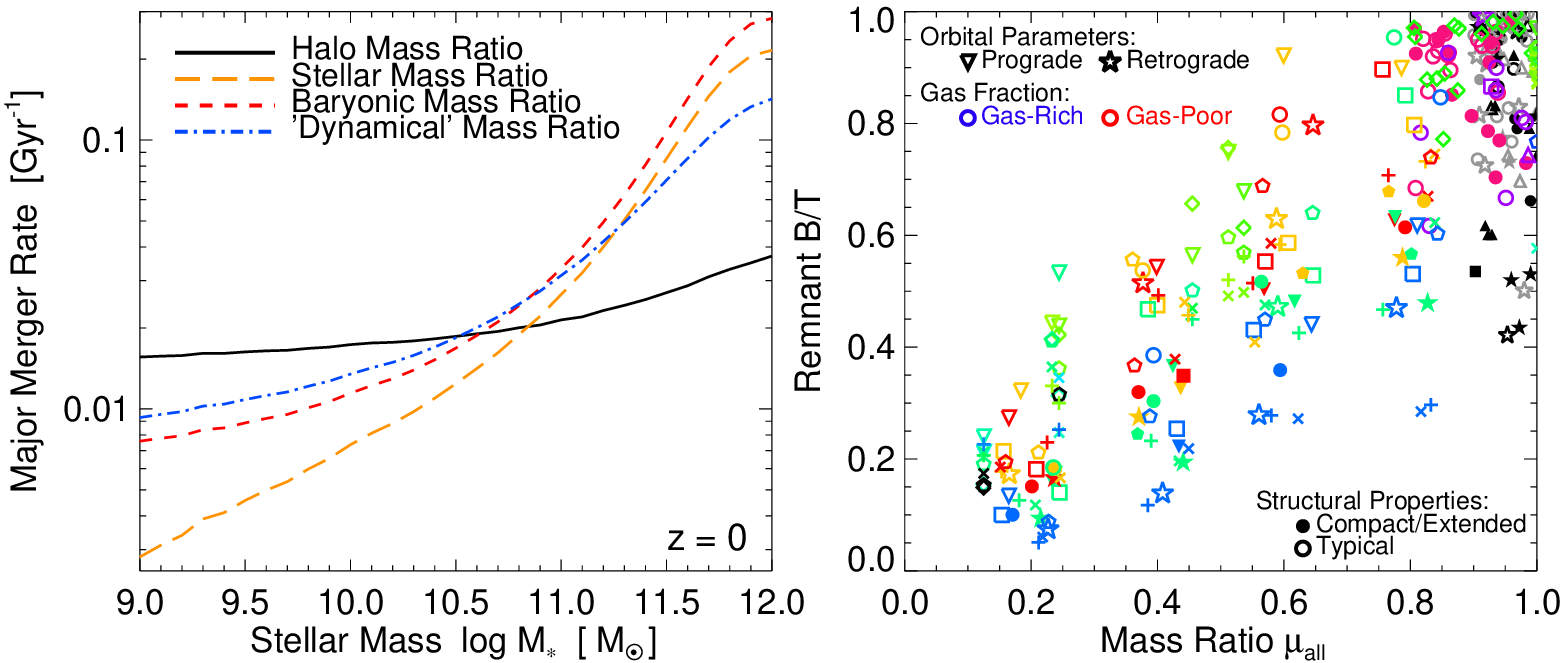}
    \caption{{\em Left:} Importance of the definition of ``mass ratio'' for merger rates. 
    We show the major ($\mu>1/3$) merger rate per galaxy at $z=0$, for galaxies 
    at a given stellar mass, in terms of the merger halo-halo mass ratio, 
    galaxy stellar-stellar mass ratio, baryonic-baryonic (stellar+cold disk gas) mass ratio, 
    and ``dynamical'' mass ratio (gravitational mass, i.e.\ baryons plus the tightly bound, 
    non-stripped dark matter inside the baryonic radii; see Equation~\ref{eqn:baryonic.massratio}). 
    The different ratios behave very differently as a function of mass. 
    Halo mass ratios are most easily predicted, but very different from galaxy mass ratios.
    The baryonic or dynamical mass ratio is most relevant for e.g.\ morphological disturbances 
    and bulge formation, but difficult to directly observe. Stellar mass ratios are accessible in 
    e.g.\ pair samples, but under-estimate the rate of ``damaging'' major mergers at low masses 
    by factors $\sim3$. 
    {\em Right:} Impact of mergers in high-resolution hydrodynamic simulations 
    from \citet{hopkins:disk.survival}. We show the remnant bulge-to-total mass ratio after a 
    merger with mass ratio $\mu$; to remove ambiguity in the merger mass ratio and rate, 
    all systems have just one merger, with the systems initialized so that 
    $\mu=\mu_{\ast}=\mu_{\rm halo}=\mu_{\rm gal}=\mu_{\rm dyn}$. 
    We show simulations with different orbital parameters (symbol types), 
    gas fractions (color, from $f_{\rm gas}=0$ in red to $f_{\rm gas}=1$ in black), 
    and structural parameters (variation in disk sizes, 
    initial $B/T$, and halo concentrations in solid). Other parameter variations include 
    surveys of stellar wind models (magenta), black hole feedback (violet), 
    and gas equations of state (grey). At fixed, perfectly-defined $\mu$, 
    there is factor of $\sim2-3$ variation in the amount of bulge formed by mergers owing 
    to these parameters -- neglecting their full distributions will inherently limit 
    any prediction of the ``impact'' of mergers. Merger impact is also clearly continuous 
    across the major/minor merger ($\mu=1/3$) distinction.     
    \label{fig:merger.fx.cautions}}
\end{figure*}

Throughout this paper, we have focused on the ``merger rate,'' with 
respect to mass ratios defined in a specific manner. 
Observationally, of course, some of these are more or 
less accessible (the galaxy-galaxy luminosity or stellar mass 
ratios being most so, the total 
dark matter halo-halo mass ratios least so). 
But {\em physically}, the same ``number'' of mergers will have 
very different implications for galaxy growth, 
morphologies, bulge formation, and star formation, depending on the mass ratio 
involved -- and the merger rate is {\em not} the same 
for different definitions of the mass ratio. 

Figure~\ref{fig:merger.fx.cautions} illustrates some of these caveats. 
First, consider the merger rate itself, within the same model, but given different 
definitions of merger mass ratio. 
There are three obvious choices of mass ratio, commonly used: 
galaxy-galaxy stellar mass ratio ($\mu_{\ast}=M_{\ast,\,2}/M_{\ast,\,1}$), 
galaxy-galaxy baryonic (stellar+cold gas) mass 
ratio ($\mu_{\rm gal}=M_{{\rm gal},\,2}/M_{{\rm gal},\,1}$), and 
halo-halo mass ratio ($\mu_{\rm halo}=M_{{\rm halo},\,2}/M_{{\rm halo},\,1}$). 
We have shown how the shape of the $M_{\rm gal}-M_{\rm halo}$ 
relation means that the number of major mergers in terms of $\mu_{\rm gal}$ 
can be quite different from the number of mergers in terms of $\mu_{\rm halo}$. 
But since galaxy gas fractions are not uniformly constant, $\mu_{\rm gal}$ 
and $\mu_{\ast}$ will also not be the same. 

In Figure~\ref{fig:merger.fx.cautions}, we plot the galaxy major merger rate 
(major mergers per galaxy per Gyr) at $z=0$, as a function of galaxy stellar mass. 
We do so for a single model -- the ``default'' model in \citet{hopkins:merger.rates} 
using the halo-halo merger rates from \citet{fakhouri:halo.merger.rates,
fakhouri:millenium.2.merger.rates} 
and abundance-matching HOD method from \citet{conroy:hod.vs.z}. But the 
results are qualitatively the same regardless of model. 
We show the major merger rate for different definitions of a 
``major'' merger: a merger with halo-halo mass ratio $\mu_{\rm halo}>1/3$, 
with galaxy stellar-stellar mass ratio $\mu_{\ast}>1/3$, and with 
galaxy baryonic-baryonic mass ratio $\mu_{\rm gal}>1/3$. 
The halo-halo rate is only weakly dependent on mass, a result 
well-known from cosmological simulations (see \S~\ref{sec:compare.dark:halos:compare}). 
The stellar-stellar rate, on the other hand, depends strongly on mass -- 
this owes to the shape of $M_{\ast}(M_{\rm halo})$. At low masses, 
galaxy stellar mass is a steep function of halo mass, 
suppressing major mergers ($M_{\ast}\propto M_{\rm halo}^{2}$, 
so a 1:3 halo-halo merger is just a 1:9 stellar-stellar merger); 
at high masses, stellar mass is a shallow function of 
halo mass, enhancing major mergers ($M_{\ast}\propto M_{\rm halo}^{0.5}$, 
so a 1:9 halo-halo merger is a 1:3 stellar-stellar merger). 
The effect is very large -- an order of magnitude at the extremes of the 
mass function -- and cannot be neglected. 
The baryonic-baryonic rate is quite similar to the stellar-stellar rate 
at high masses (not surprising, since high-mass galaxies are uniformly 
gas-poor). At low masses, however, it lies somewhere between the 
stellar-stellar and halo-halo rate; this is because low-mass galaxies 
have increasing cold disk 
gas masses, even while their stellar mass falls rapidly with halo mass. 
Thus the total baryonic mass is not as steep a function of halo mass 
as is the stellar mass, and major mergers are less strongly suppressed 
\citep[at low masses, $M_{\rm gal}\propto M_{\rm halo}^{1.0-1.5}$, 
so e.g.\ a 1:2 halo-halo merger remains a major baryonic-baryonic merger; 
see][]{avilareese:baryonic.tf,bell:baryonic.mf,mcgaugh:tf,
stark:baryonic.tg.in.gas.rich.gal}. The differences are again very non-trivial -- 
at low masses, the baryonic-baryonic merger rate is a factor of $3$ larger than the 
stellar-stellar merger rate. For further discussion of these 
distinctions, we refer to \citet{stewart:massratio.defn.conf.proc}. 

These various ratios not only relate differently to observable quantities; 
they also have a number of physical relationships to the impact of galaxy mergers. 
Using halo-halo mergers as a proxy for galaxy-galaxy mergers is clearly 
not a good approximation in either models or observations, except just at $\sim L_{\ast}$ 
where the mass function turns over. Physically, the halo-halo mass 
ratio is also not the most important quantity for e.g.\ bulge formation. This is because, by the 
time the final merger/galaxy coalescence takes place, most of the secondary halo mass will already 
have been stripped and/or mixed with the primary halo mass. In general, the outer parts of the 
halo (which contain much of the mass), being extended and low-density relative to the 
baryonic galaxy, are not able to induce significant perturbations or torques on that galaxy. 

The stellar-stellar mass ratio has the advantage of being observable, as well as 
identifying at least a portion of the tightly bound material in the galaxies that can actually induce 
structural perturbations and bulge formation. 
However, at low masses or high redshifts, where gas fractions are 
non-negligible, stellar mass alone clearly misses a very important quantity -- the amount of 
cold, rotationally supported gas in galaxy disks. 
Observations indicate that in these regimes the cold gas is gravitationally {\em dominant} 
over the stars in a large fraction of the galaxy population. 
From a gravitational perspective, in terms of the perturbation induced, 
it makes no difference whether the same merging disk is $100\%$ gas or $100\%$ 
stars \citep[although the end effects will be different and gas 
may be distributed differently from stars; see][]{hopkins:disk.survival}, 
even though one would be a stellar-stellar major merger and the other would not 
appear in any observed merger catalog. 

Thus, attempts to estimate the ``impact'' of mergers (e.g.\ the amount of bulge 
they might form), or even the total number of mergers a galaxy experiences, 
based on observed merger fractions identified with stellar mass or 
luminosity cuts, will under-estimate the true impact of 
mergers by a factor of at least $\sim3$ at low masses ($M_{\ast}\lesssim 10^{10}\,\msun$). 
More directly sampling the total baryonic content of these galaxies is 
necessary to overcome this limitation; in the meantime, empirical results 
should be corrected for this distinction or at least recognize its very large 
importance at low masses. 
And gas fraction is critical for understanding the consequences of a given 
merger; the dynamics of mergers are fundamentally different and bulge 
formation less efficient in gas-dominated, as opposed to stellar-dominated, 
systems \citep[see][]{hopkins:disk.survival}. 
At high masses, $>10^{11}\,\msun$, 
at least at low redshifts $z<1$, such a procedure does not produce a result 
very different from the baryonic-baryonic merger rate. 

But even the baryonic mass ratio is not really representative of the 
gravitational mass that induces perturbations and forms bulges, 
heats disks, and scatters stars in mergers. 
What is actually desired is the surviving, tightly bound total 
gravitational mass of material that is involved in the last couple of passages 
in the final merger. 
In high-resolution simulations of a large number of mergers, spanning 
a large parameter space, \citet{cox:massratio.starbursts} 
find that this can be {\em roughly} approximated by the sum of the 
galaxy stellar mass, the cold disk mass, 
and the dark matter mass that is tightly bound to the baryons and thus 
will be able to resist stripping -- approximately that inside a 
radius $\sim2-4\,R_{e}$ (where $R_{e}$ is the effective radius 
of the baryonic galaxy) or $\sim1\,r_{s}$ (where $r_{s}=R_{\rm vir}/c$ 
is the NFW scale radius of the halo, with $c$ the halo concentration). 
Note that since this dark matter retention at small radii is largely a baryonic effect, 
it will not appear in collisionless simulations. 

We therefore define a ``dynamical'' mass $M_{\rm dyn}$, 
and corresponding mass ratio $\mu_{\rm dyn}$: 
\begin{align}
\nonumber M_{\rm dyn} &\equiv M_{\ast}+M_{\rm gas}+M_{\rm DM}(<3\,R_{e}) \\ 
\mu_{\rm dyn} &\equiv M_{{\rm dyn},\,2}/M_{{\rm dyn},\,1}\ .
\label{eqn:baryonic.massratio}
\end{align}
This is a significantly better approximation to what is most important for 
bulge formation and the dynamical properties of galaxies 
than any of the mass ratios discussed above. 
Figure~\ref{fig:merger.fx.cautions} shows the corresponding 
major merger rate, as a function of primary stellar mass, 
where a major merger is defined by $\mu_{\rm dyn}>1/3$ 
(to calculate the dark matter content inside $3\,R_{e}$, we 
assume the halos follow NFW profiles with a concentration-halo 
mass relation from \citet{bullock:concentrations}, 
and that the galaxy $R_{e}$ follows the relation as a function of 
stellar mass from \citet{shen:size.mass}). 
Roughly, the result is similar to that for baryonic mass ratio, 
but it is slightly closer to the halo-halo merger rate at both high 
and low masses (unsurprisingly, since 
it includes baryons plus a halo contribution). 

Of course, this mass is difficult to extract observationally. 
However, it is readily accessible in most semi-empirical 
and semi-analytic models. 
Adopting such a definition for mass ratio, as opposed to the 
simpler definition of baryonic mass ratio, is important for 
predictions at low and high masses. 
In fact, many such models still adopt either 
just the stellar-stellar mass ratio or the halo-halo 
mass ratio as their proxy for the ``impact'' of a merger -- 
this is not only clearly not the most physical quantity, but as 
Figure~\ref{fig:merger.fx.cautions} illustrates, it can lead to 
an order-of-magnitude 
over or under-estimate of the consequences of mergers.

\subsection{The Importance of Other Parameters}
\label{sec:definitions:other}

Even with perfect knowledge of the merger rate and distribution of mass ratios, 
however, we stress that the impact of mergers {\em cannot} be 
determined at better than factor $\sim2$ accuracy, absent a large amount of 
additional information. 
Figure~\ref{fig:merger.fx.cautions} illustrates this. 
We consider a large suite of high-resolution, hydrodynamic galaxy 
merger simulations. They include realistic progenitor galaxy models (gas+stellar disk+bulge+halo), 
cooling, star formation, and feedback from star formation and BH growth in a multi-phase ISM
\citep{springel:multiphase,dimatteo:msigma}.
These are presented and discussed in detail in 
\citet{hopkins:disk.survival}; for now, we use them to illustrate a couple of simple points. 
Although it may not be the most representative for galaxies at extreme masses, 
we wish to remove the already-discussed ambiguities in merger rates and mass 
ratio definitions. Therefore, every galaxy undergoes exactly one isolated merger 
with the simulations all initialized such that 
the ``mass ratio'' is the {\em same} by any of the definitions above, 
i.e.\ $\mu=\mu_{\rm halo}=\mu_{\ast}=\mu_{\rm gal}=\mu_{\rm dyn}$. 
We run the simulations until they are fully relaxed, and then quantify 
the bulge-to-total mass ratio $B/T$ in the remnant stars -- the property most often 
desired in relation to mergers. We consider mass ratios 
$\mu=0.1-1$, and vary a large number of other parameters in the simulations, 
including galaxy gas fractions, orbital parameters, feedback prescriptions, 
and structural properties of the initial disks, halos, and bulges. 

First, it is very clear that there is no special ``division'' at the ``major'' 
merger threshold of $\mu=1/3$. 
In many models, it is simply assumed that major mergers completely destroy 
disks, while minor mergers leave them intact. 
Likewise, in many observational studies, simple assumptions such as this 
are used to relate the number of major mergers to some ``expected'' 
number of bulges. 
But in fact, to lowest order, bulge formation is continuous, with the fraction of the initial 
disk destroyed (remnant $B/T$) scaling as $B/T \sim \mu$. 
This is discussed in greater detail in \citet{hopkins:disk.survival}, 
with similar results found in a number of studies 
\citep{bournaud:minor.mergers,johansson:mixed.morph.mbh.sims,
younger:minor.mergers}, 
and some of the consequences are discussed in 
\citet{hopkins:disk.survival.cosmo,hopkins:merger.rates}. 
Such a simple assumption regarding the efficiency of mergers 
can lead to systematic factors of $\sim2-10$ differences in the 
total bulge mass formed (depending on e.g.\ the galaxy mass), 
and clearly skews the distribution of mergers impacting bulge formation. 

Also clear in Figure~\ref{fig:merger.fx.cautions} is a large 
scatter in remnant $B/T$ at fixed $\mu$, 
despite the fact that the mass ratio and number of mergers 
are {\em perfectly} well-known and defined for these simulations.
This stems from the other parameters varied in the simulations. 
Most important are the orbital parameters and gas 
fractions. 
We consider a wide variety of different orbital parameters 
(denoted by different symbols). 
Because stellar scattering and the gaseous angular momentum loss that 
drives dissipation, starbursts, and the formation of the bulge from the galaxy gas 
are dominated by resonant processes, 
orbital parameters have a huge impact 
\citep{barnes.hernquist.91,barneshernquist96}.
It is well-known that prograde 
mergers tend to be much more destructive (form much larger bulges) 
than retrograde mergers, and we see that in Figure~\ref{fig:merger.fx.cautions}. 
We also see a continuum of behavior in between, 
as has been noted in many numerical studies \citep{hernquist.89,
hernquist.mihos:minor.mergers,
naab:minor.mergers,cox:massratio.starbursts,
younger:minor.mergers,bournaud:minor.mergers,hopkins:disk.survival,
johansson:mixed.morph.mbh.sims,johansson:bh.scalings.in.remergers}. 
The effect is large, with e.g.\ a typical 1:3 merger 
leading to $B/T\approx 0.5$ in the most prograde mergers 
and just $B/T\approx 0.25$ in corresponding retrograde mergers. 
In \citet{hopkins:disk.survival}, it is noted that gas loses angular 
momentum and falls to the galaxy center, contributing to a 
nuclear starburst and bulge mass, within an initial critical radius 
$R \propto 1/(1-\psi\,\cos{\theta})$, where $\psi\sim0.5-0.6$ depends 
on structural properties and $\theta$ is the initial 
disk inclination ($\theta=0$ is prograde, $\theta=\pi$ retrograde) -- 
i.e.\ angular momentum loss is efficient out to a radius $\approx3-4$ 
times larger in prograde cases, relative to retrograde cases. 

Likewise, the gas fraction has a similar effect -- the most gas-poor 
mergers lead to very efficient bulge formation while the most 
gas-rich mergers lead to very little bulge formation even 
in extremely major mergers. 
Figure~\ref{fig:merger.fx.cautions} shows a large number of 
$\mu=0.5-0.8$ cases with only $B/T\sim0.2-0.3$, provided 
that the gas fractions at the time of the merger are 
high ($f_{\rm gas}\sim0.6-0.8$). 
Again, this has been seen in a number of works 
\citep{springel:spiral.in.merger,
robertson:disk.formation,governato:disk.rebuilding,
richard:gas.rich.merger.disk.properties,
johansson:mixed.morph.mbh.sims,johansson:bh.scalings.in.remergers}. In 
\citet{hopkins:disk.survival}, these simulations are used to show 
that, to lowest order, the efficiency of gas angular momentum 
loss scales with a factor $\propto (1-f_{\rm gas})$ -- i.e.\ 
in a merger with $50\%$ gas, the ultimate gas angular momentum 
loss per unit mass is (on average) half as efficient as in a merger with just 
$10\%$ gas. 

Other parameters lead to further scatter in $B/T$ at fixed $\mu$. 
Varying galaxy structural parameters, for example, can 
stabilize or destabilize systems \citep{springel:models}.
The presence of initial bulges 
in a certain mass and scale radius range suppresses 
gas inflows and disk heating on first passages
\citep{mihos:starbursts.94,mihos:starbursts.96}. 
Allowing gaseous disks to be more extended than stellar 
disks leads to less efficient angular momentum loss in the 
gas (because that angular momentum loss is usually dominated 
by transfer from the gas to the stars). Changing the prescriptions 
for star formation, stellar feedback, pressurization of the 
multi-phase interstellar medium, and cooling will lead to more or 
less efficient gas compression and star formation in the earlier phases of 
the merger -- this changes the absolute amount and spatial distribution 
of the gas at the time of the final merger, and thus the efficiency with 
which gas is channeled into a central starburst and bulge. 
In sufficiently minor mergers, tidal stripping of the external portions of the 
halo can effect some of the gas and stars, and this can be accelerated 
by ram-pressure effects if we model a halo with a very dense, high-pressure 
hot gas background (analogous to e.g.\ massive clusters). 
There is even the possibility, especially in very low-$\mu$ mergers, of 
complete tidal disruption of the secondary. This may be important for the 
overall mass budget in the most massive galaxies 
\citep[although that depends on the precise separation between `galaxy' and 
intra-cluster or intra-group light, which is poorly defined; see e.g.][]{dolag:icl.velocity.distrib}.

It is not our purpose here to address these effects in detail; 
we refer to \citet{hopkins:disk.survival} for a more thorough discussion. 
However, we wish to stress that if any physical {\em consequence} of 
mergers is desired, then even perfect knowledge of the 
merger rate and mass ratio distribution is only sufficient for 
an order-of-magnitude estimate of the impact of mergers. Further knowledge 
of e.g.\ the distribution of merger orbital parameters and gas fractions 
is necessary to improve estimates of the impact of mergers 
at the factor $\sim2-3$ level. And to improve to an accuracy much 
beyond a factor $\sim2$, knowledge of the much more detailed 
structural properties of galaxies and highly uncertain effects of feedback 
would be required.

\breaker
\section{Summary and Discussion: The Uncertainty Budget}
\label{sec:discuss}

A simple comparison of different predictions of the galaxy-galaxy merger 
rate from e.g.\ different semi-analytic models 
and simulations demonstrate that the predictions
vary by an order of magnitude 
(Figure~\ref{fig:mgr.rate.vs.sams}).
We have attempted to survey 
the sources of uncertainty and systematic differences between 
various theoretical attempts to predict the galaxy merger rate, to 
identify what drives these differences and address how 
progress can be made.


\subsection{Dark Matter and Dynamics}
\label{sec:discuss:budget:dark}

All models depend similarly on the ``background'' 
dark matter merger rate. However, 
with proper caution in adopting definitions, and modern convergence in 
high-resolution cosmological simulations, the uncertainties in this rate 
can be reduced to the factor $\sim2-3$ level. In other words, if how 
galaxies populate halos is appropriately fixed, there is relatively 
little uncertainty in the merger rate owing to the dynamics of the 
dark matter (Figure~\ref{fig:mgr.rate.vs.model}). 
The relevant quantities considered here include: 

{\bf Cosmology:} Nominally, changing the cosmology 
within the statistical uncertainty in modern observational 
constraints \citep{komatsu:wmap5} leads to factor $\sim1.5$ differences 
in the merger rate. However, most of this owes to changes in the 
mass function which can be normalized out -- if e.g.\ the galaxy population 
is normalized so as to match the observed stellar mass function and 
large-scale bias, then the resulting differences in the merger rate 
are negligible compared to the other sources of 
uncertainty below (Figure~\ref{fig:mgr.rate.vs.model}). 

{\bf Halo-Halo Merger Rate Determinations:} Again, 
nominally different halo-halo merger rate determinations differ by 
factors of $\sim2$. Some of this stems from e.g.\ the inherent ambiguity in 
masses and extents of halos. 
However, much owes to the definitions adopted when 
attempting to fit/quantify the instantaneous merger rate ``function.'' 
If these definitions are properly accounted for, or if halo merger 
trees are used directly to track galaxies (rather than defining halo merger 
rates at all), the real uncertainties are small, a factor $\sim1.5$. 
Effects of baryons contribute similarly small ($\sim10-20\%$) uncertainties in the overall 
halo-halo merger rate (Figure~\ref{fig:model.halorates}).
However, care is needed with definitions -- 
defining halo mergers simply in terms of ``instantaneous'' mass ratios, and/or applying 
cuts which can be useful for constructing {\em galaxy} merger trees, 
or using timestep-sensitive mass ratio definitions 
can lead to an apparent (artificial) suppression of major halo-halo mergers by 
an order-of-magnitude or more 
(Figures~\ref{fig:model.halorates.durham}-\ref{fig:model.halorates.timestepping}).
This requires careful consideration in simulation-based semi-analytic models. 

{\bf Subhalo versus Halo Merger Rates:} Defining the merger rate not 
when halos first merge into e.g.\ a larger friends-of-friends group, but when 
subhalos merge/are destroyed into the central (primary) group subhalo, 
may be more representative of galaxy-galaxy mergers. 
Doing so, however, introduces additional uncertainties owing to e.g.\ how 
subhalos are identified. 
It is also important, in this case, that the subhalo ``mass'' 
(for merger mass ratio purposes) be defined as 
the ``infall'' or maximum pre-accretion/pre-stripping mass (i.e.\ maximum 
mass before the system became a subhalo), otherwise the mass of all 
systems $\rightarrow0$ at merger, by definition. 
Different calculations from various
simulations, with different subhalo identification/destruction criteria, 
yield rates converged to within a factor $\sim2$ (Figure~\ref{fig:model.subhalorates}). 
Using hydrodynamic simulations 
to ``tag'' mergers (using the galaxies therein as the ideal subhalo/halo 
``tracers'') yields consistent results. 

{\bf Merger Timescales:} In models without subhalo resolution
or with Press-Schechter merger trees, galaxy-galaxy mergers are often 
assumed to be ``delayed'' with respect to the halo-halo merger by 
a timescale given by e.g.\ the dynamical friction time (this approximates the 
subhalo evolution). 
We show that calibrations of such timescales, from high-resolution 
galaxy merger simulations, are relatively well-converged 
(Figure~\ref{fig:tmerger.compare}). Adopting one of these calibrations, 
we show that the implied merger rate agrees well with that 
obtained from full tracking of subhalos, 
with similar factor $\sim2$ uncertainties 
(Figure~\ref{fig:model.subhalorates.tdf}). 
For {\em major} mergers, both methods agree well with the simple  
halo-halo merger rate -- this is because the timescale for such a 
merger to complete ($\ll t_{\rm Hubble}$) is short compared to the 
time between such mergers ($\sim t_{\rm Hubble}$). 

However, we show that adopting an artificially long merger timescale, 
or adopting e.g.\ the simple Chandrasekhar timescale 
with a normalization higher than the specific numerical calibrations 
here, can suppress high-redshift mergers by factors $\sim5-10$. 
We also caution that these calibrations are designed for the 
time from halo-halo merger -- they are {\em not} constructed for application in 
``hybrid'' models, where subhalos are followed in simulations down to 
some resolution limit, and then a merger time is applied to the 
residual ``orphan'' based on its instantaneous radius and mass. 
We show that in such cases, if the system is lost to resolution at 
large radius $>0.2\,R_{\rm vir}$, the calibrated formulae are not 
self-consistent, and can lead to over-estimates of the total merger 
timescale by factors of $\sim2-8$ (Figure~\ref{fig:tmerger.orphan.issues}).

\subsection{Baryonic Physics}
\label{sec:discuss:budget:baryons}

Controlling for some of the caveats and definitions above, 
the dominant uncertainties in predicted merger rates owe to 
variations in how galaxies populate halos. 
The differences can be identified with the shape of the assumed 
halo occupation distribution (HOD) -- essentially, the distribution of 
galaxy masses in given (sub)halo masses, $M_{\rm gal}(M_{\rm halo})$. 
Some model for this is necessary to translate halo-halo mergers 
(or halo-subhalo mergers) into galaxy-galaxy mergers. 
We therefore consider how these physics are modeled and 
how they lead to variations in the merger rate, in three
classes of models with very different approaches towards modeling 
the galaxy-halo connection:

{\bf Semi-Empirical (HOD-based) Models:} In 
semi-empirical models the HOD is adopted explicitly from observational 
constraints. As such, it is subject to the attendant
uncertainties and limited to the dynamic 
range observed. At low redshifts and galaxy masses within factors of 
several around $\sim L_{\ast}$, the constraints are tight and different 
methods agree well -- resulting uncertainties in the galaxy-galaxy 
merger rate (holding halo-halo merger rates {\em fixed}) 
are a factor $\sim1.5$ (Figures~\ref{fig:mgr.rate.vs.model} \&\ \ref{fig:model.HODs}). 
The uncertainties grow to a factor $\sim2$ at the lowest and highest 
masses, and $z\sim1-2$. Above $z\sim2$, the uncertainties grow 
very rapidly: there are not sufficient observational constraints on the HOD 
to make strong statements about galaxy merger rates. 
Predicted merger rates and pair counts from such models agree 
well with direct observations over a stellar mass range 
$\sim10^{9.5}-10^{11.5}\,\msun$ and 
redshifts $z=0-2$ \citep[see Figures above and][]{stewart:merger.rates,
hopkins:merger.rates}. At this level, it appears, there is no tension 
between $\Lambda$CDM merger rates and merger/pair counts. 
To the extent that other models disagree significantly with the observed 
merger fractions, it should owe to issues in the model baryonic physics 
leading to a $M_{\rm gal}(M_{\rm halo})$ distribution different from that observed.

{\bf Cosmological Hydrodynamic Simulations:}
In hydrodynamic simulations, the $M_{\rm gal}(M_{\rm halo})$ 
distribution is predicted in an 
{\em a priori} manner based on the cooling, star formation, and feedback 
models implemented in the simulation. Unfortunately, it is not yet 
possible to run large-volume cosmological hydrodynamic simulations (needed to 
quantify the galaxy-galaxy merger rate) with the spatial and mass resolution 
and detailed prescriptions for feedback from stars and black holes that 
it is becoming clear are necessary to form ``realistic'' galaxies. 
Simulations available that do not include feedback 
yield poor agreement with the observed HOD and stellar mass function, 
predicting a relationship closer to the efficient star formation limit 
($M_{\rm gal}=f_{b}\,M_{\rm halo}$). Relative to what the merger rates would be,
taking the same dynamics and merger locations but re-populating galaxies 
with masses chosen to fit the observed HOD (stellar mass and clustering), 
these simulations tend to over-predict merger rates at low masses 
and high redshifts, and under-predict rates at high masses and low redshifts 
by factors $\sim3-5$, as well as over-predicting the relative importance of 
minor versus major mergers at all masses (Figure~\ref{fig:model.HODs}). 
Because it is the {\em shape} of the $M_{\rm gal}-M_{\rm halo}$  
relation that is most important, simply re-normalizing predicted masses by a 
uniform factor will not correct for these effects. Re-normalizing 
all masses to their ``correct'' masses given some observed HOD is 
an improvement, but care is still needed, since the incorrect masses 
and morphologies will affect quantities such as the dynamical friction time. 

{\bf Semi-Analytic Models:} 
In semi-analytic models, the difficulties and expense of simulations 
are replaced by use of analytic prescriptions, given some background 
dark matter population, to predict ultimate galaxy properties. 
Such models are adjusted to give good agreement with the 
galaxy stellar mass function (and clustering) at $z=0$; 
as such, some agreement with the $M_{\rm gal}(M_{\rm halo})$ 
distribution of {\em central} galaxies (which dominate the 
stellar mass function at all masses) is implicitly guaranteed. 
Indeed, we find that adopting the predicted SAM HODs 
for central galaxies, instead of the empirically determined HOD, 
yields no systematic difference and scatter within the same factor $\sim2$ allowed 
by different observational constraints (Figures~\ref{fig:model.HODs}). 
At high redshifts, the uncertainties in both grow. Some SAMs yield 
growing discrepancies relative to semi-empirical models, directly related to 
issues such as e.g.\ the known tendency of SAMs to over-predict the 
abundance of low-mass galaxies at high redshift, 
but these are still within the factor $\sim3-5$ level at $z<3$ 
(Figure~\ref{fig:model.HODs.z}). 

However, the HOD of {\em satellite} galaxies in semi-analytic models is 
extremely sensitive to prescriptions for cooling, stellar feedback, 
and halo mixing/stripping in satellites. Moreover, satellite masses 
are not strongly constrained by the stellar mass function, so there is 
no implicit guarantee/check that these are correct, and more detailed 
comparison with e.g.\ observed group catalogs and small-scale clustering 
must be used to calibrate the models. 

In detail, it is well-known that most SAMs have difficulty 
reproducing the observed properties of 
satellite galaxy populations: they tend to predict satellite galaxies 
that are under-massive and over-quenched, relative to observations 
\citep[see Figures~\ref{fig:model.HODs.sat} \&\ \ref{fig:sat.overquenching} 
and e.g.][and references therein]{
weinmann:obs.hod,weinmann:group.cat.vs.sam,
wang:sat.shutdown.slower.vs.sam,kimm:passive.sats.vs.centrals}. 
This can occur even in models where the small-scale clustering of 
satellites is over-predicted, relative to observations 
\citep[see e.g.][who find both effects, although the 
clustering discrepancy is probably due to the large 
value of $\sigma_{8}$ used in the simulation]{guo:2010.millenium.2.update}. 
In most of the models, satellites lose their entire halo gas reservoir at the 
moment they become such -- even in major mergers (and even 
when the moment of ``becoming'' a satellite occurs at several times the primary virial 
radius). Moreover, the combination of simple stellar wind feedback and 
cooling models often leads to the satellites {\em also} losing almost all of their 
cold/disk gas reservoir. 

As such, at low masses 
where gas fractions are important, initially major and even equal-mass 
mergers can easily become minor mergers in the model, potentially 
suppressing the predicted merger rate by a large factor. 
Models which yield more ``over-quenched'' satellite populations tend to yield 
correspondingly smaller merger rates (Figures~\ref{fig:model.HODs.sat} 
\&\ \ref{fig:font.vs.bower}). 
Correcting for these differences, for example by enforcing that satellite 
galaxies obey a similar HOD to central galaxies, 
or by adjusting the prescriptions for cooling onto satellite galaxies such that 
they better reproduce the observed color and star formation rate 
distributions of satellites, leads to larger merger rates that converge 
with the merger rates predicted from semi-empirical models  
(Figure~\ref{fig:font.vs.bower}). 
Whether or not this is necessary for matching e.g.\ close pair counts and 
small-scale clustering is unclear \citep[see e.g.\, the comparison in][]{kitzbichler:mgr.rate.pair.calibration,
guo:2010.millenium.2.update}, 
but it demonstrates an important 
uncertainty in predictions of merger rates.

\subsection{Impact of Mergers on Galaxies}
\label{sec:discuss:budget:impact}

Of course, even with perfect knowledge of the merger rate and 
distribution of mass ratios under some definition, it is not trivial to 
relate this to either observable or physical quantities such as merger 
fractions or the amount of bulge mass formed in mergers. 
We consider how two basic aspects of this relate to uncertainties 
in merger rates and their consequences: 

{\bf Mass Ratio Definitions:} 
As outlined in \citet{stewart:massratio.defn.conf.proc}, 
a halo-halo major merger is not necessarily a 
galaxy-galaxy major merger, and vice versa. 
But there are also several means of defining galaxy-galaxy 
mass ratio, including e.g.\ the stellar-stellar mass ratio $\mu_{\ast}$ 
and baryon (stellar mass plus cold gas within the galaxy disk)-baryon mass ratio 
$\mu_{\rm gal}$. 
At high (low) masses, the merger rate in terms of stellar-stellar 
mass ratio is enhanced (suppressed) by an order of magnitude 
relative to the halo-halo merger rate (Figure~\ref{fig:merger.fx.cautions}). 
In comparing e.g.\ model predictions to stellar mass ratio-selected pair 
samples, clearly the stellar mass ratio is most applicable. 
However, in comparing to luminosity-ratio selected pair samples, 
or to morphological studies, where the 
total baryonic mass is what matters, 
the merger rate in terms of the baryon-baryon mass ratio is more relevant. 
At high masses, this behaves similarly to the merger 
rate in terms of the stellar-stellar mass ratio. 
At low masses, however, galaxies are increasingly gas rich -- 
thus many mergers with minor stellar-stellar mass ratios 
in fact have major baryon-baryon mass ratios, and the baryon-baryon 
major merger rate is a factor $\sim3$ higher than the stellar-stellar merger rate. 
This may, in fact, explain at least part of the well-established 
fact that morphology-inferred merger rates at low masses 
tend to be systematically higher than pair count-inferred merger rates 
\citep[see e.g.][and references therein]{lopezsanjuan:merger.fraction.to.z1,
lopezsanjuan:mgr.rate.pairs}.

Although often a superior definition, even 
the baryonic mass ratio misses the tightly-bound dark matter 
that is important for the dynamics of final merger, 
which will not be stripped because it is within the baryonic radii. 
We therefore propose a new 
``dynamical'' mass and mass ratio which approximates the 
most important quantity in high-resolution simulations, and includes 
both baryons and some dark matter. Merger rates in terms of this 
quantity behave similarly to baryonic major mergers, 
but with somewhat ($\sim20-50\%$) higher (lower) rates at low (high) masses 
(Figure~\ref{fig:merger.fx.cautions}). 

If one wishes to infer the amount of bulge or other ``damage'' done by 
major mergers but uses {\em either} the stellar-stellar mass 
ratio or halo-halo mass ratio, the estimate can easily be systematically 
incorrect at the factor $\sim3-10$ level. 
This is critical for observational studies seeking to infer the 
amount of bulge formed by major mergers, using 
a stellar-mass-ratio selected sample \citep[see e.g.][]{bundy:merger.fraction.new}; 
at low masses, there might be $\sim3$ times as many ``damaging'' 
mergers as given by this statistic. 
Also, many analytic models simply use the stellar (or even halo) mass 
ratio as a criterion for determining the impact of a merger -- 
the systematic errors introduced by this 
assumption can be {\em larger} than those from 
halo mis-identification, complete ignorance of subhalos, assignment of 
incorrect merger timescales, or discrepancies between model and observed 
stellar mass functions. Yet despite the considerable literature on those 
problems, there has been relatively little focus on the 
adoption of better mass ratio proxies.

{\bf Other Parameters and Merger `Impact':} 
Even when controlling for the above uncertainties, we 
show that at perfectly well-defined merger rates and merger mass ratios, 
there is a large variation in the physical effects of mergers, 
owing to other parameters such as the merger 
dynamics and orbit, merger gas fractions, and initial structural 
properties of the merging systems 
(Figure~\ref{fig:merger.fx.cautions}). 
At the same merger mass ratio, a prograde orbit can build twice as 
much bulge as a retrograde orbit, and will lead to much more dramatic 
tidal features and distortions observable in morphology-selected samples.
Also, the merger timescale will be significantly different at moderate/small 
radii $<100\,$kpc, so pair-selected samples will also see biased 
distributions. 
A gas-poor major merger will typically violently relax the 
entire stellar disk and funnel the gas entirely into a starburst that builds 
central bulge mass, but a very gas-rich merger will experience very inefficient 
angular momentum transfer, suppressing the burst mass by a factor 
$\sim (1-f_{\rm gas})$, a factor $\sim3-5$ at low masses and high redshifts. 
More subtle properties lead to scatter and offsets at a smaller, but 
non-negligible level. Many of these are discussed in detail in 
\citet{hopkins:disk.survival}. 

From a purely empirical perspective, without knowledge of these properties, it is 
difficult to assess the impact of mergers (in terms of e.g.\ the bulge mass formed) 
at better than the factor $\sim2-3$ level. 
From the theoretical perspective, neglecting quantities such as orbital parameters
and, especially, gas fractions, in forward-modeling merger remnants again 
will introduce systematic uncertainties that are larger than 
any of the uncertainties from modeling the dark matter distribution, 
subhalos, merger timescales, and the like. 
Many of these are outlined in \citet{hopkins:disk.survival.cosmo}; 
at high masses, the systematic uncertainties are less severe, but at low 
masses, where systems tend to be gas-rich, they can be factors $\sim5-10$ 
in the total bulge mass formed. 

Another important point is that mergers are continuous -- there is 
no special division at the traditional 1:3 distinction between 
``major'' and ``minor'' mergers. 
In many models, and in many observational assessments of merger 
``effects,'' it is assumed that major mergers are completely 
destructive, while minor mergers do no damage. 
This simple assumption introduces systematic factor $\sim2$ 
errors in the total merger ``damage budget'' and total bulge 
formation efficiency \citep[see][]{hopkins:merger.rates}; 
the effect in terms of skewing which mergers do ``more'' or ``less'' 
for bulge formation can obviously be severe.

\subsection{Observational Constraints and Outlook}
\label{sec:discuss:obs}

Testing these models and determining the true merger rate in a robust 
manner will of course ultimately depend on observations. 
Continued observations of the merger rate, bearing 
in mind the caveats above, are of obvious importance. In improving such 
estimates, calibration of specific samples to high-resolution $N$-body 
simulations, {\em specifically 
with mock observations matched to the exact selection and methodology adopted}, 
will be critical \citep[see e.g.][]{lotz:merger.selection}. 
And even with such a calibration, the discussion above makes it clear why 
the relation between observed merger rates and bulge buildup (depending 
on a number of secondary properties not directly observed), in detail, 
must be {\em forward-modeled}. 

Tighter observational constraints on the halo occupation distribution -- 
from e.g.\ group catalogs, kinematics, weak lensing, 
and clustering -- 
in particular at low masses and at 
high redshifts, will directly improve the semi-empirical models, and 
put strong constraints on the a priori galaxy formation models in the 
areas that have greatest effect on predicted merger rates. 
As we have shown, constraints on satellite populations specifically 
will be valuable. However, the satellite populations of particular interest are 
not the historically well-studied extreme cases of e.g.\ dwarfs in the local 
group or low-mass galaxies in Virgo and massive clusters 
(from which much of our intuition regarding dynamical friction, 
stripping, and satellite gas exhaustion comes). 
The case of interest for merger rates 
is that of major mergers (i.e.\ near equal-mass galaxies) in field or 
loose group environments (i.e.\ systems analogous to the local group, but where 
either the Milky Way or Andromeda is the ``satellite'' of interest). 
Also, at low stellar masses and high redshifts, large uncertainties remain in galaxy gas masses, 
and these matter as much or more relative to the stellar mass 
in assessing the ``impact'' of mergers. 

\acknowledgments
We thank Simon White, Gabriella de Lucia, 
Owen Parry, Carlos Frenk, Andrew Benson, Shardha Jogee, 
Thorsten Naab, Eyal Neistein, Simone Weinmann, Volker Springel, 
Martin White, Joanne Cohn, Carrie Bridge, Jennifer Lotz, T.~J.\ Cox, 
and Eliot Quataert
for helpful discussions throughout the development of this work.
Support for PFH was provided by the Miller Institute for Basic Research 
in Science, University of California Berkeley.

\bibliography{/Users/phopkins/Documents/lars_galaxies/papers/ms}

\end{document}